%% file: draft.tex
\documentclass[twocolumn,dvipsnames]{aastex62}
\usepackage{xcolor}
\usepackage{amsmath}
\usepackage{appendix}
\usepackage{makecell}
\usepackage{relsize}
\setcellgapes{4pt}
\usepackage[caption=false, position=top]{subfig}
\usepackage{hyperref}
\usepackage{xspace}
\usepackage{bm}
\usepackage{multirow}
\usepackage{array}

\newcommand{\new}[1]{\ignorespaces#1}

\newcommand{\chiEff}{\ensuremath{\chi_{\rm eff}}\xspace}
\newcommand{\chieff}{\ensuremath{\chi_{\rm eff}}\xspace}
\newcommand{\chiP}{\ensuremath{\chi_\mathrm{p}}\xspace}

\newcommand{\zetaSpike}{\zeta_\mathrm{spike}}

\input{macros}
\input{spin_summary_macros.txt}

\newcommand{\CCA}{\affiliation{Center for Computational Astrophysics, Flatiron Institute, 162 Fifth Avenue, New York, NY 10010, USA}}
\newcommand{\CIT}{\affiliation{Department of Physics, California Institute of Technology, Pasadena, California 91125, USA}}
\newcommand{\CITLab}{\affiliation{LIGO Laboratory, California Institute of Technology, Pasadena, California 91125, USA}}
\newcommand{\SBU}{\affiliation{Department of Physics and Astronomy, Stony Brook University, Stony Brook NY 11794, USA}}

\shortauthors{Callister et al.}

\begin{document}

\title{No evidence that the majority of black holes in binaries have zero spin}

\author[0000-0001-9892-177X]{Thomas A. Callister}
\CCA{}
\email{tcallister@flatironinstitute.org}
\author[0000-0001-5670-7046]{Simona J. Miller}
\email{smiller@caltech.edu}
\CIT
\CITLab
\author[0000-0002-5833-413X]{Katerina Chatziioannou}
\email{kchatziioannou@caltech.edu}
\CIT
\CITLab
\author[0000-0003-1540-8562]{Will M. Farr}
\email{will.farr@stonybrook.edu}
\CCA{}
\SBU{}

\begin{abstract}
The spin properties of merging black holes observed with gravitational waves can offer novel information about the origin of these systems.
The magnitude and orientations of black hole spins offer a record of binaries' evolutionary history, encoding information about massive stellar evolution and the astrophysical environments in which binary black holes are assembled.
Recent analyses of the binary black hole population have yielded conflicting portraits of the black hole spin distribution.
Some work suggests that black hole spins are small but non-zero and exhibit a wide range of misalignment angles relative to binaries' orbital angular momenta.
Other work concludes that the majority of black holes are non-spinning while the remainder are rapidly rotating and primarily aligned with their orbits.
We revisit these conflicting conclusions, employing a variety of complementary methods to measure the distribution of spin magnitudes and orientations among binary black hole mergers.
We find that the existence of a sub-population of black hole with vanishing spins is not required by current data.
Should such a sub-population exist, we conclude that it must contain $\lesssim 60\%$ of binaries.
Additionally, we find evidence for significant spin-orbit misalignment among the binary black hole population, with some systems exhibiting misalignment angles greater than $90^{\circ}$, and see no evidence for an approximately spin-aligned sub-population. \\
\end{abstract}

\section{Introduction}
\label{sec:intro}

The spins of black holes in merging binaries detected with gravitational waves promise to illuminate open questions in massive stellar evolution and compact binary formation.
The orientations of component black hole spins may differentiate between binaries formed via isolated stellar evolution and those formed dynamically in clusters or the disks of active galactic nuclei, and additionally offer a means of measuring natal kicks that black holes receive upon their formation~\citep{Rodriguez2016,Farr2017,Gerosa2017,oshaughnessy_inferences_2017,2017CQGra..34cLT01V,Gerosa2018,Liu2018,wysocki_explaining_2018,fragione_effective_2019,mckernan_monte-carlo_2019,Steinle2021,O3a-pop,2021ApJ...920..157C}.
Spin magnitudes, meanwhile, are determined by poorly-understood angular momentum processes operating in stellar cores, and may be further affected by binary processes such as tidal torques or mass transfer~\citep{qin_spin_2018,bavera_origin_2020,bavera_impact_2021,Steinle2021,Zevin:2022wrw}.
Black holes with large spin magnitudes might also point to hierarchical assembly in dense environments, involving component black holes that are themselves the products of previous mergers~\citep{Gerosa2017,Doctor2019,Rodriguez2019,2020ApJ...900..177K,2021ApJ...915L..35K,2021NatAs...5..749G}.

Despite the large astrophysical interest, spin measurements are hampered by the fact that spin dynamics have a relatively weak imprint on the gravitational-wave signal.
The main effect of spins (anti)parallel to the Newtonian orbital angular momentum is to (speed-up) slow-down the binary inspiral and merger.
Spins in the plane of the orbit, on the other hand, give rise to precession that modulates the amplitude and phase of emitted gravitational waves.
Even with informative measurements of these effects, however, it is not straightforward to constrain all six degrees of freedom independently.

Recent work~\citep{O3a-pop,Roulet2021,Galaudage2021} has yielded conflicting conclusions regarding the distribution of spins among the binary black hole population witnessed by Advanced LIGO~\citep{aasi_advanced_2015} and Virgo~\citep{acernese_advanced_2015}.
Specifically:
\begin{itemize}
\item Do binary black holes have small but non-zero spins that may be misaligned significantly with the Newtonian orbital angular momentum, i.e., with spin-orbit misalignments $>90^\circ$?
\item Or, do a majority of binaries have spins that are identically zero or, if nonzero, preferentially aligned with the Newtonian orbital angular momentum, i.e., with misalignments $<90^\circ$?
\end{itemize}
These questions highlight the subtleties and difficulties inherent in statistical analysis of weakly informative measurements.
This debate also hinges on a variety of \emph{technical difficulties} related to hierarchical inference of narrow population features using discretely sampled data.

The two possibilities listed above carry considerably different astrophysical implications for the assembly and evolution of binary black hole mergers.
Systems arising from isolated binary evolution are traditionally expected to have spins preferentially parallel to their orbital angular momenta~\citep{belczynski_black_2008,qin_spin_2018,zaldarriaga_expected_2018,bavera_origin_2020}.
Significant spin-orbit misalignment, in contrast, is considered difficult to achieve under canonical isolated binary evolution and so would indicate either the presence of
alternative formation channels or a change in the paradigm of isolated binary
evolution~\citep{Rodriguez2016,Farr2017,Steinle2021,2021ApJ...920..157C,2022arXiv220502541T}. The degree to which black holes are observed to be primarily spinning
or non-spinning, meanwhile, would support or refute theories positing highly efficient
angular momentum transport in stellar interiors \citep{Spruit1999,Fuller2015,Fuller2019}.

Our goal in this paper is to revisit these incompatible conclusions.
In Sec.~\ref{sec:measurement} we review recent literature and describe in more detail what is known, unknown, and still debated about binary black hole spins.
We then employ a string of increasingly sophisticated analyses to study the population of binary black hole spins and determine what we can and cannot robustly conclude about the magnitudes and orientations of black hole spins.
We begin in Sec.~\ref{sec:counting} with a simple counting argument, which demonstrates that the fraction of black holes that are non-spinning is consistent with zero.
We then turn to full hierarchical analyses of the binary black hole population, studying the distributions of effective inspiral spins (Sec.~\ref{sec:spike}) and component spin magnitudes and orientations (Sec.~\ref{sec:results}).
In all cases, we find that the fraction of non-spinning black holes can comprise up to $60-70\%$ of the total population, but that this fraction cannot be confidently bounded away from zero.
Overall, the inferred spin magnitude distribution is consistent with a single population extending smoothly from zero up to magnitudes of approximately $0.4$.
Additionally, we find a preference for considerable spin-orbit misalignments among the binary black hole population, with some spins inclined by more than $90^\circ$ relative to their orbits.

\section{The spins of black holes in binaries}
\label{sec:measurement}

Each component black hole in a binary has dimensionless spin vectors $\bm{\chi}_1$ and $\bm{\chi}_2$.
Six parameters are needed to fully specify these two spins, with each spin vector characterized by a magnitude $0\leq\chi_i\leq1$, tilt angle $\theta_i$, and azimuthal angle $\phi_i$, where $i\in\{1,2\}$ in a coordinate system where the z-axis is aligned with the Newtonian orbital angular momentum.

Not all spin degrees of freedom are as dynamically important in a gravitational-wave signal, however.
While the two spin magnitudes are conserved throughout the binary evolution~\citep[modulo horizon absorption effects;][]{Poisson:2004cw,Chatziioannou:2012gq}, the four spin angles vary due to spin-precession. A combination known as the \emph{effective inspiral spin} is conserved under spin-precession to at least the 2PN order\footnote{An $N$PN order is defined as being proportional to $(u/c)^{2N}$ compared to its leading order term, where $u$ is a characteristic velocity of the system and $c$ is the speed of light.}~\citep{Racine:2008qv}
\begin{equation}
    \chi_\mathrm{eff} = \frac{\chi_1 \cos\theta_1 + q\chi_2 \cos\theta_2}{1+q}\in [-1,1]\,,
\end{equation}
where $q = m_2/m_1<1$ is the mass ratio.
Though not conserved under spin-precession, the \emph{effective precessing spin}
    \begin{equation}
    \chi_\mathrm{p} = \mathrm{max}\left[\chi_1 \sin\theta_1,
        \left(\frac{3+4q}{4+3q}\right) q \,\chi_2 \sin\theta_2\right]\in [0,1]\,,
    \end{equation}
reflects the degree of in-plane spin and characterizes spin-precession dynamics~\citep{Schmidt:2014iyl}.~\footnote{
Recent work has also explored alternative parameters which better capture the imprint of spin-precession in gravitational-wave signals~\citep{Gerosa2021,Thomas2021}.
}
Although modern waveform models make use of the full 6-dimensional spin parameter space~\citep{Khan:2018fmp,Varma:2019csw,Ossokine:2020kjp,Pratten:2020ceb}, earlier versions were constructed in terms of $\chiEff$ and $\chiP$, leveraging their relevance in binary dynamics.

At current signal strengths it is not possible to meaningfully constrain all $6$ spin parameters.  
When exploring the population of compact binary spins, we therefore generally work in one of two lower-dimensional spaces.
\begin{itemize}
\item The most straightforward approach is to constrain the distribution of ``effective" spin parameters.
Though the $\chiEff$ and $\chiP$ distributions do not unambiguously reveal information about individual component spins, they do enable \textit{categorical} conclusions to be made regarding compact binary spins.
A non-vanishing $\chiP$ distribution, for example, indicates that spins are not perfectly aligned with binary orbits, while the identification of negative $\chiEff$ requires at least some component spins to be inclined by more that $\theta_i=90^\circ$.
A common approach, and the baseline model that we will extend below, is to treat the marginal distribution of $\chiEff$ as a truncated Gaussian~\citep{Roulet2019,Miller2020}. We refer to this as the \texttt{Gaussian} model; see Appendix~\ref{app:effectivemodels}.
\item Going one step beyond the effective spin parameters, we can directly model the distribution of component spin magnitudes and tilt angles.
A popular choice is to treat the component spin magnitude distribution as a Beta distribution, while spin tilts are drawn from a mixture between two components, an isotropic component and a preferentially aligned component~\citep{Talbot:2017yur,Wysocki2019}. The azimuthal spin angles are ignored and presumed to be uniformly distributed as $\phi_1,\phi_2\sim U[0,2\pi]$~\citep[this assumption was relaxed in][]{Varma:2021xbh}. We assume that component spin magnitudes and tilts are independently and identically distributed within this model and refer to it as the {\tt Beta+Mixture} model,\footnote{A closely related model in which the spin tilts are not independently and identically distributed but instead \textit{both} originate from either the isotropic or the aligned component is called the \texttt{Default} spin model in~\citet{O3a-pop} and~\citet{O3b-pop}.} discussed further in Appendix~\ref{app:componentmodels}.
\end{itemize}

\subsection{What we know about black hole spins}

Before proceeding to explore the prevalence of zero-spin events and extreme spin-orbit misalignment (i.e., spin tilts larger than $90^{\circ}$), it is useful to first examine the conclusions that are broadly and robustly recovered by different analyses and authors.

\textit{i. Black hole spins are not all maximal}.
Following the first four binary black hole detections by Advanced LIGO, \citet{Farr2017} and \citet{Farr2018} determined that if all component spins are aligned ($\cos\theta_i=1$), then their magnitudes must be small, $\chi_i \lesssim 0.3$.
If spins were assumed to be isotropically-oriented, though, near-extremal spins were still allowed.
\citet{Tiwari2018} conducted a similar analysis using an expanded catalog, and also found large spins to be disfavored, regardless of their orientation.
The degeneracy between spin magnitude and orientation was later broken by efforts to simultaneously measure these two properties.
Using the \texttt{Default} model, \cite{Wysocki2019} and \citet{O2-pop} found that typical spin magnitudes are small, with 50\% of black holes having $\chi\lesssim 0.3$.
\citet{O2-pop} furthermore revisited the analysis of \citet{Farr2018}, now finding that large spin magnitudes are moderately disfavored \emph{even when} assuming isotropic orientations.
\citet{Roulet2019}, meanwhile, studied the $\chiEff$ distribution, leveraging the \texttt{Gaussian} model to conclude that effective spins are concentrated about zero.
They argued that, if component spins are aligned, then the measured $\chiEff$ distribution implies that component spin magnitudes are $\chi\lesssim 0.1$.
\citet{Roulet2019} additionally measured the fraction of binaries whose secondaries have maximal spins due to tidal spin-up; they found the fraction to be consistent with zero and bounded to $<0.3$.
To date, no confident detection has exhibited unambiguously large $\chi \gtrsim 0.5$.

\textit{ii. Black hole spins are not all zero}.
Despite evidence pointing towards preferentially small spin magnitudes, not \textit{all} black holes can be non-spinning.
Using the \texttt{Gaussian} effective spin model, \citet{Roulet2019} and \citet{Miller2020} concluded that the $\chiEff$ distribution is \textit{inconsistent} with a delta-function at zero; hence the $\chiEff$ distribution possessed either a non-zero mean or non-zero width.
This conclusion was bolstered by \citet{O3a-pop}, who found that the $\chiEff$ distribution is centered at $\sim0.05$ with a non-zero width $\geq 0.08$, and that the component spin magnitude distribution peaks at small values but also with a non-vanishing width $\sim0.15$. 
Several individual events such as GW151226~\citep{LIGOScientific:2016sjg} and GW190517~\citep{LIGOScientific:2020ibl} are also confidently known to possess spin, although their component spins are individually poorly measured (see e.g. Fig.~\ref{fig:good_versus_bad_Neff_events}).

\textit{iii. Black holes exhibit a range of spin-orbit misalignment angles}.
Since spin-precession is a subtle effect, the spin tilts of individual binaries are highly uncertain.
Analyses of the population, however, indicate that spins are not purely aligned but instead exhibit a range of misalignment angles.
In their analysis of the $\chiEff$ distribution, \citet{Tiwari2018} reported evidence against pure spin-orbit alignment.
\citet{O3a-pop} later employed the \texttt{Default} model to directly measure the distribution of misalignment angles, recovering a possible preference for alignment but ruling out perfect alignment at high credibility.
\citet{O3a-pop} furthermore extended the \texttt{Gaussian} model to jointly measure the mean and variance of both $\chiEff$ and $\chiP$.
They found a delta-function at $\chiP=0$ to be disfavored, indicating the presence of spin-orbit misalignment in the population.
\citet{Galaudage2021}, meanwhile, used an extended version of the \texttt{Default} model to measure the \textit{maximum} spin-orbit misalignment angle among the binary black hole population, finding that the observed population requires some misalignment angles exceeding $\theta\gtrsim65^\circ$ ($\cos{\theta}\leq 0.43$) at 99\% credibility.
Evidence for spin precession identified in individual events is also under active investigation~\citep{GW190412,2021arXiv210506486C,2021PhRvD.103j4027I,2022arXiv220313406V,Mateu-Lucena:2021siq,2022ApJ...924...79E,2021arXiv211211300H,hoy_evidence_2022}.

\subsection{What is under debate about black hole spins}
\label{subsec:questions}

\textit{i. Do most black holes have zero spin?}
Although it is agreed that \textit{not all} black holes can possess zero spin, a debated question is whether \textit{most} do.
Using both the {\tt Default} and {\tt Gaussian} models, \citet{O3a-pop} found no indication of an excess of zero-spin systems;
predictive checks designed to test the goodness-of-fit of these models found these continuous unimodal distributions to be good descriptors of the observed population.
In the context of hierarchical black hole formation,~\citet{2020ApJ...900..177K} and~\citet{2021ApJ...915L..35K} directly measured the fraction of ``first-generation'' black holes with zero spin, finding this fraction to be consistent with zero.
A different conclusion was drawn in~\citet{Roulet2021}, who modeled the $\chiEff$ distribution not as a single Gaussian but via a mixture of three components,
    \begin{equation}
    \begin{aligned}
    &p(\chiEff|\zeta_\mathrm{pos},\zeta_0,\sigma)
        = \zeta_\mathrm{pos}\, {\cal{N}}_{[0,1]}(\chiEff|0,\sigma) \\
        &\hspace{2cm} + \zeta_0 \, {\cal{N}}_{[-1,1]}(\chiEff|0,0.04) \\
        &\hspace{2cm} + (1-\zeta_\mathrm{pos}-\zeta_0)\, {\cal{N}}_{[-1,0]}(\chiEff|0,\sigma)\,,
    \end{aligned}
    \label{eq:roulet}
    \end{equation}
corresponding to a half-Gaussian encompassing $\chiEff>0$, a half-Gaussian encompassing $\chiEff<0$, and a Gaussian centered at zero.
This third component was intended to capture systems with $\chiEff=0$; its finite standard deviation (fixed to 0.04) was chosen to mitigate sampling effects, a technical issue we discuss further below.
\citet{Roulet2021} argued that a significant fraction of observed binaries are possibly associated with this zero-spin sub-population.
They reported a maximum likelihood value of $\zeta_0 \approx 0.5$, although $\zeta_0$ remained consistent with zero.
A similar but stronger conclusion was forwarded by~\citet{Galaudage2021}.
Working in the component spin domain, they extended the {\tt Default} model to include an additional sub-population whose spin magnitudes are identically zero for both binary components.
\citet{Galaudage2021} concluded that $54^{+21}_{-25}\%$ of binaries are members of the zero-spin sub-population at 90\% credibility.\footnote{During the late stages of preparation of this study, the exact numerical results of \citet{Galaudage2021} were updated to account for an analysis bug, though their main conclusions remain unchanged.
Numbers in this manuscript correspond to their updated results.}

\textit{ii. Do some black holes have large spin magnitudes?}
\new{\citet{O3a-pop} performed a series of predictive checks, testing the goodness of fit of their models against observation.
They concluded that black hole spins are well described by a single unimodal distribution concentrated at small-but-nonzero values.
In contrast,~\citet{Galaudage2021} argued that, although they infer most binary black holes to be non-spinning, the remaining binaries are members of a distinct rapidly-spinning sub-population.
This secondary population is claimed to exhibit a broad range of spin magnitudes, centered at $\chi\approx0.45$ but extending to maximal spins.
\citet{hoy_understanding_2022} noted that some individual events exhibit more confidently positive spin than others, speculating that they comprise a secondary population of more rapidly spinning events.
However, their results are based on inspection of individual posteriors and not hierarchical inference of the underlying population, and so it is unclear how their conclusions compare to those of~\citet{Galaudage2021}.}

\textit{iii. Do extreme spin-orbit misalignments exist?}
Although all analyses agree that at least a moderate degree of spin-orbit misalignment exists, the question of \textit{extreme} misalignments, i.e., $\theta>90^\circ$, remains.
Using the {\tt Gaussian} model, \citet{O3a-pop} inferred that at least 12\% of binaries have negative $\chiEff$, possible only if one or both component spins have $\theta>90^\circ$.
To determine whether this conclusion was a proper \textit{measurement} or simply an extrapolation of their model, \cite{O3a-pop} introduced a variable lower bound $\chi_\mathrm{eff,min}$ on the Gaussian $\chiEff$ distribution, finding the data to require $\chi_\mathrm{eff,min}<0$ at 99\% credibility.
\citet{Roulet2021}, however, found that this support for negative effective spins is diminished when allowing for the possibility of a zero-spin population as in Eq.~\eqref{eq:roulet}.
\citet{Galaudage2021}, meanwhile, explored another variant of the {\tt Default} component spin model, now introducing a variable truncation bound on the spin-tilt distribution.
They found this minimum $\cos\theta$ to be consistent with zero.
Motivated by these analyses, \citet{O3b-pop} further extended the {\tt Gaussian} model to include both a variable truncation bound and a possible zero-spin sub-population.
They recovered diminished evidence for negative effective spins, but still recovered a preference for $\chi_\mathrm{eff,min}<0$, now at 90\% credibility.
To date, no individual events discovered by the LIGO-Virgo-KAGRA Collaboration have confidently negative $\chiEff$ or component spins unambiguously inclined by more than $90^\circ$~\citep{LIGOScientific:2021djp}. 
Independent reanalyses of LIGO/Virgo data have identified several candidates with confidently negative $\chiEff$~\citep{2020PhRvD.101h3030V,2022arXiv220102252O}, although most of these candidates do not pass the significance threshold adopted in~\citet{Roulet2021}.

\section{A counting experiment}
\label{sec:counting}

The central question of whether the majority of detected black hole binaries have vanishing spins admits a quick back-of-the-envelope estimate. 
Fully marginalized likelihoods\footnote{Sometimes known as ``evidence," though in this paper we reserve this term for non-technical use.}
have been obtained for every event in GWTC-2~\citep{LIGOScientific:2020ibl,2021ApJ...915L..35K} under two different prior hypotheses: (i) both binary components are non-spinning (NS), with spin magnitudes fixed to $\chi_{1,2} = 0$, and (ii) the black holes are spinning (S), with \new{spin magnitudes and cosine tilts distributed uniformly across the intervals} $0 \leq \chi_{1,2} \leq 1$ and $-1 \leq \cos\theta_{1,2} \leq 1$. 
The ratio of the fully marginalized likelihoods gives the Bayes factor $\mathcal{B}^\mathrm{NS}_\mathrm{S}$ between non-spinning and spinning hypotheses.
Such Bayes factors serve as a primary input in the analysis of~\citet{Galaudage2021}, which makes use of parameter estimation samples obtained under both the non-spinning and spinning priors; the Bayes factors between hypotheses is critical in determining how to properly combine these samples.

We start by considering a simple one-parameter model for the fraction of non-spinning binaries, $\zeta$.
Given a catalog of $N_\mathrm{obs}$ observations and data $\{d\}$, the likelihood of $\{d\}$ is
    \begin{equation}
    \begin{aligned}
    p(\{d\} | \zeta)
        &= \prod_{i=1}^{N_\mathrm{obs}} p(d_i|\zeta) \\
        &= \prod_{i=1}^{N_\mathrm{obs}} \big[p(d_i|\mathrm{NS})\,p(\mathrm{NS}|\zeta)
            + p(d_i|\mathrm{S})\,p(\mathrm{S}|\zeta)\big]\,,
    \end{aligned}
    \label{eq:mixture-model}
    \end{equation}
where in the second line we have written the likelihood for each individual event as the sum of two terms corresponding to the non-spinning (NS) and the spinning (S) hypothesis.
By definition, $p(\mathrm{NS}|\zeta) = \zeta$ and $p(\mathrm{S}|\zeta) = 1-\zeta$.
Substituting these expressions into Eq.~\eqref{eq:mixture-model} gives
    \begin{equation}
    \begin{aligned}
    p(\{d\} | \zeta)
        &= \prod_{i=1}^{N_\mathrm{obs}} \big[p(d_i|\mathrm{NS})\,\zeta
            + p(d_i|\mathrm{S})\,(1-\zeta)\big] \\
        &= \prod_{i=1}^{N_\mathrm{obs}} p(d_i|\mathrm{S}) \left[\frac{p(d_i|\mathrm{NS})}{p(d_i|\mathrm{S})}\,\zeta
            + (1-\zeta)\right] \\
        &\propto \prod_{i=1}^{N_\mathrm{obs}} \big[\mathcal{B}^\mathrm{NS}_{\mathrm{S},i}\,\zeta
            + (1-\zeta)\big]\,,
    \end{aligned}
    \label{eq:mixture-model2}
    \end{equation}
where the likelihood ratio $p(d_i|\mathrm{NS})/p(d_i|\mathrm{S})$ is the non-spinning vs. spinning Bayes factor $\mathcal{B}^\mathrm{NS}_{\mathrm{S},i}$.

\begin{figure}
    \includegraphics[width=0.49\textwidth]{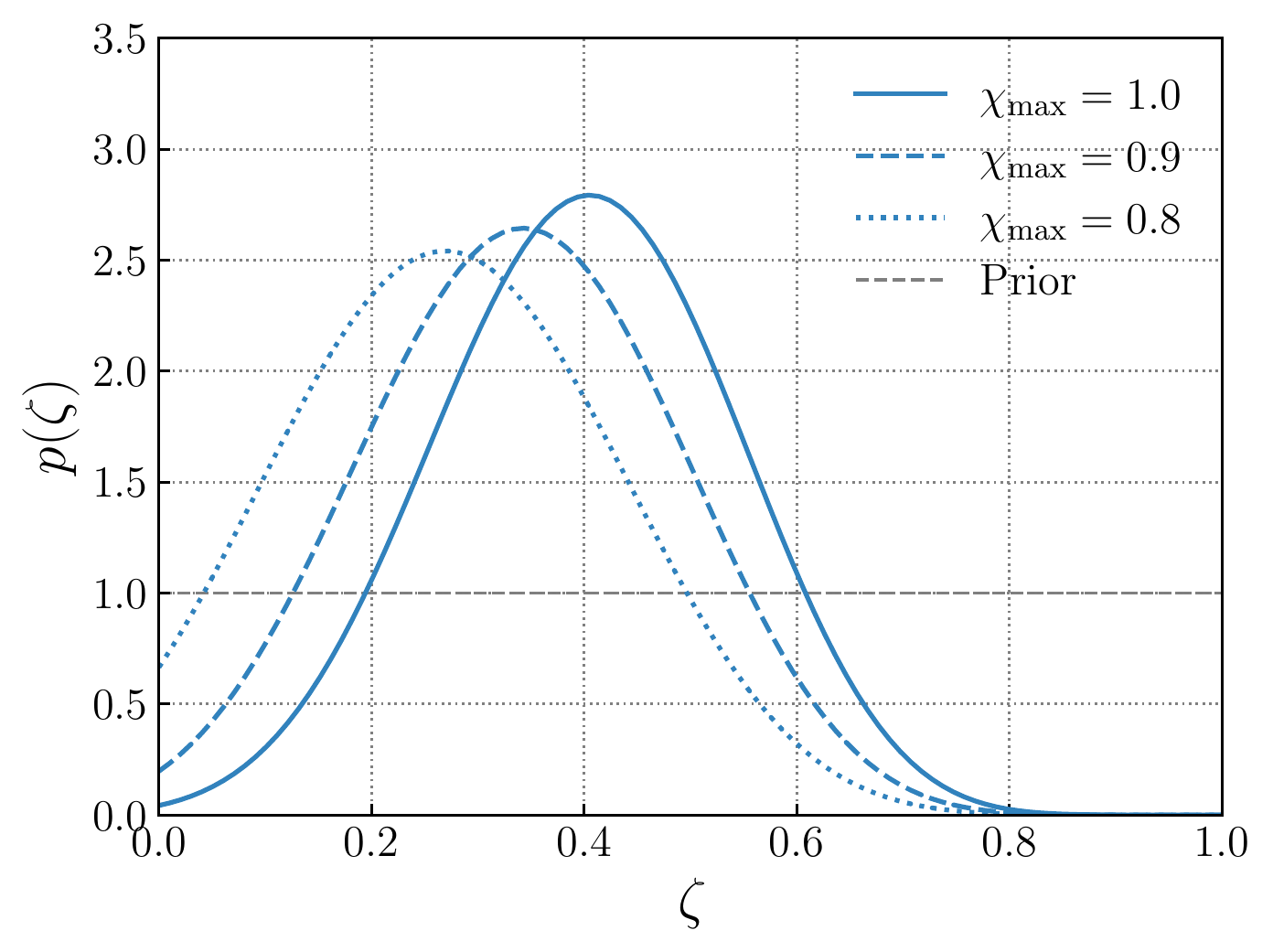}
    \caption{Posterior on the fraction $\zeta$ of binary black holes in which both components have zero spin, as obtained in the simple counting experiment of Sec.~\ref{sec:counting}.
    This counting experiment uses only events detected through GWTC-2~\citep{LIGOScientific:2020ibl}, and furthermore only relies on the fully-marginalized likelihoods for each binary under spinning ($0\leq \chi_1,\chi_2 \leq \chi_\mathrm{max}$) and non-spinning ($\chi_1=\chi_2=0$) priors.
    Values of $\zeta\gtrsim0.8$ are definitively ruled out.
    Whether or not $\zeta$ is consistent with zero, however, depends more sensitively on assumptions regarding the distribution of black hole spin magnitudes.
    If we assume that black hole spins range uniformly up to $\chi_\mathrm{max}=1.0$, $\zeta=0$ is disfavored.
    At the same time, relaxing $\chi_\mathrm{max}$ to slightly smaller values yields posteriors increasingly consistent with zero, which would indicate no distinct sub-population of non-spinning systems.}
    \label{fig:bayes-test}
\end{figure}

The Bayes factors computed in~\citet{LIGOScientific:2020ibl} and used by~\citet{Galaudage2021} were obtained via nested sampling~\citep{Skilling2004,Skilling2006,Speagle2020}.
To evaluate Eq.~\eqref{eq:mixture-model2}, we instead use posterior samples under the spinning hypothesis to calculate $\mathcal{B}^\mathrm{NS}_\mathrm{S}$ via a Savage-Dickey density ratio.
The Bayes factors we compute generally agree with those used as inputs in~\citet{Galaudage2021}, although with some notable exceptions that may contribute to the discrepancies between their results and our own; see Appendix~\ref{appendix:BFs}.

The solid curve in Fig.~\ref{fig:bayes-test} shows our resulting posterior on $\zeta$ using only GWTC-2 events for which such Bayes factors are available. We find that zero-spin fractions $\zeta \gtrsim 0.8$ are excluded at high credibility.
It also appears that $\zeta = 0$ is disfavored, which would imply the presence of at least a few zero-spin systems.  
However, note that the spinning hypothesis (S) \textit{requires} that spin magnitudes be distributed uniformly up to $\chi_\mathrm{max} = 1$, a possibility that is heavily disfavored as discussed in Sec.~\ref{sec:measurement}.
What happens if we adopt a more plausible prior distribution for the spinning hypothesis?
To answer this question, the dashed and dotted curves in Fig.~\ref{fig:bayes-test} show the $\zeta$ posterior given by Eq.~\eqref{eq:mixture-model2} if we recompute $\mathcal{B}^\mathrm{NS}_\mathrm{S}$ but now assume maximum spin magnitudes of $\chi_\mathrm{max} = 0.9$ and $0.8$, respectively, among the ``spinning'' population.
We see that even these small adjustments to $\chi_\mathrm{max}$ further rule out large $\zeta$ and increasingly support $\zeta = 0$.

If, rather than our Savage-Dickey estimates, we instead use the same nested sampling Bayes factors adopted by~\citet{Galaudage2021}, we now much more strongly rule out $\zeta=0$.
In Appendix~\ref{appendix:BFs} we track the origin of this behavior to one event, GW190408\_181802, whose nested sampling Bayes factor appears significantly overestimated.
This system is reported to have a large Bayes factor $\mathcal{B}^\mathrm{NS}_\mathrm{S}\sim 130$ in favor of the non-spinning hypothesis, but this conclusion is not supported by the posterior on this system's spins (and thus the Savage-Dickey density ratio); see Fig.~\ref{fig:chiposterior}.
When we exclude GW190408\_181892 from our sample, we obtain consistent $p(\zeta)$ posteriors from both sets of Bayes factors.
When including this event, however, the nested sampling Bayes factors cause $p(\zeta=0)$ to be underestimated by a factor of $\sim10^2$ relative to the result obtained with Savage-Dickey ratios (again see Fig.~\ref{fig:chiposterior}).
This could account for the nonzero fraction of non-spinning events found in~\citet{Galaudage2021}.

The initial check presented in Fig.~\ref{fig:bayes-test} suggests that the conclusion that most binary black holes comprise a distinct non-spinning sub-population is inconsistent with the parameter estimates for individual binary black systems.
At the same time, however, we saw that exact quantitative conclusions depend sensitively on assumptions regarding other aspects of the binary black hole population.
We therefore need to undertake a more complete hierarchical analysis, measuring the zero-spin fraction $\zeta$ while simultaneously fitting the distribution of black hole spin magnitudes and orientations.

\section{No sharp features in the effective spin distribution}
\label{sec:spike}

%
\begin{figure}
    \centering
    \includegraphics[width=0.5\textwidth]{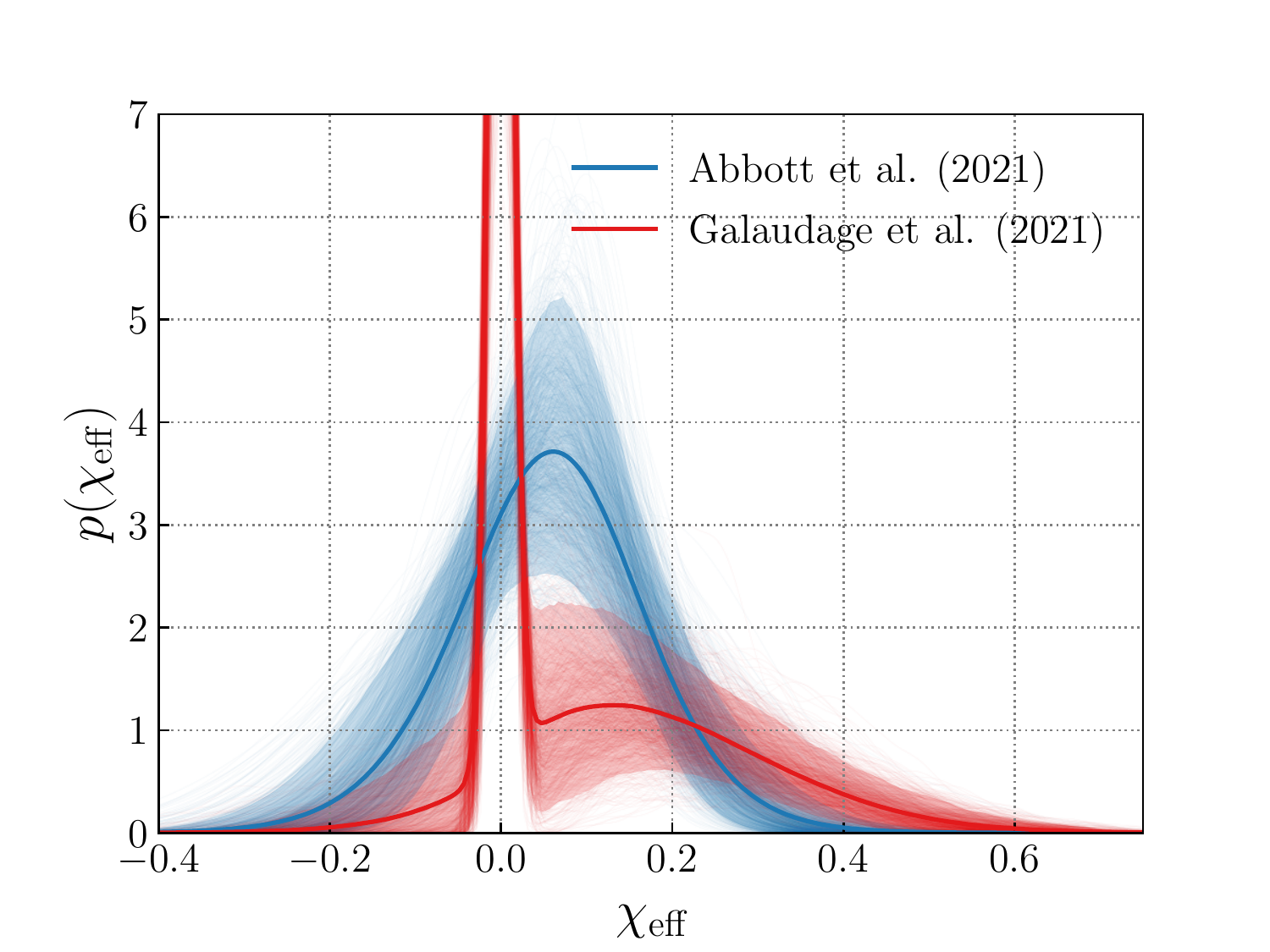}
    \caption{Marginal posterior distribution for $\chiEff$ using results from~\cite{O3b-pop} that have no excess of zero-spin events (blue) and results from~\citet{Galaudage2021} that imply that $\sim50\%$ of binaries have zero spins (red).
    The solid lines trace the mean distribution inferred by each model while shaded regions denote the 90\% credible regions; the light traces trace individual draws from the population posterior under each model.
    We see that, if a significant fraction of binary black holes are indeed non-spinning, this should be clearly identifiable through their $\chiEff$ distribution.
    This figure is modeled after Fig. 5 of~\citet{Galaudage2021}.}
    \label{fig:chi-eff-comparison}
\end{figure}

Going beyond our simple counting experiment, we next consider hierarchical inference of the \textit{effective} spin distribution.
An excess of events with vanishing spins would have stark implications for the distribution of effective spin parameters.
In Fig.~\ref{fig:chi-eff-comparison} we compare the $\chi_\mathrm{eff}$ distribution implied by the results of~\citet{Galaudage2021} and~\citet{O3b-pop}.
If most black holes are indeed non-spinning, we should correspondingly see a prominent spike at $\chiEff=0$, and if such a spike exists it should be robustly measurable using an appropriate model.

There are two benefits to searching for this excess of zero-spin systems in the $\chiEff$ domain, before proceeding to even more carefully explore the distribution of component spins.
First, inference of the $\chiEff$ distribution offers an independent and complementary check on the existence of a prominent zero-spin population: such a feature should be detectable either in the space of component spins or effective spins.
Second, by analyzing $\chiEff$ and not the higher-dimensional space of spin magnitudes and tilts, we can more easily avoid systematic uncertainties due to finite sampling effects.
As detailed in Appendix~\ref{appendix:inference}, the core ingredients in any hierarchical analysis are the posteriors $p(\lambda_i|d_i)$ on the properties $\lambda_i$ of our individual binaries (labeled by $i\in[1,N_\mathrm{obs}]$).
In general, however, we do not have direct access to these posteriors, but instead have discrete \textit{samples} $\{\lambda_{i,j}\}$ drawn from the posteriors, where $j\in [1,N_i]$ enumerates the samples from posterior $i$ and $N_i$ is the total number of samples for this event.
We therefore ordinarily replace integrals over $p(\lambda_i|d_i)$ with Monte Carlo averages over these discrete samples.
The fundamental assumption underlying this approach is that the posterior samples are sufficiently dense relative to the population features of interest.
In this paper, however, we are concerned with very narrow features in the binary black hole spin distribution, see Fig.~\ref{fig:chi-eff-comparison}.
In this case, we cannot automatically assume that Monte Carlo averages over posterior samples will yield accurate results.

Analysis of the $\chiEff$ distribution allows us to circumvent these sampling issues by alternatively representing each event's posterior as a Gaussian kernel density estimate (KDE) over its posterior samples.
This approach effectively imparts a finite ``resolution'' to each posterior sample, and allows us to assess the likelihood of arbitrarily narrow population features that would otherwise cause the typical Monte Carlo procedure to break down.
Further details about the KDE representation of posteriors are provided in Appendix~\ref{appendix:kde}.

\begin{figure}
    \centering
    \includegraphics[width=0.48\textwidth]{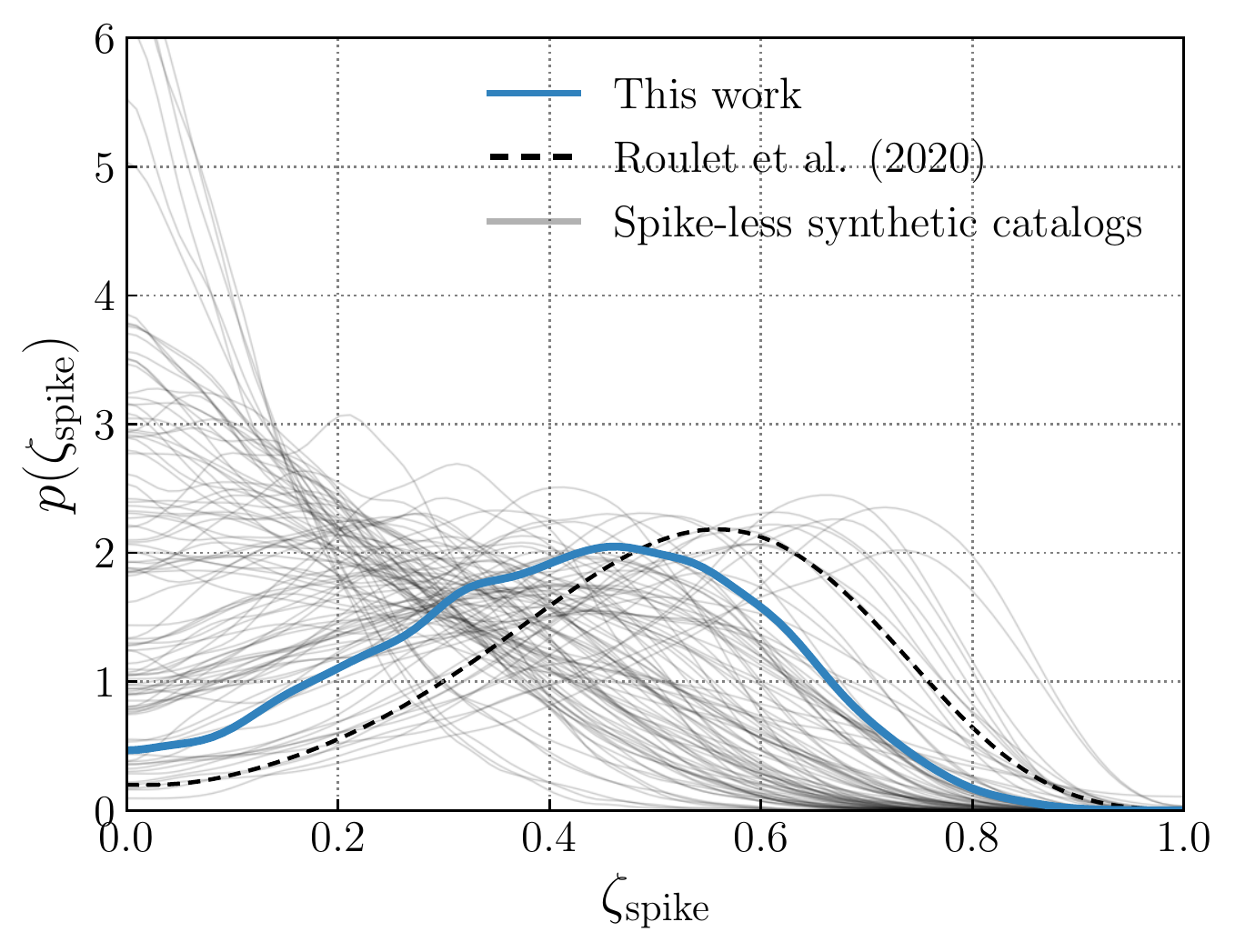}
    \caption{Marginalized posterior on the fraction $\zetaSpike$ of binary black holes comprising a distinct sub-population with $\chiEff=0$.
    The blue curve shows results obtained through analysis of the binary black holes in GWTC-3~\citep{LIGOScientific:2021djp}.
    We find $\zetaSpike$ to be consistent with zero, indicating no evidence for an excess of zero-spin systems.
    Nevertheless, $p(\zetaSpike)$ peaks suggestively at $\zetaSpike\approx0.5$.
    We demonstrate that this is not unexpected by repeatedly generating and analyzing mock catalogs of $\chiEff$ measurements, drawn from an intrinsically spike-less population described by a simple Gaussian.
    These catalogs yield posteriors similar to our own, often (and incorrectly) disfavoring $\zetaSpike=0$.
    As discussed in the main text, this behavior is due to a degeneracy between $\zetaSpike$ and the inferred mean of the spinning ``bulk'' population.
    For reference, the black dashed line shows the posterior on the zero-spin fraction $\zeta_0$ inferred in~\citet{Roulet2021}, which is qualitatively consistent with both our GWTC-3 measurement and the simulated spike-less measurements.
    }
    \label{fig:spike-posterior}
\end{figure}

Motivated by Fig.~\ref{fig:chi-eff-comparison}, we attempt to measure the presence of any zero-spin sub-population by modeling the $\chiEff$ distribution as a mixture between a broad ``bulk'' population, centered at $\mu_\mathrm{eff}$ with width $\sigma_\mathrm{eff}$, and a narrow ``spike'' centered at zero
    \begin{equation}
    \begin{aligned}
    &p(\chiEff|\zeta_\mathrm{spike},\epsilon,\mu_\mathrm{eff},\sigma_\mathrm{eff}) = \zeta_\mathrm{spike} {\cal{N}}_{[-1,1]}(\chiEff|0,\epsilon) \\
    &\hspace{2cm} + (1 - \zeta_\mathrm{spike}) {\cal{N}}_{[-1,1]}(\chiEff|\mu_\mathrm{eff},\sigma_\mathrm{eff})\,.
    \end{aligned}
    \label{eq:gaussianSpike-intext}
    \end{equation}
We call this the {\tt GaussianSpike} model; see App.~\ref{app:effectivemodels} for further details.
\new{Leveraging the KDE posterior representation introduced above, we take $\epsilon=0$, such that the spike is infinitely narrow at $\chiEff=0$.}
We hierarchically infer the parameters of this model using every binary black hole detection in GWTC-3~\citep{LIGOScientific:2021djp} with a false alarm rate below $1\,\mathrm{yr}^{-1}$; see Appendix~\ref{appendix:inference} for further details on the data we use.

The blue curve in Fig.~\ref{fig:spike-posterior} shows our resulting marginal posterior on the fraction $\zetaSpike$ of binary black holes comprising a zero-spin spike.
We find that $\zetaSpike$ remains consistent with zero, indicating no evidence for an over-density of events at $\chiEff=0$.
For reference, the ``zero-spin'' fraction measured by~\citet{Roulet2021} [$\zeta_0$ in Eq.~\eqref{eq:roulet}] is shown as a dashed black curve.
The results from~\citet{Roulet2021} are qualitatively consistent with our own; exact agreement isn't expected due to the different models and different data employed in each analysis.

Although $\zetaSpike$ is not bounded away from zero, a non-zero value is nevertheless preferred, with a maximum posterior value of $\zetaSpike \approx 0.5$.
\textit{We demonstrate that this behavior is not unexpected from intrinsically spike-less populations.}
We perform a series of null tests, repeatedly simulating and analyzing catalogs of events drawn from a spike-less population and with uncertainties typical of current detections; see Appendix~\ref{appendix:mock-injections} for further details on the simulations.
The grey curves in Fig.~\ref{fig:spike-posterior} show the posteriors on $\zetaSpike$ given by these synthetic catalogs.
Despite being drawn from a spike-less population, the simulated catalogs generically yield posteriors morphologically similar to the those obtained using actual observations, with some posteriors that even more extremely (and incorrectly) disfavor $\zetaSpike=0$.

\begin{figure}
    \centering
    \includegraphics[width=0.5\textwidth]{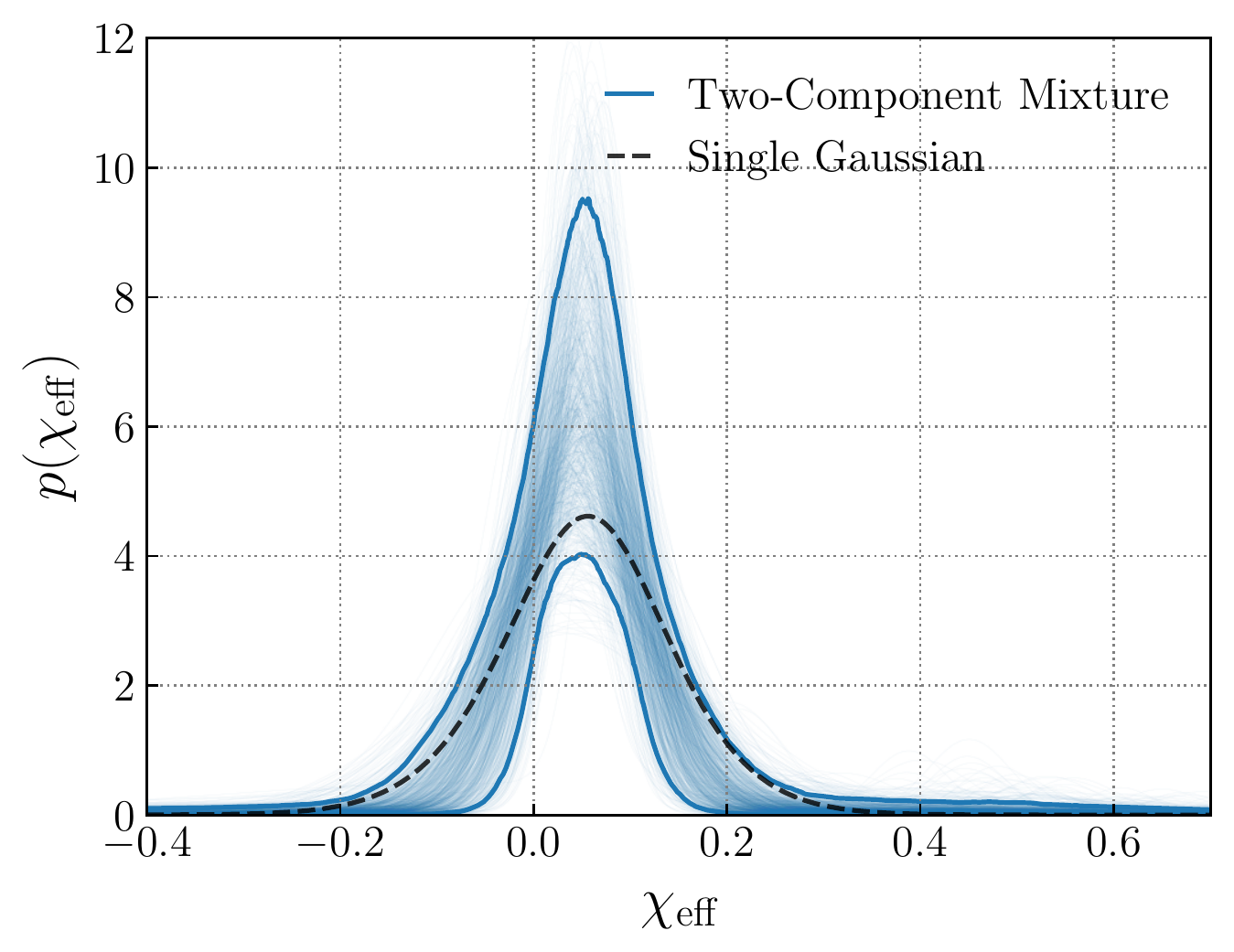}
    \caption{
    The $\chiEff$ distribution as inferred by a {\tt BimodalGaussian} effective spin model, defined as the sum of two Gaussians (see Appendix~\ref{appendix:models}).
    Solid blue lines denote the $90$\% credible intervals, while blue light curves are select individual draws from the posterior.
    For reference, the dashed black curve shows the mean $\chiEff$ distribution as inferred using a \textit{single} Gaussian.
    Both results are consistent with one another, indicating no evidence for bimodal features, narrow or otherwise, in the binary black hole $\chiEff$ distribution.
    }
    \label{fig:double-gaussian}
\end{figure}

The cause of this behavior is a degeneracy between $\mu_\mathrm{eff}$ and $\zeta_\mathrm{spike}$.
The mock catalogs that strongly disfavor $\zeta=0$ are typically those that have events with moderately high, well-measured effective spins.
The presence of such events increases the inferred mean $\mu_\mathrm{eff}$ of the ``bulk'' population, pulling the bulk away from those events near $\chiEff\approx 0$ and leaving them to be absorbed into a zero-spin sub-population.
We have verified that removing the most rapidly-spinning events from these mock catalogs indeed acts to resolve the apparent tension in Fig.~\ref{fig:spike-posterior}; see Appendix~\ref{appendix:mock-injections}.
This demonstration further emphasizes the need for additional caution when drawing strong astrophysical conclusions based on narrow population features, particularly in the regime when uncertainties on individual events are much larger than the features of interest.

Going beyond the question of a narrow spike at zero, we now more generally ask if there is evidence for \textit{any} bimodality in the $\chiEff$ distribution of binary black holes.
We explore this question using the {\tt BimodalGaussian} model (see Appendix~\ref{app:effectivemodels}) in which the ``spike'' in Eq.~\eqref{eq:gaussianSpike-intext} is replaced with a second, independent Gaussian with a variable mean and width.
The $\chiEff$ distribution inferred under this model is shown in Fig.~\ref{fig:double-gaussian}, where light traces show individual draws from our population posterior and the slide lines mark $90\%$ credible bounds.
We find no evidence that the effective spin distribution deviates from a simple unimodal shape.
For reference, the dashed black line marks the mean distribution inferred using a simple {\tt Gaussian} model.
Inference using the more complex \texttt{BimodalGaussian} remains extremely consistent with this simple result, \new{with both models yielding consistent means ($\gaussianChiEffMean$ and $\doubleGaussianChiEffMean$ under the \texttt{Gaussian} and \texttt{BimodalGaussian} models, respectively) and standard deviations ($\gaussianChiEffStd$ and $\doubleGaussianChiEffStd$).}

\new{
An alternative way to test for the presence of additional structure in the $\chi_\mathrm{eff}$ distribution is to ask about the \textit{predictive power} of our models: are there systematic residuals between the $\chi_\mathrm{eff}$ values we observe and those predicted by each model?
In Appendix~\ref{appendix:ppc} we subject the \texttt{Gaussian}, \texttt{GaussianSpike}, and \texttt{BimodalGaussian} models to predictive tests, finding that all three models, once fitted, successfully predict the observed $\chi_\mathrm{eff}$ values among GWTC-3.
The fact that the simple \texttt{Gaussian} models passes this check once more points to the lack of observational evidence for additional structure in the binary black hole $\chi_\mathrm{eff}$ distribution, whether a spike, a secondary mode, or still other feature.
}

\section{The population of spin magnitudes and tilts}
\label{sec:results}

Preliminary results so far, based on the Bayes factor counting experiment in Sec.~\ref{sec:counting} and hierarchical $\chiEff$ analyses in Sec.~\ref{sec:spike}, do not point to an excess of zero-spin events.
Here we confirm these conclusions via a more complete inference of the component spin magnitude and tilt distributions, under a series of increasingly complex models.

As a baseline model~\citep{Talbot:2017yur,Wysocki2019,O3a-pop,O3b-pop}, we take each component spin magnitude to be distributed following a Beta distribution,
    \begin{equation}
    p(\chi_i | \alpha, \beta) = \frac{\chi_i^{1-\alpha} \, (1-\chi_i)^{1-\beta}}{c(\alpha, \beta)}\,,
    \label{eq:beta}
    \end{equation}
    where $c(\alpha, \beta)$ is a normalization constant.
Every spin tilt, meanwhile, is distributed as a mixture between an isotropic component and a preferentially-aligned component, modeled as a half-Gaussian centered at $\cos\theta=1$:
    \begin{equation}
    \begin{aligned}
    &p(\cos\theta_i|f_\mathrm{iso},\sigma_t) = \frac{f_\mathrm{iso}}{2} \\
    &\hspace{2cm} + (1-f_\mathrm{iso})
            {\cal {N}}_{[-1,1]}(\cos\theta_i|1,\sigma_t)\,.
    \end{aligned}
    \label{eq:tilt}
    \end{equation}
We refer to Eqs.~\eqref{eq:beta} and~\eqref{eq:tilt} as the \texttt{Beta+Mixture} model.
A related version of this model in which both component spin tilts are together drawn from either the isotropic or the aligned component is also called the \texttt{Default} spin model in~\citet{O3a-pop} and~\citet{O3b-pop}.

\citet{Galaudage2021} explored an extension to the \texttt{Default} model that allowed for an excess of systems with zero spin and a variable bound on the spin tilt distribution; it was termed the {\tt Extended} model in that study.
Motivated by these extensions, we introduce the possibility of a zero-spin ``spike'' in the spin magnitude distribution, modeled as a half-Gaussian with finite width $\epsilon_\mathrm{spike}$ centered at zero:
    \begin{equation}
    \begin{aligned}
    &p(\chi_i | \alpha, \beta, f_\mathrm{spike}, \epsilon_\mathrm{spike}) = f_\mathrm{spike} \, {\cal{N}}_{[0,1]}(\chi_i|0,\epsilon_\mathrm{spike}) \\
    &\hspace{2cm} + (1-f_\mathrm{spike})  \frac{\chi_i^{1-\alpha} \, (1-\chi_i)^{1-\beta}}{c(\alpha, \beta)}\,.
    \end{aligned}
    \label{eq:betaspike-intext}
    \end{equation}
Following~\citet{Galaudage2021} we also introduce a variable truncation bound $z_\mathrm{min}$ on the spin-tilt distribution:
    \begin{equation}
    \begin{aligned}
    &p(\cos\theta_i|f_\mathrm{iso},\sigma_t,z_\mathrm{min}) = \frac{f_\mathrm{iso}}{1-z_\mathrm{min}} \\
    &\hspace{2.5cm} + (1-f_\mathrm{iso})
            {\cal {N}}_{[z_\mathrm{min},1]}(\cos\theta_i|1,\sigma_t)\,,
    \end{aligned}
    \label{eq:tiltzmin-intext}
    \end{equation}
for $z_\mathrm{min}\leq\cos\theta_i\leq 1$.
We refer to Eqs.~\eqref{eq:betaspike-intext} and~\eqref{eq:tiltzmin-intext} together as the \texttt{BetaSpike+TruncatedMixture} model.
To better understand how conclusions regarding a zero-spin excess and the prevalence of spin-orbit misalignment are related, we also consider variants of this model in which only one extended feature is present: a zero-spin spike but no $\cos\theta$ truncation (\texttt{BetaSpike+Mixture}) or a $\cos\theta$ truncation but no spike (\texttt{Beta+TruncatedMixture}).
See Appendix~\ref{app:componentmodels} for further details.

Our full \texttt{BetaSpike+TruncatedMixture} model differs from the \texttt{Extended} model of \citet{Galaudage2021} in two ways.
First, we do not fix the width $\epsilon_\mathrm{spike}$ of the vanishing spin subpopulation, but instead treat it as a free parameter to be inferred from the data.
This allows us to \textit{test} whether a narrow sub-population is actually preferred by the data, similar to our investigation with the \texttt{BimodalGaussian} model in Sec.~\ref{sec:spike} above.
We adopt a prior requiring $\epsilon_\mathrm{spike}\geq 0.03$.
This lower limit is motivated by tracking our number of effective posterior samples per event (see further discussion below), and by the effective population resolution of our catalog, which we estimate as
\begin{equation}
    \frac{1}{\sigma^2_{\mathrm{obs}}}=\sum_{i=1}^{N_{\mathrm{obs}}}\left(\frac{1}{\sigma^2_{\chi_1,i}} + \frac{1}{\sigma^2_{\chi_2,i}}\right)\,,
\end{equation}
where $\sigma^2_{\chi_1,i}$ and $\sigma^2_{\chi_2,i}$ are the variances of the spin magnitude posteriors for each event $i$.
We find $\sigma_{\mathrm{obs}}= 0.02$, approximately equal to our lower bound on $\epsilon_\mathrm{spike}$.

Second, the \texttt{Extended} model does not allow for independently and identically distributed spin magnitudes and orientations.
Instead, component spin magnitudes are either \textit{both} vanishing or \textit{both} non-vanishing in a given binary.
This choice precludes astrophysical scenarios such as tidal spin-up, which is expected to affect only one component spin in a given binary.
Similarly, within the \texttt{Extended} model the spin tilts are both either members of the isotropic component or the preferentially aligned component in Eq.~\eqref{eq:tiltzmin-intext}.
Here, we instead assume that all component spin magnitudes and tilts are independently drawn from Eqs.~\eqref{eq:betaspike-intext} and~\eqref{eq:tiltzmin-intext}.

When hierarchically analyzing the $\chieff$ distribution above, we relied on a KDE representation of individual event likelihoods $p(d_i|\lambda_i)$ to mitigate sampling uncertainties and evaluate population models with narrow features.
This trick cannot be straightforwardly applied to inference of the component spin distribution, due to both the increased dimensionality and the fact that the sharp feature of interest (a spike at $\chi_1 = \chi_2 = 0$) lies on the boundary of parameter space, rather than the center.
We will therefore return to standard Monte Carlo averaging over posterior samples when hierarchically inferring the component spin distribution.
To diagnose possible breakdowns in our inference due to finite sampling effects, we monitor the effective number of posterior samples, $N_\mathrm{eff}$, informing our Monte Carlo estimates for each event's likelihood.
As discussed in Appendix~\ref{appendix:Neff}, we explicitly track $\mathrm{min} \left[N_{\mathrm{eff},i}(\Lambda)\right]$, the minimum effective sample count across all events for a proposed population $\Lambda$, and use this quantity to define which regions of parameter space we can and cannot make claims about.
In particular, we find that we expect reliable population inference when $\epsilon_\mathrm{spike}>0.03$.

\begin{figure*}
\centering
    \includegraphics[width=0.9\textwidth]{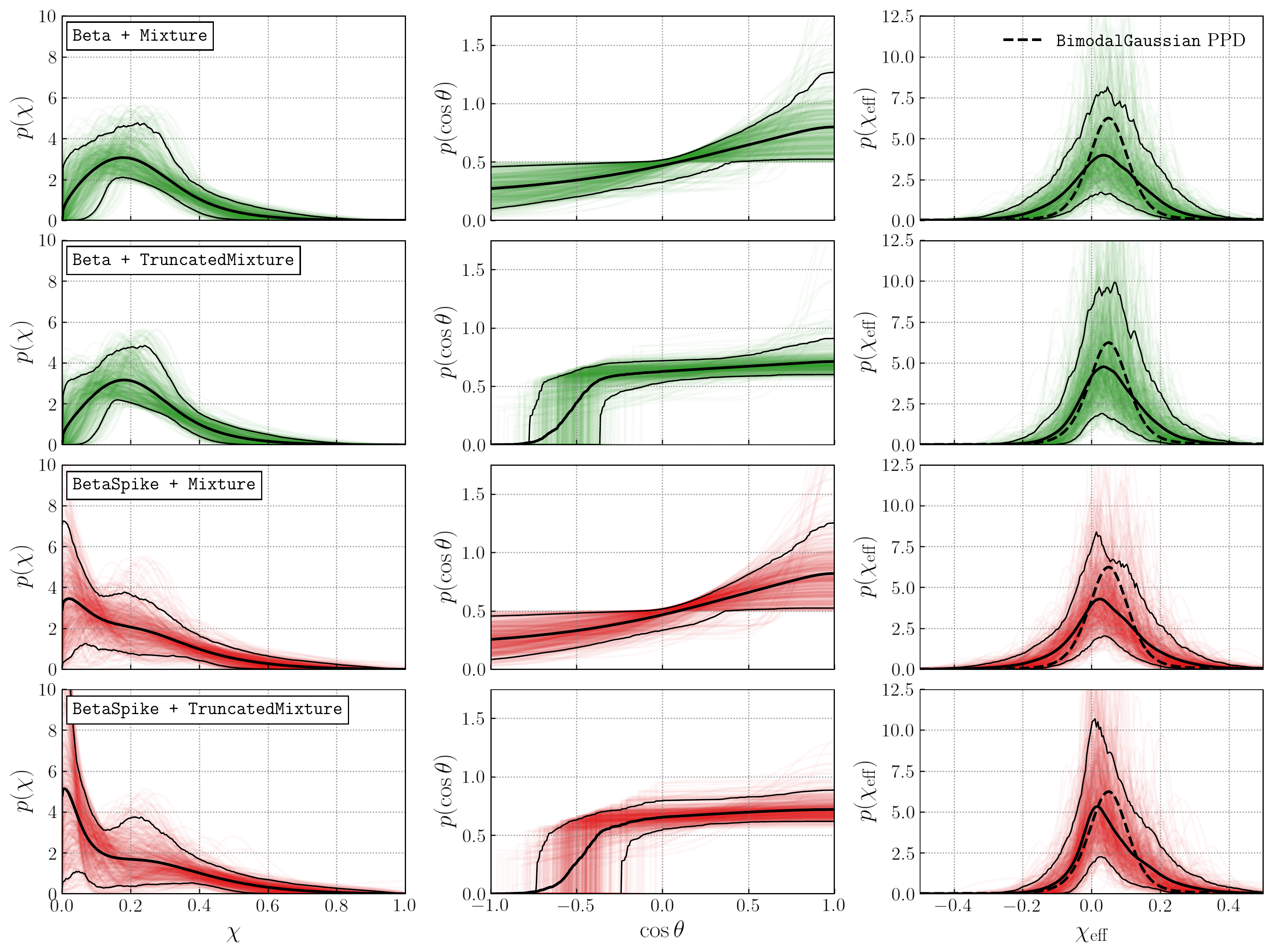}
    \caption{
    Inferred distributions of component spin magnitudes (left), spin-orbit tilts (middle), and effective inspiral spins (right) of binary black holes under the various component spin models considered in this paper.
    Solid black lines denote the median and the $90$\% credible intervals, while red/green light curves correspond to individual draws from the posterior for each model.
    Among models that allow for a distinct sub-population of non-spinning systems (bottom two rows), we infer that the fraction of binaries in such systems is consistent with zero.
    Meanwhile, we find that the $\cos\theta$ distribution confidently extends to negative values; models that allow for a sharp truncation in the $\cos\theta$ distribution require this truncation to occur at $z_\mathrm{min}<0$.
    Despite the varying component spin magnitude and tilt distributions recovered by these different models, all yield similar $\chiEff$ distributions.
    For reference, the dashed black curves in the right-hand column show the mean $\chiEff$ distribution obtained by direct inference with the {\tt BimodalGaussian} model of Sec.~\ref{sec:spike}; all four component spin models give $\chiEff$ distributions that are consistent with this result.}
    \label{fig:PPDs_component_models}
\end{figure*}

\begin{table*}
    \centering
    \begin{tabular}{r|c|c|c|c|c|c|c}
    
         Model & $f_\mathrm{spike}$ & $z_{\mathrm{min}}$ & $z_{1\%}$ & $\chi_{1\%}$ & $\chi_{99\%}$ &  $\chi_{\mathrm{eff},1\%}$ & $\chi_{\mathrm{eff},99\%}$ \\
         \hline
         \hline
         {\tt Beta+Mixture} & -- & -- & $\betaPlusMixturezOne$ & $\betaPlusMixturechiOne$ & $\betaPlusMixturechiNinetyNine$ & $\betaPlusMixturechieffOne$ & $\betaPlusMixturechieffNinetyNine$ \\
         {\tt BetaSpike+Mixture} & $\betaSpikePlusMixturefspike$ & -- & $\betaSpikePlusMixturezOne$ & $\betaSpikePlusMixturechiOne$ & $\betaSpikePlusMixturechiNinetyNine$ & $\betaSpikePlusMixturechieffOne$ & $\betaSpikePlusMixturechieffNinetyNine$ \\
         {\tt Beta+TruncatedMixture} & -- & $\betaPlusTruncatedMixturezmindata$ & $\betaPlusTruncatedMixturezOne$ & $\betaPlusTruncatedMixturechiOne$ & $\betaPlusTruncatedMixturechiNinetyNine$ & $\betaPlusTruncatedMixturechieffOne$ & $\betaPlusTruncatedMixturechieffNinetyNine$ \\
         {\tt BetaSpike+TruncatedMixture} & $\betaSpikePlusTruncatedMixturefspike$ & $\betaSpikePlusTruncatedMixturezmindata$ & $\betaSpikePlusTruncatedMixturezOne$ & $\betaSpikePlusTruncatedMixturechiOne$ & $\betaSpikePlusTruncatedMixturechiNinetyNine$ & $\betaSpikePlusTruncatedMixturechieffOne$ & $\betaSpikePlusTruncatedMixturechieffNinetyNine$ \\
         \hline
         \hline
    \end{tabular}
    \caption{Median and 90\% credible intervals on various physical quantities of interest with the component spin models. From left to right we present: the fraction of black holes that belong in the (finite-width) zero-spin spike ($f_\mathrm{spike}$; see Fig.~\ref{fig:component_cornerplot} for posterior), the minimum value of the spin tilt ($z_{\mathrm{min}}$; Fig.~\ref{fig:component_cornerplot} for posterior), 1\% lower value of the spin tilt posterior distribution ($z_{1\%}$), the 1\%/99\% lower/upper values of the spin magnitude posterior distribution ($\chi_{1\%}$/$\chi_{99\%}$), and the 1\%/99\% lower/upper values of the effective spin distribution ($\chi_{\mathrm{eff},1\%}$/$\chi_{\mathrm{eff},99\%}$).
    We choose to report 1\% and 99\% values since $1/N_{\mathrm{obs}} = 0.014\simeq 1\%$ for our $N_{\mathrm{obs}}=69$ events.}
    \label{tab:results-quant}
\end{table*}
\begin{figure*}
    \centering
    \includegraphics[width=0.9\textwidth]{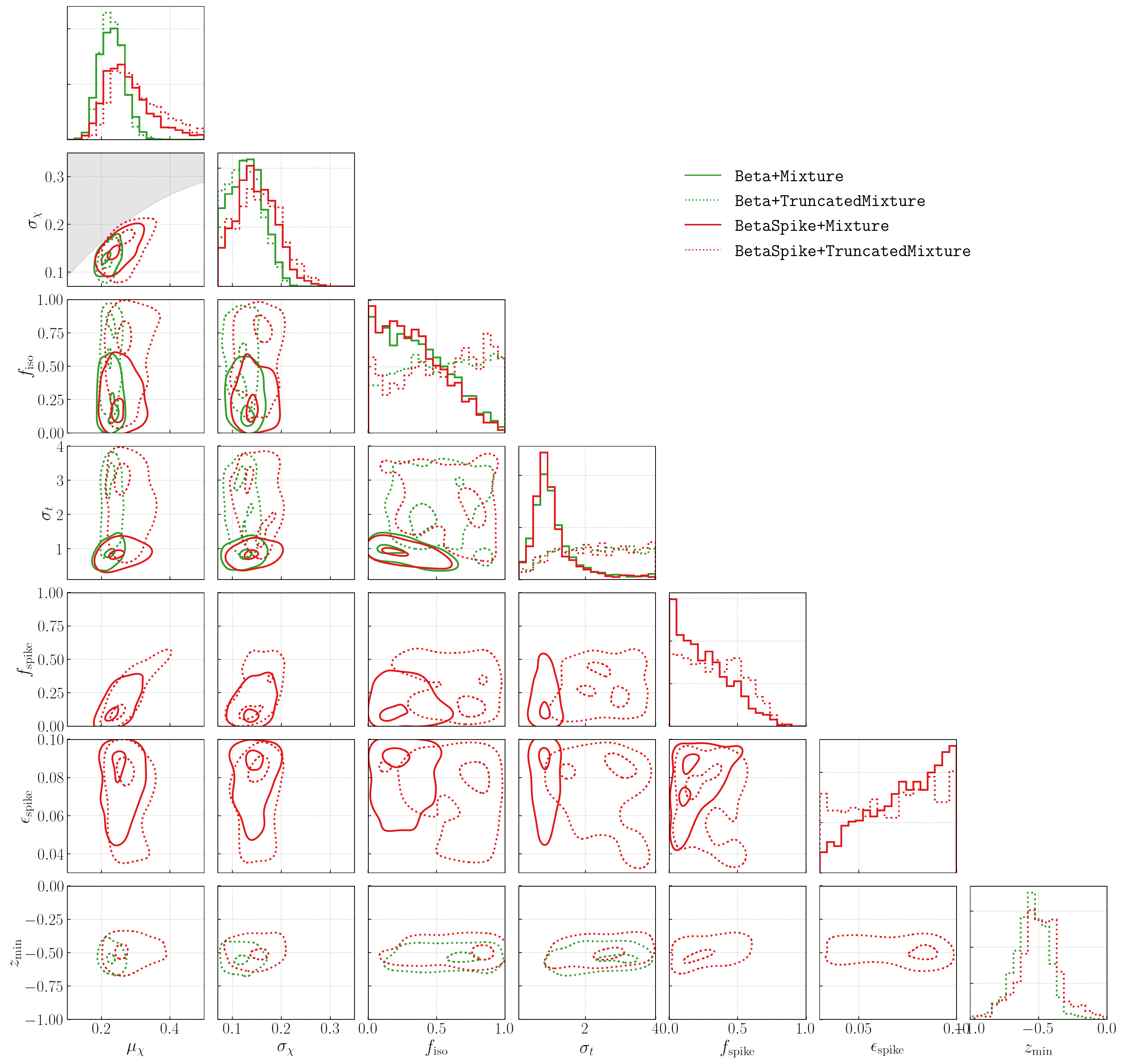}
    \caption{
    One- and two-dimensional marginalized posteriors for the parameters of each component spin model we consider (see Appendix~\ref{app:componentmodels}).
    These posteriors correspond to the measured distributions of component spin magnitudes and tilts shown above in Fig.~\ref{fig:PPDs_component_models}.
    Some parameters are defined only for a subset of component spin models.
    The shaded region in the joint $\mu_\chi$--$\sigma_\chi$  posterior is the region excluded by the prior cut on the shape parameters of the spin magnitude Beta distribution, see Appendix~\ref{app:componentmodels}.
    We find that the fraction $f_\mathrm{spike}$ of black holes comprising a zero-spin sub-population is consistent with zero.
    The data also require that at least some spins are misaligned by more than $90^\circ$ relative to their Newtonian orbital angular momentum; among models that include a variable truncation bound $z_\mathrm{min}$ on the $\cos\theta$ distribution, this bound is inferred to be confidently $\leq 0$, regardless of assumptions about a possible zero-spin sub-population.
    Allowing such a truncation bound, meanwhile, significantly impacts constraints on $f_\mathrm{iso}$, the fraction of binaries with isotropically-oriented spins, as well as $\sigma_t$, the width of a preferentially spin-aligned mixture component (see Eq.~\eqref{eq:tilt}).
    As seen in Fig.~\ref{fig:PPDs_component_models}, introduction of a truncation bound yields a significantly flatter $p(\cos\theta)$ distribution, corresponding here to larger values of $f_\mathrm{iso}$ and $\sigma_t$.}
    \label{fig:component_cornerplot}
\end{figure*}

Figure~\ref{fig:PPDs_component_models} shows the measured spin magnitude and tilt distributions under our four component spin models, and we show the posteriors obtained on the hyperparameters of each model in Fig.~\ref{fig:component_cornerplot}.
To hierarchically measure the component spin distribution, we again use all binary black holes in GWTC-3 with false alarm rates below $1\,\mathrm{yr}^{-1}$; see Appendix~\ref{appendix:inference} for details.

For those models allowing a distinct zero-spin spike, the left panels of Fig.~\ref{fig:PPDs_component_models} indicate that such a feature is not required to exist.
We instead see inferred spin magnitude distributions consistent with a single, smooth function that remains finite at $\chi=0$ and most events have spin magnitudes in the $\chi\in(0,0.3)$ range.
Correspondingly, the posteriors on $f_\mathrm{spike}$ in Fig.~\ref{fig:component_cornerplot} are consistent with zero.
Compared to the fraction $\zeta_{\mathrm{spike}}$ with $\chiEff=0$ (see Fig.~\ref{fig:spike-posterior}), we find that $f_{\mathrm{spike}}$ is more consistent with zero.
We interpret this variation as reflective of the systematic uncertainty in exactly how the question ``\textit{Does there exist an excess of zero-spin systems?}'' is posed: whether in the component spin or effective spin domains, with a delta-function at zero or a finite-width spike, etc.
Despite these differences, all results indicate that the presence of a zero-spin population is not required by the current data.
A zero-spin population is not \textit{precluded}, though: both sets of results bound any zero-spin fraction to $\lesssim 60\%$, suggesting that a distinct zero-spin population could yet emerge in future data.

The fraction $f_\mathrm{spike}$ is primarily correlated with $\mu_{\chi}$, the mean of the ``bulk" spin magnitude population.
Larger $\mu_\chi$ values reduce the capability of this ``bulk" Beta distribution to accommodate events with small spin; these events will therefore necessarily be assigned to the ``spike'' and thus increase the value of $f_\mathrm{spike}$.
This is similar to the phenomenon identified in Sec.~\ref{sec:spike} above.
Additionally, all four component spin models yield similar spin magnitude distributions above $\chi \gtrsim 0.4$, suggesting that the data are robustly consistent with the absence of large spin magnitudes.
Table~\ref{tab:results-quant} lists the 1st and 99th spin magnitude percentiles ($\chi_{1\%}$ and $\chi_{99\%}$ respectively) under each model; our most conservative estimate gives $\chi_{99\%}=\betaSpikePlusTruncatedMixturechiNinetyNine$.

In contrast to \citet{Galaudage2021}, we left the width $\epsilon_\mathrm{spike}$ of our ``spike" population as a free parameter in order to test whether the data indeed require a narrow feature at $\chi = 0$.
Although we recover largely uninformative constraints on $\epsilon_\mathrm{spike}$, Fig.~\ref{fig:component_cornerplot} in fact shows a slight preference for \textit{large} $\epsilon_\mathrm{spike}$, further indicating that no narrow features are confidently present in the spin magnitude distribution.
Our conclusions regarding small $\epsilon_\mathrm{spike}$, however, are limited by the finite sampling effects discussed above.
In Appendix~\ref{appendix:Neff} we study the number $N_\mathrm{eff}$ of effective posterior samples as a function of $\epsilon_\mathrm{spike}$.
We find that $N_\mathrm{eff}$ depends sensitively on $\epsilon_\mathrm{spike}$, and we caution that values of $\epsilon_\mathrm{spike} \lesssim 0.04$ are possibly subject to significant Monte Carlo uncertainty.
Even with this restriction, however, we find no preference for a small $\epsilon_\mathrm{spike}$ in the posterior.
This conclusion is further bolstered by the analysis of the effective spin distribution in Sec.~\ref{sec:spike} above, which was \textit{not} subject to finite sampling effects and which similarly found no evidence for an excess of zero-spin systems. 

Irrespective of modeling assumptions regarding a zero-spin sub-population, all four component spin models exhibit significant support at negative $\cos\theta$ in Fig.~\ref{fig:PPDs_component_models}.
Models that allow for a truncation in the spin tilt distribution consistently infer this truncation to be at negative values; in Fig.~\ref{fig:component_cornerplot} we correspondingly see that $z_\mathrm{min}\leq0$ at high significance.
This result signals the presence of events with spins misaligned by more than $90^{\circ}$ from their Newtonian orbital angular momentum.
Table~\ref{tab:results-quant} gives the 1st percentile ($z_{1\%}$) on $\cos\theta$ inferred by each model, with $z_{1\%} = \betaSpikePlusTruncatedMixturezOne$ in the most conservative case.
Notably, the addition of a truncation in the spin tilt distribution significantly ``flattens" the recovered $\cos\theta$ distribution, no longer requiring any peak at $\cos\theta=1$.
This behavior is due to the fact that the data disfavor an equal density of events at $\cos{\theta}=+1$ and $\cos{\theta}=-1$.
Since an isotropic distribution is necessarily symmetric at $\cos{\theta}=\pm 1$,  models that allow for a mixture of isotropic and aligned spin tilts must compensate by adding more weight to the ``primarily-aligned'' Gaussian component in Eq.~\eqref{eq:tilt}.
No such compensation is necessary if the spin tilt distribution is allowed to truncate.
In models including a $z_\mathrm{min}$ truncation, all events are either assigned to the ``isotropic'' (but now truncated) component, or the Gaussian component itself is stretched to near-isotropy; in Fig.~\ref{fig:component_cornerplot} we see that $f_\mathrm{iso}$ becomes consistent with $1$ and $\sigma_t$ shifts to prefer large values when a truncation is included in the model.

\begin{figure}
    \centering
    \includegraphics[width=0.48\textwidth]{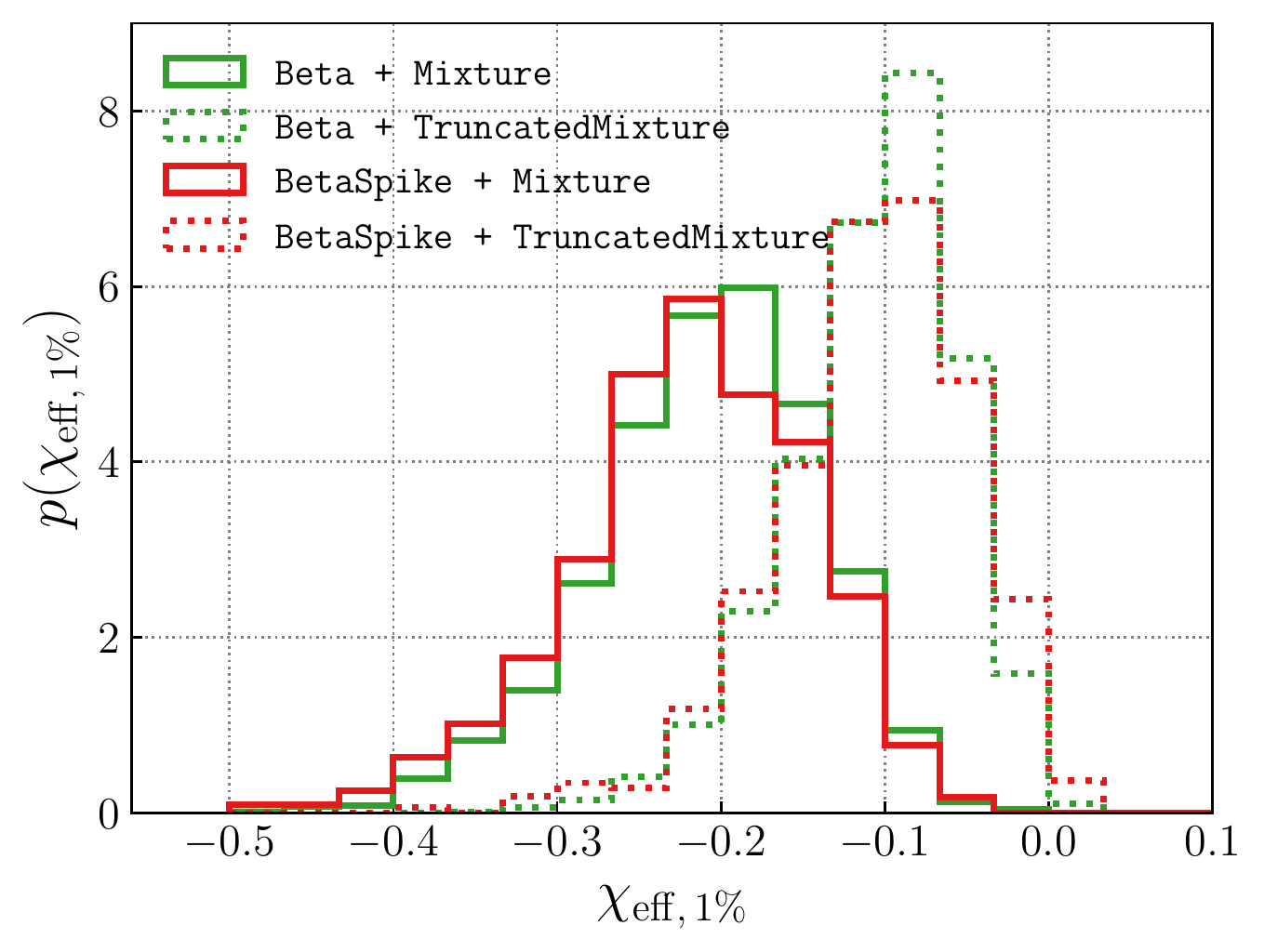}
    \caption{
    Posterior for the 1st percentile ($\chi_\mathrm{eff,1\%}$) of the $\chiEff$ distributions implied by each of our component spin models, as shown in the right-hand column of Fig.~\ref{fig:PPDs_component_models}.
    The median values and 90\% credible intervals for $\chi_{\mathrm{eff},1\%}$ are also reported in Table~\ref{tab:results-quant}.
    Each of our component spin models require the binary black hole population to contain spins misaligned by more than $90^\circ$ relative to their Newtonian orbital angular momentum.
    Correspondingly, $\chi_{\mathrm{eff},1\%}$ is inferred to likely (although not \textit{necessarily}) be negative.
    The $\chi_{\mathrm{eff},1\%}$ values recovered here are consistent with the minimum $\chi_\mathrm{eff}$ values measured by~\citet{O3a-pop,O3b-pop}.
    }
    \label{fig:chi_eff_mins_from_component_models}
\end{figure}

This result signals the presence of events with tilt angles greater than $90^{\circ}$ and hence the possibility of binaries with effective spins $\chi_\mathrm{eff}<0$.
The third column in Fig.~\ref{fig:PPDs_component_models} shows the $\chi_\mathrm{eff}$ distributions implied by each component model.
Despite the wide range of possible features possessed by each model, they all yield similar $\chiEff$ distributions that exhibit asymmetry about zero but that extend to negative values.
As a consistency check, we also compare these implied $\chiEff$ distributions to our result obtained through direct inference on $\chiEff$.
The dashed curve in each panel shows the mean obtained under the {\tt BimodalGaussian} effective spin model.
Each of the four component spin models yields consistent $\chiEff$ distributions, although they are suggestive of possible additional skewness~\citep{O3a-pop,O3b-pop}. 
\new{
As an additional consistency check and safeguard against erroneous conclusions due to poorly-fitting models, we subject each of the component spin models to predictive checks, as introduced in Sect.~\ref{sec:spike} above.
The results, shown in Appendix~\ref{appendix:ppc}, indicate that all four models provide a good fit to observation.
}

As a proxy for the minimum $\chiEff$ values implied by our component spin results, Fig.~\ref{fig:chi_eff_mins_from_component_models} shows posteriors on the 1st percentile ($\chi_{\mathrm{eff},1\%}$) of the effective spin distribution corresponding to each model.
Under the {\tt BetaSpike+TruncatedMixture} model, for example, we find $\chi_{\mathrm{eff},1\%}=\betaSpikePlusTruncatedMixturechieffOne$, and bound $\chi_{\mathrm{eff},1\%}<0$ at $\betaSpikePlusTruncatedMixturechieffOnePercentile$ credibility.
This finding is consistent with the conclusions drawn by~\citet{O3a-pop} and~\citet{O3b-pop}, who added a lower truncation in the $\chiEff$ distribution and inferred $\chi_{\mathrm{eff},\mathrm{min}}<0$ at $\sim 90\%$ credibility.
Our conclusions are also consistent with~\citet{Roulet2021}, who modeled the $\chiEff$ distribution as the sum of a positive component, a negative, and a ``near-zero'' component whose $2\sigma$ width spanned $-0.08< \chiEff <0.08$; see Eq.~\eqref{eq:roulet}.
This near-zero component is broad enough to likely encompass the small but negative $\chi_{\mathrm{eff},1\%}$ values we report here.

Finally, Figs.~\ref{fig:PPDs_component_models} and \ref{fig:component_cornerplot} reveal how questions regarding zero-spin sub-populations and spin-orbit misalignment are potentially related.
The inclusion or exclusion of a zero-spin sub-population (\texttt{Beta+TruncatedMixture} vs. \texttt{BetaSpike+TruncatedMixture}) has a negligible effect on our conclusions regarding the location of $z_\mathrm{min}$.
On the other hand, introduction of a spin tilt truncation (\texttt{BetaSpike+Mixture} vs. \texttt{BetaSpike+TruncatedMixture}) more noticeably impacts the inferred $f_\mathrm{spike}$ posterior, resulting in less support for zero.
As discussed above, though, the spin tilt truncation has a much larger effect on $f_{\mathrm{iso}}$, which becomes consistent with $1$ (i.e., no excess of spin-aligned events) when a spin tilt truncation is included in the model.
This same effect can be seen in Fig. 4 of~\cite{Galaudage2021}.

\section{Discussion and Conclusions}
\label{sec:conclusions}

In this paper we have sought to explore the three open questions posed in Sec.~\ref{subsec:questions} via three complementary routes: a counting experiment using only the fully-marginalized likelihoods for each event, hierarchical analysis of the binary black hole effective spin distribution, and hierarchical analysis of the binaries' component spin distribution.
Each of these three routes yielded consistent conclusions, which we summarize below.

\textit{i.}
\textbf{\textit{We find no evidence for an excess of zero-spin events.}}
Using each of our three methods we find that the fraction of black holes in a distinct zero-spin sub-population is consistent with zero.
We furthermore verified that this behavior is common among analyses of synthetic populations lacking a zero-spin excess.
We therefore conclude that the observational data do not presently support the notion that the majority of black holes in merging binaries are non-spinning.
Given current observations, a non-spinning population can comprise at most $\lesssim 60-70\%$ of merging binary black holes.

\textit{ii.}
\textbf{\textit{The inferred population is consistent with less than 1\% of spin magnitudes above $\mathbf{\chi\sim0.6}$.}}
In each of the component spin models explored in Sec.~\ref{sec:results}, the inferred population is nearly identical for spin magnitudes $\chi\gtrsim 0.4$, falling rapidly, with negligible support for $\chi \gtrsim 0.6$.
This conclusion is robust against  modeling systematics, including a possible truncation in the spin tilt distribution or a zero-spin sub-population.

\textit{iii.}
\textbf{\textit{Binary black holes exhibit a broad range of spin-orbit misalignment angles, with some angles greater than $\mathbf{90^\circ}$.}}
In all models, we find a preference for a flat and broad distribution of spin-orbit misalignment angles.
When we introduce a lower truncation bound on the $\cos\theta$ distribution, we confidently infer that this truncation bound must be negative, thus requiring the presence of systems with anti-aligned spins among the binary black hole population.
Additionally, the minimum $\chiEff$ among the binary black hole population is inferred to be confidently negative under each component spin model.

Our conclusions are consistent with the findings of~\citet{O3a-pop} and~\citet{O3b-pop}, who reported evidence for spins misaligned by more than $90^\circ$ and no modeling tension that would indicate the existence of a large zero-spin sub-population.
Our conclusions are moreover in broad agreement with~\citet{Roulet2021}. Figure~\ref{fig:spike-posterior} shows the inferred fraction of events with $\chiEff=0$ from this work and the inferred fraction of events in the ``zero-spin'' sub-population [$\zeta_0$ in Eq.~\eqref{eq:roulet}] from~\citet{Roulet2021}.
Both results are in qualitative agreement, consistent with a zero-spin fraction of zero but peaking at $\sim0.5$.
As demonstrated in Sec.~\ref{sec:spike}, posteriors of this form arise generically when analyzing mock catalogs of events drawn from a population lacking a zero-spin sub-population.

\citet{Roulet2021} find that the fraction of events in their ``negative-spin" sub-population is consistent with zero; this too is consistent with our result.
As discussed in Sec.~\ref{sec:results}, we infer $\chi_{\mathrm{eff},1\%}$ to be negative but small in magnitude.
Because \citet{Roulet2021}'s ``zero-spin'' sub-population has a broad standard deviation of $0.04$, events with negative but small-in-magnitude $\chiEff$ are counted as members of this zero-spin sub-population, rather than associated with the ``negative-spin'' sub-population.
Indeed, \citet{Roulet2021} conclude that the \textit{sum} of their ``zero-spin" and ``negative-spin" sub-populations is non-zero. This is evident from Fig.~3 of~\citet{Roulet2021} where $\zeta_{\mathrm{pos}}$ is inconsistent with $1$.

\new{
Shortly after the initial circulation of our study, an independent and complementary study of binary black hole spins was presented by~\citet{mould_which_2022}.
They sought to explore evidence for mass ratio reversal in compact binary formation, employing models in which component spins are \textit{not} independently and identically distributed.
As in our study, though, they additionally considered the possibility of zero-spin spikes appearing in the spin magnitude distribution.
Their findings corroborate our own, indicating that $<46\%$ of black hole primaries and $<36\%$ of secondaries have vanishing spins.
}

At the same time, our conclusions are generally inconsistent with those reported by~\citet{Galaudage2021}.
Although our posterior distributions do have overlap with those of~\citet{Galaudage2021} to within statistical uncertainty, differences between them result in \emph{qualitatively different conclusions regarding the nature of binary black hole spins}.
Below, we detail several differences between our analyses and those of~\citet{Galaudage2021} and comment on whether each could be contributing to the discrepancy.

First, an important categorical difference between our analyses and those conducted in~\citet{Galaudage2021} is the sample of gravitational-wave events analyzed.
While our counting experiment in Sec.~\ref{sec:counting} uses only GWTC-2~\citep{LIGOScientific:2020ibl} events, the hierarchical analyses in Sec.~\ref{sec:spike} and Sec.~\ref{sec:results} make use of the full GWTC-3 catalog (see Appendix~\ref{appendix:inference} for the exact event selection criteria).
\citet{Galaudage2021}, meanwhile, analyzed only GWTC-2 events, as GWTC-3 had not yet been released at the time of their study.
To gauge whether our differing data sets contribute to the disagreement between our own conclusions and those of~\citet{Galaudage2021}, in Appendix~\ref{appendix:gwtc2} we show results obtained using only GWTC-2 events.
Our results are qualitatively unchanged, with GWTC-2 yielding no evidence for a distinct sub-population of non-spinning black holes.
Additionally, GWTC-2 contains spin-orbit misalignment greater than $90^\circ$, though at a slightly reduced credibility compared to GWTC-3 (\gwtctwozminpercentile vs \gwtcthreezminpercentile quantiles).

Unlike our analyses, which use a single set of posterior samples for each event, \citet{Galaudage2021} make use of two distinct sets of samples for each binary, obtained under spinning and non-spinning parameter estimation priors.
In order to perform inference with both sets of samples, it is necessary to quantify the Bayes factors between these priors for each event.
As we mentioned in Sec.~\ref{sec:counting} and illustrate further in Appendix~\ref{appendix:BFs}, there is at least one event whose fully-marginalized likelihood under the non-spinning hypothesis is significantly overestimated in the data set used by \citet{Galaudage2021}. This could cause \citet{Galaudage2021} to spuriously confirm the existence of a distinct non-spinning sub-population.

Another factor may be the demand by \citet{Galaudage2021} that the zero-spin sub-population have vanishing width.
In Sec.~\ref{sec:results}, we did not fix the width $\epsilon_\mathrm{spike}$ of our approximate zero-spin spike, but let it vary as another free parameter to be inferred from the data.
In addition to concluding that the fraction $f_\mathrm{spike}$ of events occupying this spike is consistent with zero, we also found no preference for the spike to be narrow, even if it \textit{were} to exist.
If we nevertheless \textit{require} $\epsilon_\mathrm{spike}$ to be small, however, we do see an increased preference for larger $f_\mathrm{spike}$.
It is therefore possible that the strict delta-function model adopted by \citet{Galaudage2021} is systematically affecting their results.
We note, however, that our posteriors in the region $\epsilon_\mathrm{spike}\leq 0.04$ may be subject to elevated Monte Carlo variance (see Appendix~\ref{appendix:Neff}) and so we cannot confidently conclude that the demand of $\epsilon_\mathrm{spike}=0$ is responsible for the discrepant conclusions.

Although we may be able to reproduce \citet{Galaudage2021}'s measurement of $f_\mathrm{spike}$ via artificially demanding small $\epsilon_\mathrm{spike}$, we are unable to reproduce their conclusions regarding $z_\mathrm{min}$, the lower truncation bound on the component spin $\cos\theta$ distribution.
While we infer $z_\mathrm{min}$ to be confidently negative, indicating spins misaligned by more than $90^\circ$, \citet{Galaudage2021} infer this parameter to be consistent with zero or positive values.
Qualitatively, binaries with large in-plane spins are difficult to distinguish from binaries with very small or vanishing aligned spins -- both give the same $\chiEff$ values.
Therefore, if \citet{Galaudage2021} are overestimating the fraction $f_\mathrm{spike}$ of non-spinning systems, they may correspondingly be \textit{underestimating} the prevalence of systems with significant spin-orbit misalignments.
This said, neither \citet{Galaudage2021}'s results nor our own exhibit any significant correlation between $z_\mathrm{min}$ and $f_\mathrm{spike}$.

Finally, although our spin models were heavily inspired by those of~\citet{Galaudage2021}, ours \textit{do not} reduce to theirs in the limit $\epsilon_\mathrm{spike}=0$.
Our component spin models assume that individual spins are independently and identically distributed between spike and bulk sub-populations, whereas~\citet{Galaudage2021} assume that \textit{both} spins in a given binary together lie in the spike or the bulk of the component spin distribution.
Similarly, ~\citet{Galaudage2021} require that both spins are together members of the isotropic or preferentially-aligned components of the spin-tilt distribution.
We have verified that this modeling choice is not responsible for the different conclusions regarding $f_\mathrm{spike}$, as can be seen in Fig.~\ref{fig:iid_vs_noniid} in Appendix~\ref{appendix:gwtc2}.
In fact, we find that $f_\mathrm{spike}$ is constrained to even \textit{smaller} values if we alternatively adopt \citet{Galaudage2021}'s modeling convention.
This is expected: under the convention of \citet{Galaudage2021}, a component spin can only contribute to $f_\mathrm{spike}$ if its companion is \textit{also} consistent with zero spin.
Under our assumption of independent and identically distributed spins, in contrast, a component spin can be identified as a member of the zero-spin spike regardless of its companion's spin measurement.
Such differing conventions and how they impact population measurements are further elaborated upon in Appendix~\ref{appendix:gwtc2}.

Overall, we find a component spin distribution that is consistent with a single, smooth function.
Our results suggest that the spin distribution is possibly non-zero at $\chi=0$, but without a requirement for a distinct and discontinuous sub-population of non-spinning black holes (see Fig.~\ref{fig:PPDs_component_models}).
Such a behavior cannot be easily captured by the common modeling choice of a non-singular Beta distribution, which is constrained to vanish at $\chi=0$.
This tension is illustrated in the joint $\mu_\chi$-$\sigma_\chi$ posterior of Fig.~\ref{fig:component_cornerplot}, which shows the inferred mean and variance of the spin magnitude distribution railing against the prior cut that ensures non-singular behavior.
When the spin magnitudes are modeled with a single Beta distribution (green posteriors in Fig.~\ref{fig:PPDs_component_models}), $\mu_{\chi}$ must be small in order to accommodate events with small but finite spins.
Events with larger spins, meanwhile, can nominally be captured with a large $\sigma_\chi$.
However, $\sigma_\chi$ \textit{cannot} be large when $\mu_\chi$ is small, as seen by the prior cut in Fig.~\ref{fig:component_cornerplot}.
The introduction of a zero-spin ``spike" relieves this tension (even if no such spike is actually present in the data).
Small-spin events can now be captured by the spike, allowing the Beta distribution to shift to higher $\mu_\chi$ and $\sigma_\chi$ to accommodate higher-spin events.
This phenomenon can also be seen in the correlation, noted above, between $f_{\mathrm{spike}}$ and $\mu_{\chi}$: the spin magnitude distribution is best described either as a single Beta distribution that peaks at low values (small $f_{\mathrm{spike}}$ and $\mu_{\chi}$), or a ``spike" combined with a Beta distribution that peaks at higher values (large $f_{\mathrm{spike}}$ and $\mu_{\chi}$).
The combination of these two effects yields a smooth spin magnitude distribution that does not exhibit distinct sub-populations.
This discussion suggests that a Beta distribution might be a sub-optimal model for spin magnitudes going forward.

The observed presence or absence of a distinct zero-spin sub-population, rapidly spinning black holes, and/or significantly misaligned black hole spins would all carry considerable theoretical implications for the formation channels and astrophysical processes driving compact binary evolution.
At the same time, hierarchical inference of the compact binary spin distribution is technically difficult, relying on a large number of highly uncertain measurements that are themselves finitely sampled.
When drawing observational conclusions about the nature of compact binary spins, ensuring that results are reproducible under different complementary approaches can help mitigate these concerns and determine which conclusions are driven by priors, which by modeling systematics, and which by the data itself.

It certainly remains \textit{possible} that the binary black hole mergers witnessed by Advanced LIGO and Virgo contain sub-populations of non-spinning and/or nearly-aligned binaries, but this scenario is not presently required by gravitational-wave observations.
If such sub-populations were to appear in an analysis of future data, they could be taken as evidence towards  the hypothesis that merging black holes are born in the stellar field, with spins acquired primarily through tidal spin-up~\citep{Galaudage2021,Olejak2021,Belczynski2021,Stevenson:2022hmi,Mandel2022,Shao2022}.
Such astrophysical conclusions are currently unsupported by the data, however.
Further detections made in the upcoming O4 observation run and beyond will  shine further light on the nature of compact binary spins and, consequently, the evolutionary origin of the black hole mergers we see today.

\section*{Data \& Code Availability}

Code used to produce the results of this study is available via Github (\url{https://github.com/tcallister/gwtc3-spin-studies}) or Zenodo (\url{https://doi.org/10.5281/zenodo.6555167}).
Our data, meanwhile, can be obtained at~\url{https://doi.org/10.5281/zenodo.6555145}.

\section*{Acknowledgments}

We thank our anonymous referee, the AAS Data Editors, and Ilya Mandel for their valuable comments on our manuscript, as well as Eric Thrane, Colm Talbot, and Shanika Galaudage for numerous discussions about these results. We also thank Javier Roulet for sharing data from~\citet{Roulet2021} with us and for insightful feedback on this study, and Christopher Berry, Charlie Hoy, Vaibhav Tiwari, and Mike Zevin for their thoughtful comments.
This research has made use of data, software and/or web tools obtained from the Gravitational Wave Open Science Center (https://www.gw-openscience.org), a service of LIGO Laboratory, the LIGO Scientific Collaboration and the Virgo Collaboration.
Virgo is funded by the French Centre National de Recherche Scientifique (CNRS), the Italian Istituto Nazionale della Fisica Nucleare (INFN) and the Dutch Nikhef, with contributions by Polish and Hungarian institutes.
This material is based upon work supported by NSF's LIGO Laboratory which is a major facility fully funded by the National Science Foundation.
The authors are grateful for computational resources provided by the LIGO Laboratory and supported by NSF Grants PHY-0757058 and PHY-0823459.
Software: {\tt astropy}~\citep{astropy1,astropy2}, {\tt emcee}~\citep{2013PASP..125..306F}, {\tt h5py}~\citep{h5py}, {\tt jax}~\citep{jax}, {\tt matplotlib}~\citep{Hunter:2007}, {\tt numpy}~\citep{numpy}, {\tt numpyro}~\citep{numpyro1,numpyro2}, {\tt scipy}~\citep{scipy}.

\appendix
\section{Spin population models}
\label{appendix:models}

We employ two broad categories of parametrized models for the black hole spins: models on the effective spin parameters and models on the spin components. A summary of all models, their corresponding parameters, and their priors is presented in Tables~\ref{tab:chi-eff-models} and~\ref{tab:component-models}.  All component spin models ignore the azimuthal spin angles, assuming they are distributed according to the uniform prior used during the original parameter estimation~\citep{LIGOScientific:2020ibl}.
As measurements of compact binary spins and mass ratios are generally correlated, we hierarchically measure the distribution of binary mass ratios alongside spins, assuming that the secondary mass distribution follows
    \begin{equation}
    p(m_2|m_1) \propto m_2^{\beta_q} \quad \left(m_\mathrm{min} \leq m_2 \leq m_1\right)\,,
    \end{equation}
and inferring the power-law index $\beta_q$.
We adopt a Gaussian prior $N(0,3)$ on $\beta_q$.
The distributions of primary masses and redshifts, in contrast, have a negligible impact on conclusions regarding spin.
We assume that primary masses follow the \texttt{PowerLaw+Peak} model \citep{Talbot2018}, with parameters fixed to their (one-dimensional) median values as inferred in~\citet{O3b-pop}.
Following the notation of~\citet{O3b-pop}, we take $\alpha=3.5$, $m_\mathrm{min} = 5.0$, $m_\mathrm{max} = 88.2$, $\lambda_\mathrm{peak} = 0.03$, $\mu_m = 33.6$, $\sigma_m = 4.7$, and $\delta_m = 4.9$.
Meanwhile, we assume that the source-frame binary black hole merger density rate evolves as
    \begin{equation}
    R(z) \propto \frac{dV_c}{dz} \left(1+z\right)^{2.7}\,,
    \end{equation}
where $\frac{dV_c}{dz}$ is the differential comoving volume per unit redshift.

\subsection{Effective spin models}
\label{app:effectivemodels}

In this subsection, we list the various models considered when exploring the distribution of effective inspiral spins among binary black hole mergers.
See Table~\ref{tab:chi-eff-models} for illustrations of each model, as well as the priors adopted for each parameter.\\

\begin{table}[]
    \centering
    \begin{tabular}{l c|c|c|c}
         Model Name & $p(\chiEff)$ & Parameter & Prior & Comments\\
         \hline
         \hline
         &&& \\[-12pt]
         \multirow{2}{*}{\tt Gaussian} &    
         \multirow{2}{*}{\includegraphics[width=25mm,height=10mm]{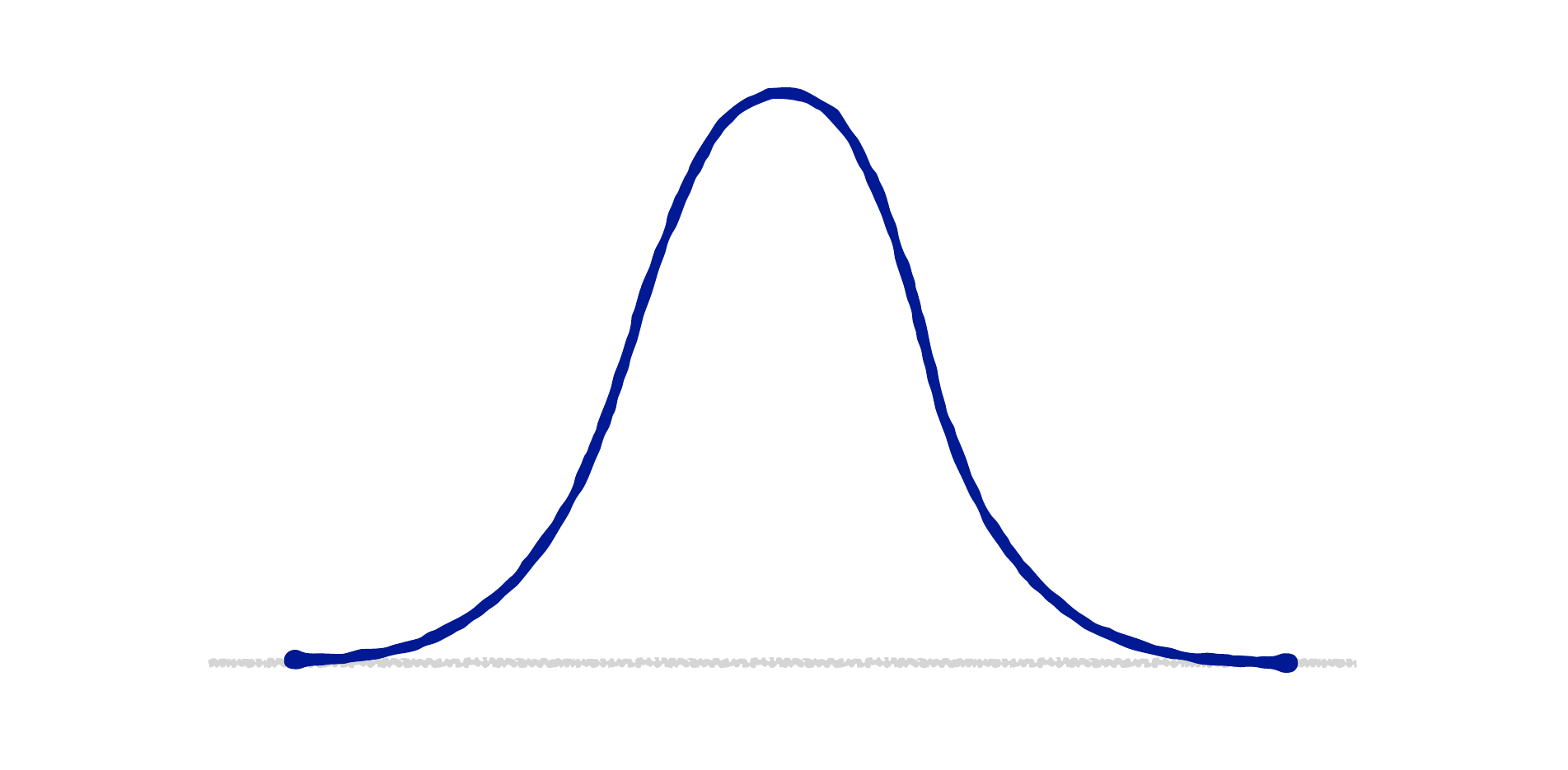}} & $\mu_\mathrm{eff}$ & U(-1,1) &  \multirow{2}{4.5cm}{Simplest Gaussian model}\\
         & & $\sigma_\mathrm{eff}$ & LU(0.01,1) &  \\[2pt]
         \hline
         &&& \\[-10pt]
         \multirow{4}{*}{\tt GaussianSpike} &
         \multirow{4}{*}{\includegraphics[height=14mm]{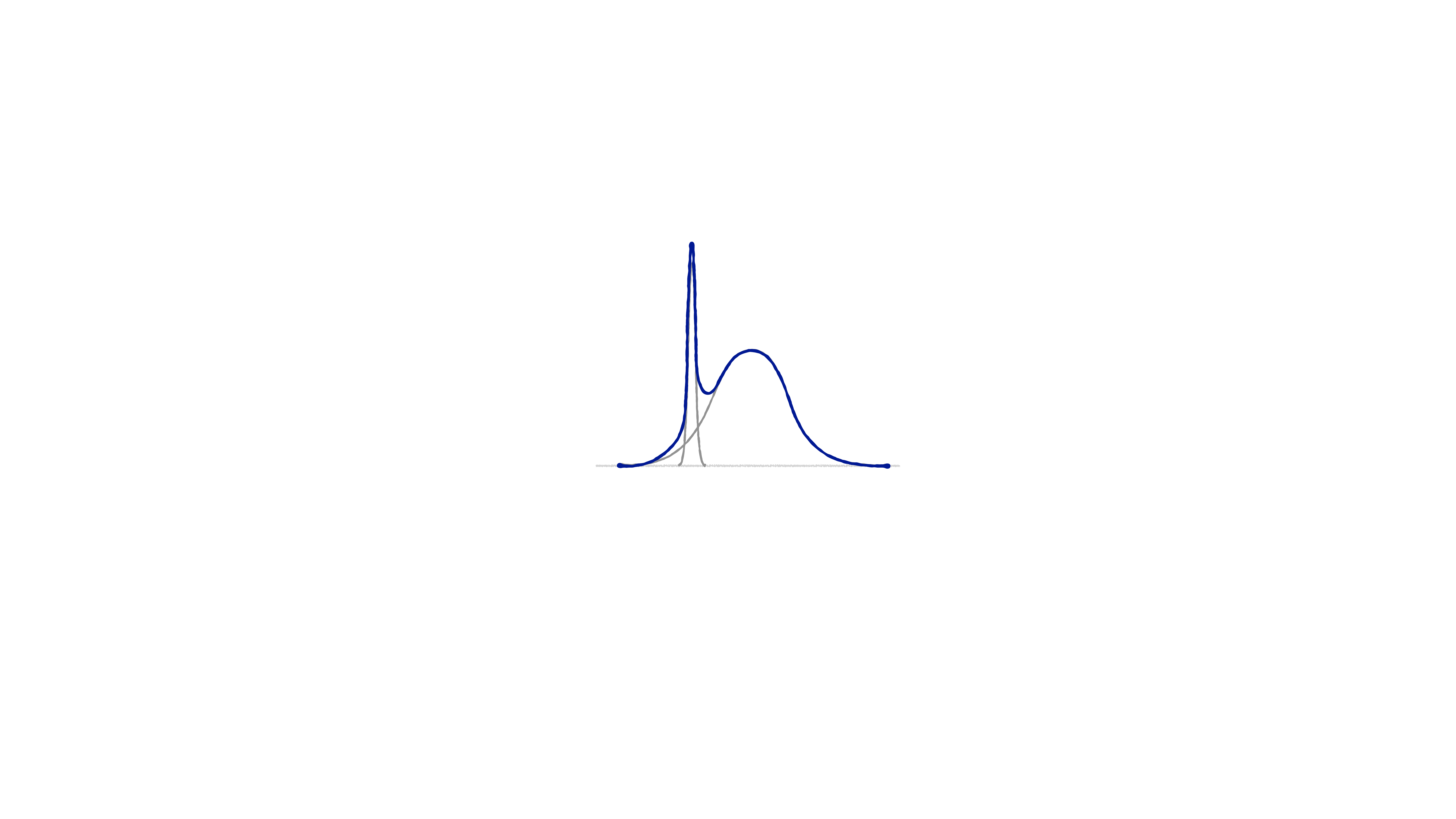}} & $\mu_\mathrm{eff}$ & U(-1,1) & \multirow{4}{4.5cm}{Mixture of a Gaussian ``bulk" with a distinct ``spike" sub-population of non-spinning systems} \\
                & & $\sigma_\mathrm{eff}$ & LU(0.01,1) & \\
                & & $\zeta_\mathrm{spike}$ & U(0,1) & \\
                & & $\epsilon$ & $\delta(0)$ & \\[2pt]
         \hline
         \multirow{5}{*}{\tt BimodalGaussian} &
         \multirow{5}{*}{\includegraphics[width=25mm,height=16mm]{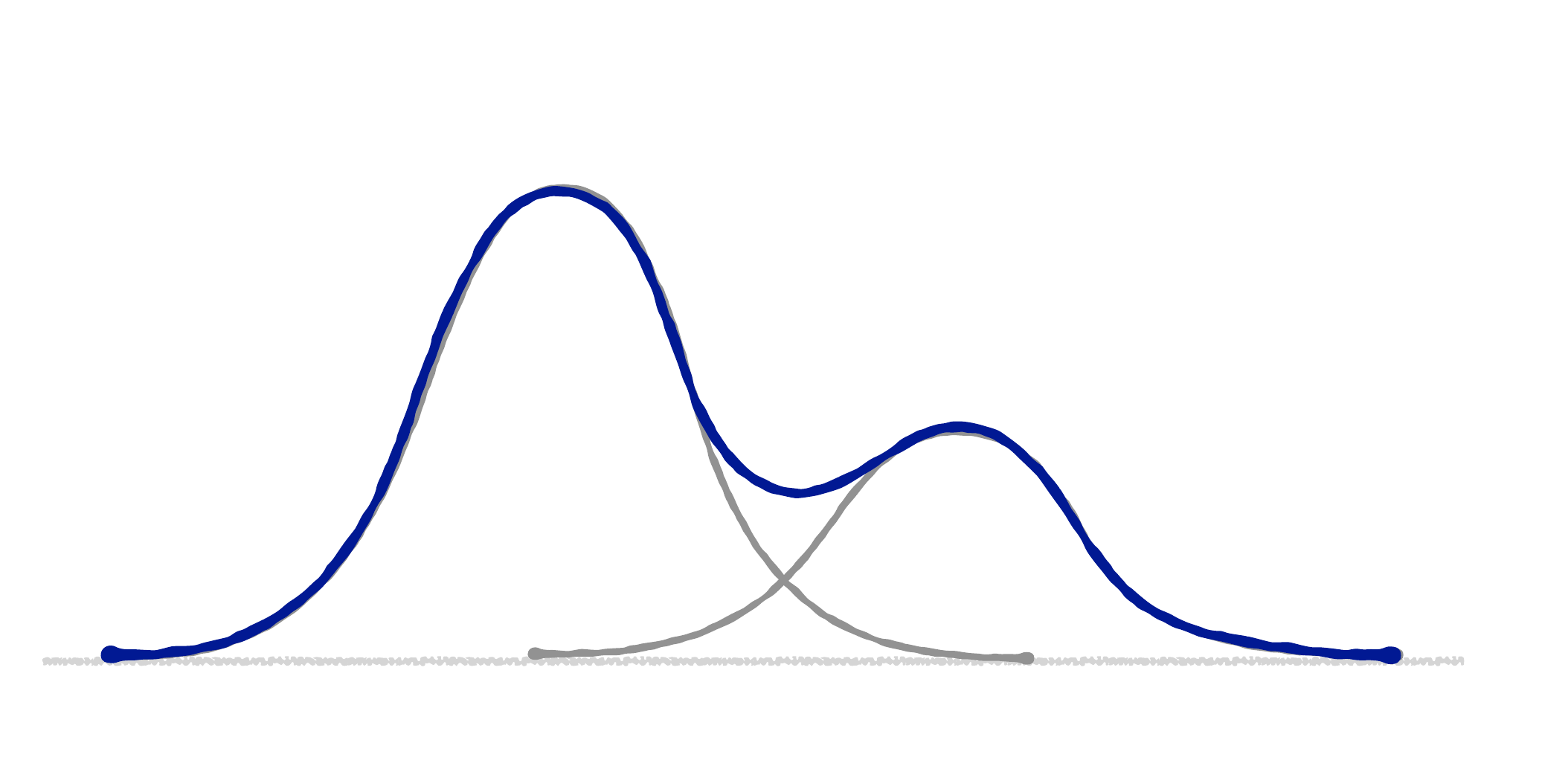}}& $\mu_{\mathrm{eff},a}$ & U(-1,1) & \multirow{5}{4.5cm}{Mixture of two Gaussians with variable means and widths} \\[-1pt]
                & & $\mu_{\mathrm{eff},b}$ & U$(-1,1)$ & \\[-1pt]
                & & $\sigma_{\mathrm{eff},a}$ & LU(0.01,1) & \\[-1pt]
                & & $\sigma_{\mathrm{eff},b}$ & LU(0.01,1) & \\[-1pt]
                & & $\zeta_a$ & U(0.5,1) & \\[2pt]
         \hline
         \hline
    \end{tabular}
    \caption{Summary of effective spin models we employ, including their names, an example $p(\chiEff)$ plot, parameters, priors, and comments.
    $\mathrm{U}(x_\mathrm{min},x_\mathrm{max})$ denotes a uniform prior between $x_\mathrm{min}$ and $x_\mathrm{max}$, $\mathrm{LU}(x_\mathrm{min},x_\mathrm{max})$ signifies a log-uniform prior across the same bounds, and $\delta(0)$ a delta-function prior at zero.}
    \label{tab:chi-eff-models}
\end{table}
%

\noindent \underline{\tt Gaussian}.
Our simplest model assumes that effective inspiral spins as Gaussian-distributed with mean $\mu_\mathrm{eff}$ and variance $\sigma^2_\mathrm{eff}$,
\begin{equation}
p(\chiEff|\mu_\mathrm{eff},\sigma_\mathrm{eff})
    = {\cal{N}}_{[-1,1]}(\chi_\mathrm{eff}|\mu_\mathrm{eff},\sigma_\mathrm{eff})\,,
\label{eq:gaussian-chi-eff}
\end{equation}
where ${\cal{N}}_{[a,b]}$ denotes a Gaussian distribution truncated within $[a,b]$.
Our priors on $\mu_\mathrm{eff}$ and $\sigma_\mathrm{eff}$ are listed in Table~\ref{tab:chi-eff-models}.
This model was proposed in~\citet{Roulet2019} and~\citet{Miller2020} and has been employed and extended in~\citet{O3a-pop},~\citet{O3b-pop},~\citet{Callister2021},~\citet{Biscoveanu2022}, and~\citet{Bavera2022}.\\

\noindent \underline{\tt GaussianSpike}.
To initially assess the prevalence of identically non-spinning black holes, we extend the {\tt Gaussian} model to include a narrow ``spike'' centered at $\chiEff=0$ with width $\epsilon \ll 1$
\begin{equation}
p(\chiEff|\zeta_\mathrm{spike},\epsilon,\mu_\mathrm{eff},\sigma_\mathrm{eff}) = \zeta_\mathrm{spike}\, {\cal{N}}_{[-1,1]}(\chiEff|0,\epsilon)+ (1 - \zeta_\mathrm{spike}) {\cal{N}}_{[-1,1]}(\chiEff|\mu_\mathrm{eff},\sigma_\mathrm{eff})\,.
\label{eq:chi-eff-model-1}
\end{equation}
The fraction of binary black holes comprising the zero-spin spike is given by $\zeta_\mathrm{spike}$.
A similar model was studied by~\citet{O3b-pop}, who imparted a finite width to the zero-spin spike population.
In this work, we fix $\epsilon=0$ following the modeling choice of~\citet{Galaudage2021}, leveraging the kernel density estimation trick discussed in Appendix~\ref{appendix:kde} to accurately and stably evaluate the likelihood for this delta-function spike.
\\

\noindent \underline{\tt BimodalGaussian}.
Given the inferred absence of a zero-width spike with {\tt GaussianSpike}, we introduce this model to more generally assess \textit{any} potential multimodality. We model the distribution of $\chiEff$ as a mixture of two Gaussians with arbitrary means and standard deviations
\begin{equation}
\begin{aligned}
p(\chiEff|\zeta_a,\mu_{\mathrm{eff},a},\sigma_{\mathrm{eff},a},
\mu_{\mathrm{eff},b},\sigma_{\mathrm{eff},b}) =& \,\zeta_a\, {\cal{N}}_{[-1,1]}(\chiEff|\mu_{\mathrm{eff},a},\sigma_{\mathrm{eff},a}) \\
&+ (1 - \zeta_a) {\cal{N}}_{[-1,1]}(\chiEff|\mu_{\mathrm{eff},b},\sigma_{\mathrm{eff},b})\,.
\end{aligned}
\label{eq:chi-eff-model-2}
\end{equation}
The mixing fraction $\zeta_a$ is constrained to be $\geq 0.5$, solving the ``label-switching'' problem by demanding that $\mu_{\mathrm{eff},a}$ and $\sigma_{\mathrm{eff},a}$ are the mean and standard deviation of the dominant sub-component.
No further prior constraints are placed on $\mu_{\mathrm{eff},a}$ and $\sigma_{\mathrm{eff},a}$ or $\mu_{\mathrm{eff},b}$ and $\sigma_{\mathrm{eff},b}$.

\subsection{Component spin models}
\label{app:componentmodels}

We next list the various models used to study the distribution of component spin magnitudes and orientations among binary black holes.
See Table~\ref{tab:component-models} for illustrations and priors.\\

\begin{table}[]
    \centering
    \begin{tabular}{l c|c|c|c}
         Model Name & $p(\chi)$ \qquad \qquad \quad $p(\cos \theta)$ & Parameter & Prior & Comments\\
         \hline
         \hline
         &&& \\[-12pt]
         \multirow{4}{*}{\tt Beta+Mixture} &    
         \multirow{4}{*}{\includegraphics[width=40mm,height=13mm]{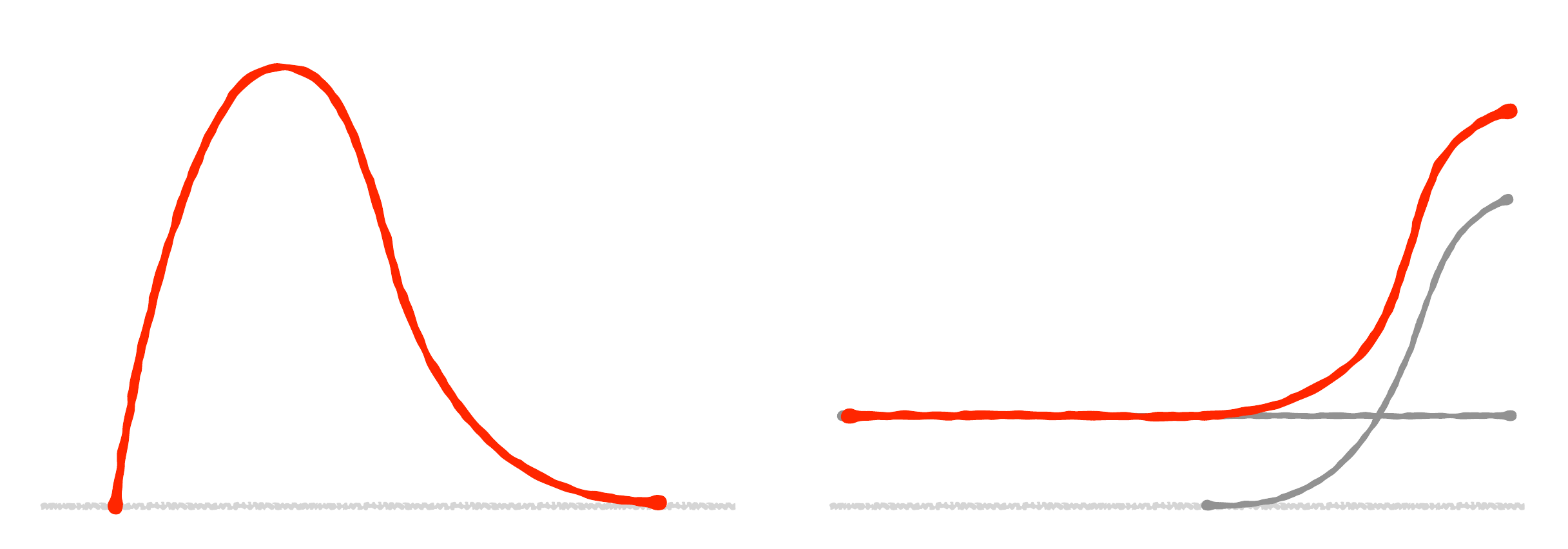}} & $\mu_\chi$ & U(0,1) & \multirow{4}{4cm}{Simplest component spin model; additional prior requirement that $\alpha, \beta>1$, see Eq.~\eqref{eq:mu-sigma-to-alpha-beta}} \\
         & & $\sigma_\chi$ & U(0.07,0.5) & \\
         & & $\sigma_t$ & U(0.1,4) & \\
         & & $f_\mathrm{iso}$ &  U(0,1) & \\[2pt]
         \hline
         \multirow{5}{*}{\tt Beta+TruncatedMixture} &    
         \multirow{5}{*}{\includegraphics[width=40mm,height=15mm]{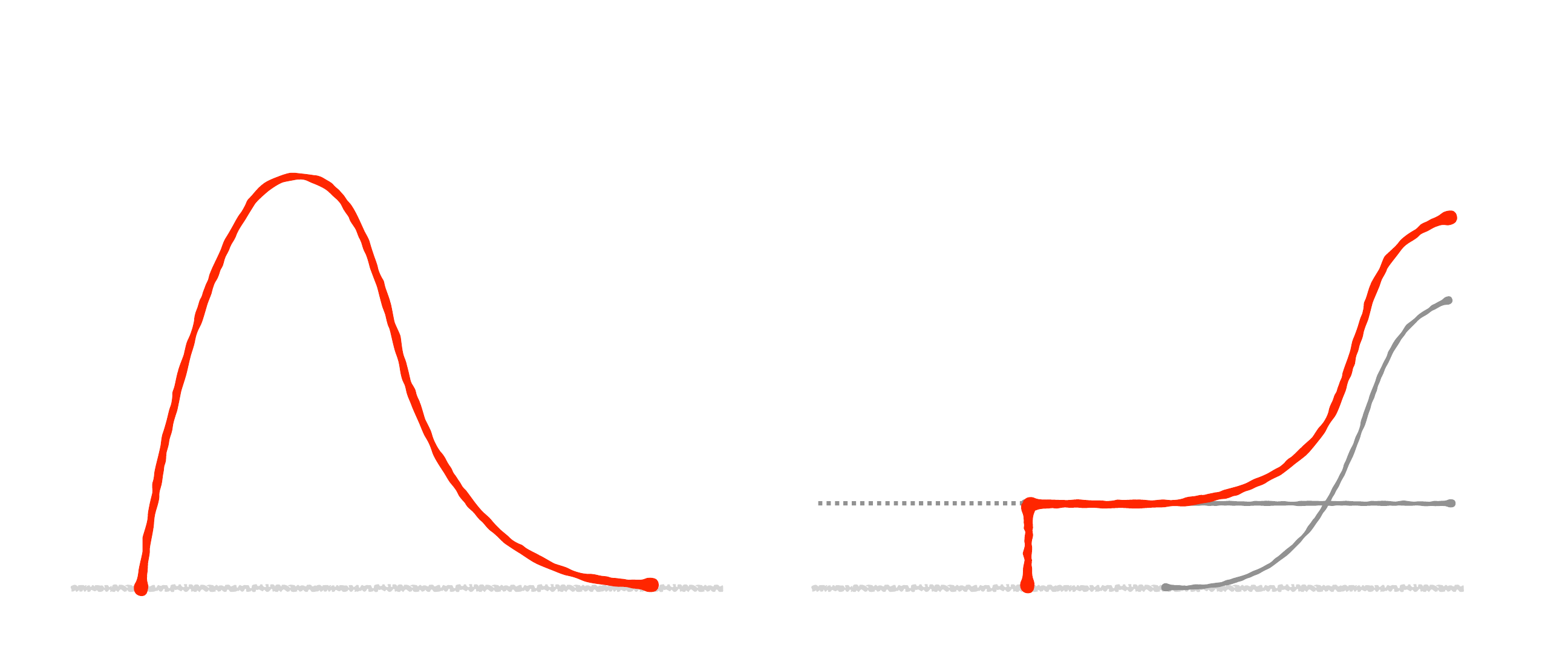}} & $\mu_\chi$ & U(0,1)  & \multirow{5}{4cm}{Introduces a truncation in the $\cos\theta$ distribution at some minimum value (maximum misalignment angle)} \\
         & & $\sigma_\chi$ & U(0.07,0.5) & \\
         & & $\sigma_t$ & U(0.1, 4) & \\
         & & $f_\mathrm{iso}$ &  U(0,1) & \\
         & & $z_\mathrm{min}$ &  U(-1,1) & \\[2pt]
         \hline
         \multirow{6}{*}{\tt BetaSpike+Mixture} &    
         \multirow{6}{*}{\includegraphics[width=40mm,height=15mm]{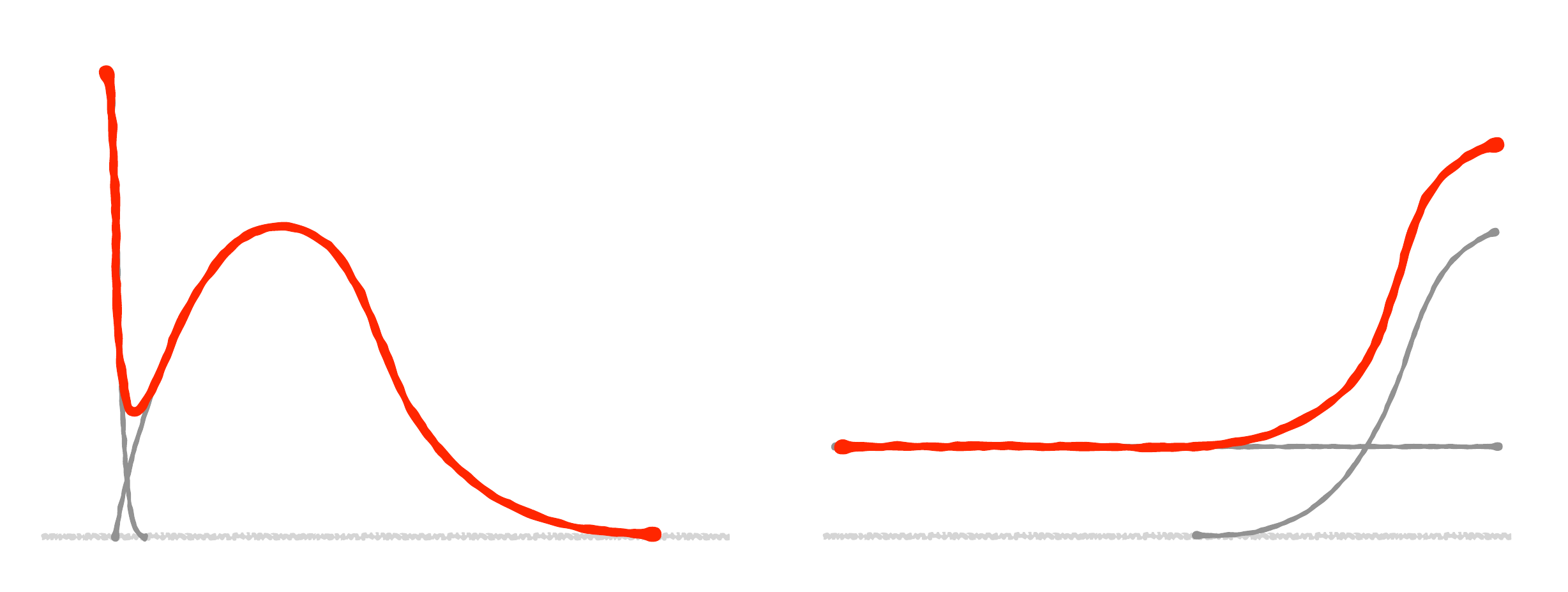}} & $\mu_\chi$ & U(0,1) & \multirow{6}{4cm}{Spin magnitude distribution modified to include Beta-distribution ``bulk" and a finite-width ``spike" at approximately zero spin} \\
         & & $\sigma_\chi$ & U(0.07,0.5) & \\
         & & $f_\mathrm{spike}$ &  U(0,1) & \\
         & & $\epsilon_\mathrm{spike}$ & U(0.03, 0.1) & \\
         & & $\sigma_t$ & U(0.1,4) & \\
         & & $f_\mathrm{iso}$ &  U(0,1) & \\[2pt]
         \hline
         \multirow{7}{*}{\tt BetaSpike+TruncatedMixture} & 
         \multirow{7}{*}{\includegraphics[width=40mm,height=15mm]{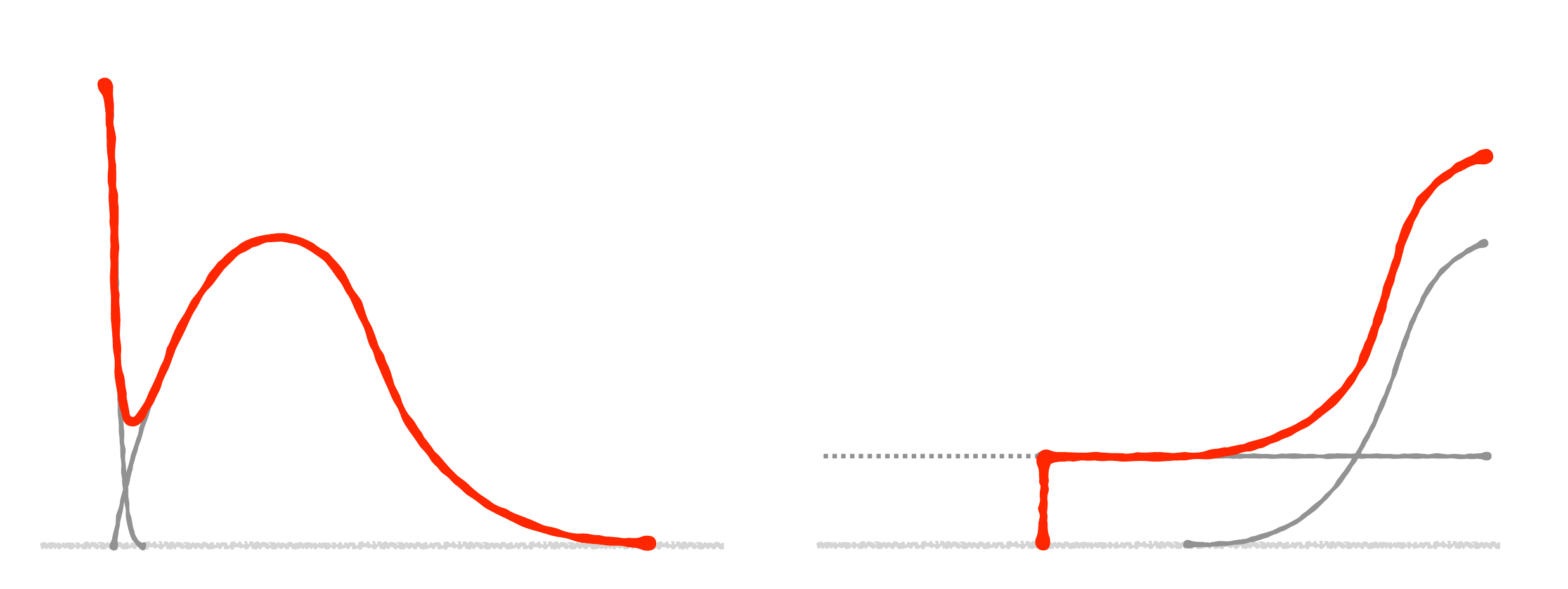}} & $\mu_\chi$ & U(0,1) & \multirow{7}{4cm}{Most complex component spin model; includes both the multiple two spin magnitude sub-populations and the truncation in the spin tilt distribution} \\
         & & $\sigma_\chi$ & U(0.07,0.5) & \\
         & & $f_\mathrm{spike}$ & U(0,1) & \\
         & & $\epsilon_\mathrm{spike}$ & U(0.03, 0.1) & \\
         & & $\sigma_t$ & U(0.1, 4) & \\
         & & $f_\mathrm{iso}$ &  U(0,1) & \\
         & & $z_\mathrm{min}$ &  U(-1,1) & \\[2pt]
         \hline
         \hline
    \end{tabular}
    \caption{Summary of component spin models we employ, including their names, an example $\chi$ and $\cos \theta$ plot, parameters, priors, and comments. $\mathrm{U}(x_\mathrm{min},x_\mathrm{max})$ denotes a uniform prior between $x_\mathrm{min}$ and $x_\mathrm{max}$, while $\mathrm{LU}(x_\mathrm{min},x_\mathrm{max})$ signifies a log-uniform prior.}
    \label{tab:component-models}
\end{table}
%

\noindent \underline{\tt Beta+Mixture}:
In our simplest component spin model, we assume that spin magnitudes are independently draw from identical Beta distributions,
\begin{equation}
p(\chi_i|\alpha,\beta) = \frac{\chi_i^{\alpha-1} (1-\chi_i)^{\beta-1}}{c(\alpha,\beta)},
\label{eq:default-magnitude}
\end{equation}
where $c(\alpha,\beta)$ normalizes the distribution to unity.
The two shape parameters $\alpha$ and $\beta$ are constrained such that $\alpha, \beta > 1$ to ensure that the distribution is bounded.
This choice of parameters \textit{requires} that $p(\chi_i=0)=p(\chi_i=1)=0$, thus \emph{a priori} assuming that there are no systems with exactly vanishing spin.
We sample not in $\alpha$ and $\beta$ directly but rather in the mean $\mu_\chi$ and standard deviation $\sigma_\chi$ of the Beta distribution.
The shape parameters $\alpha$ and $\beta$ are calculated from $\mu_\chi$ and $\sigma_\chi$ through 
\begin{equation}
    \alpha =\mu_\chi \nu ~ , \quad
      \beta = (1-\mu_\chi) \nu , 
      \quad \nu = \frac{\mu_\chi(1-\mu_\chi)}{\sigma_\chi^2} - 1 \,.
\label{eq:mu-sigma-to-alpha-beta}
\end{equation}
The region of the $\mu_\chi$ and $\sigma_\chi$ parameter space excluded by restriction $\alpha,\beta>1$ can be seen in Figure~\ref{fig:component_cornerplot}.
The spin tilt distribution is a mixture between two components: a uniform isotropic component and a preferentially-aligned component
\begin{equation}
\begin{aligned}
p(\cos\theta_i|f_\mathrm{iso},\sigma_t) = \frac{f_\mathrm{iso}}{2} + (1-f_\mathrm{iso})
    {\cal{N}}_{[-1,1]}(\cos\theta_i|1,\sigma_t)\,.
\end{aligned}
\label{eq:default-tilt}
\end{equation}
Here, $f_\mathrm{iso}$ is the fraction of events comprising the isotropic subpopulation, while the second term is a half-Gaussian peaking at $\cos\theta = 1$.
Each component spin tilt is independently drawn from Eq.~\eqref{eq:default-tilt}.
A model of this form was introduced in~\citet{Talbot:2017yur} and \citet{Wysocki2019}, with the restriction that both spin tilts together lie in either the isotropic or the preferentially-aligned component, and is referred to as the \texttt{Default} model in~\citet{O3a-pop}, \citet{O3b-pop}, and \citet{Galaudage2021} \\

\noindent \underline{\tt BetaSpike+Mixture}.
In order to assess the presence of non-spinning black hole binaries, we emulate~\citet{Galaudage2021} and extend the \texttt{Beta+Mixture} model include a half-Gaussian ``spike" that peaks at $\chi_i=0$ but has a finite width $\epsilon_\mathrm{spike}$:
\begin{equation}
    p(\chi_i | \alpha, \beta, f_\mathrm{spike}, \epsilon_\mathrm{spike}) = f_\mathrm{spike}  {\cal{N}}_{[0,1]}(\chi_i|0,\epsilon_\mathrm{spike}) + (1-f_\mathrm{spike})  \frac{\chi_i^{1-\alpha} \, (1-\chi_i)^{1-\beta}}{c(\alpha, \beta)}.
\label{eq:betaspike}
\end{equation}
A fraction $f_\mathrm{spike}$ of the population falls in this (approximately) zero-spin spike while $1 - f_\mathrm{spike}$ lives in the bulk, parametrized again by a Beta distribution.
Unlike~\citet{Galaudage2021}, we do not require both black holes in a given binary to occupy the same sub-population, but instead assume that component spins are independently and identically drawn from Eq.~\eqref{eq:betaspike}.
The spin tilt distribution is the same as that in the {\tt Beta+Mixture} model, given by Eq.~\eqref{eq:default-tilt}. \\

\noindent \underline{\tt Beta+TruncatedMixture}.
Further motivated by~\citet{Galaudage2021}, we alternately extend the \texttt{Beta+Mixture} model by instead modifying the $\cos\theta$ distribution, augmenting it with a tunable lower truncation bound $z_\mathrm{min}$:
\begin{equation}
    p(\cos\theta_i|f_\mathrm{iso},\sigma_t) = \frac{f_\mathrm{iso}}{1-z_\mathrm{min}} + (1-f_\mathrm{iso})
        \mathcal{N}_{[z_\mathrm{min},1]}(\cos\theta_i|1,\sigma_t) \,,
\label{eq:tiltzmin}
\end{equation}
with $p(\cos\theta_i)=0$ for $\cos\theta_i \leq z_\mathrm{min}$.
The spin magnitude distribution is the same as that in the {\tt Beta+Mixture} model, given in Eq.~\eqref{eq:default-magnitude}. \\

\noindent \underline{\tt BetaSpike+TruncatedMixture}.
Finally, we consider both extensions together, incorporating both the (finite-width) zero-spin sub-population in Eq.~\eqref{eq:betaspike} and the truncated spin tilt distribution in Eq.~\eqref{eq:tiltzmin}.

\section{Hierarchical inference and data analyzed}
\label{appendix:inference}

In this appendix we briefly describe our hierarchical inference framework as well as the exact data we use in this study.
Given a catalog of gravitational-wave events with data $\{d_i\}_{i=1}^{N_\mathrm{obs}}$, the likelihood that such events arise from a population with parameters $\Lambda$ is~\citep{Mandel2019,Loredo2004,Fishbach2018,Vitale2020}
    \begin{equation}
    p(\{d\}|\Lambda) \propto \xi(\Lambda)^{-N_\mathrm{obs}} \prod_i
        \int d\lambda_i \,p(d_i|\lambda_i) p(\lambda_i|\Lambda)\,,
    \label{eq:likelihood}
    \end{equation}
where $\lambda_i$ denotes the parameters (masses, spins, etc.) of each individual binary, and we have marginalized over the overall rate of binary mergers assuming a logarithmically-uniform prior.
The detection efficiency $\xi(\Lambda)$ is the fraction of events that we expect to successfully detect given the proposed population $\Lambda$.
If $P_\mathrm{detection}(\lambda)$ is the probability that an event with parameters $\lambda$ is successfully recovered by detection pipelines, then
    \begin{equation}
    \xi(\Lambda) = \int d\lambda \,p(\lambda|\Lambda) P_\mathrm{detection}(\lambda)\,.
    \label{eq:detection-efficiency}
    \end{equation}

In order to evaluate Eq.~\eqref{eq:likelihood}, we require the likelihood $p(d_i|\lambda_i)$ for each detection.
In most cases, however, we do not have access to these likelihoods, but rather the \textit{posterior} $p(\lambda_i|d_i)$ obtained using some default prior $p_\mathrm{pe}(\lambda)$.
In this case we rewrite Eq.~\eqref{eq:likelihood} as
    \begin{equation}
    p(\{d\}|\Lambda) \propto \xi(\Lambda)^{-N_\mathrm{obs}} \prod_i
        \int d\lambda_i \,\frac{p(\lambda_i|d_i)}{p_\mathrm{pe}(\lambda_i)}  p(\lambda_i|\Lambda)\,.
    \label{eq:likelihood-post}
    \end{equation}
Furthermore, we generally do not know $p(\lambda_i|d_i)$ itself, but instead have a discrete set of \textit{independent samples} $\{\lambda_{i,j}\}_{j=1}^{N_i}$ drawn from $p(\lambda_i|d_i)$, where $N_i$ is the number of posterior samples used for event $i$.
The standard course of action is to approximate the integral within Eq.~\eqref{eq:likelihood-post} via a Monte Carlo average over these posterior samples,
    \begin{equation}
    p(\{d\}|\Lambda) \propto \xi(\Lambda)^{-N_\mathrm{obs}}
        \prod_i
        \frac{1}{N_i}\sum_{j=1}^{N_i} \frac{p(\lambda_{i,j}|\Lambda)}{p_\mathrm{pe}(\lambda_{i,j})}\,.
    \label{eq:likelihood-mc}
    \end{equation}
We can similarly recast the detection efficiency [Eq.~\eqref{eq:detection-efficiency}] as a Monte Carlo average.
Given a number $N_\mathrm{inj}$ of injected signals drawn from some reference distribution $p_\mathrm{inj}(\lambda)$, the detection efficiency may be approximated via
    \begin{equation}
    \xi(\Lambda) = \frac{1}{N_\mathrm{inj}} \sum_{i=1}^{N_\mathrm{found}} \frac{p(\lambda_i|\Lambda)}{p_\mathrm{inj}(\lambda_i)}\,,
    \label{eq:detection-efficiency-mc}
    \end{equation}
summing over the $N_\mathrm{found}$ injections that pass our detection criteria.
    
The fundamental assumption made in Eqs.~\eqref{eq:likelihood-mc} and \eqref{eq:detection-efficiency-mc} is that posterior samples (and recovered injections) are sufficiently dense relative to the population features of interest.
In this paper, however, we are concerned with very narrow features in the binary black hole spin distribution: Gaussians of width $\epsilon \ll 1$ or even true delta functions.
We cannot therefore be automatically assured that Eqs.~\eqref{eq:likelihood-mc} and \eqref{eq:detection-efficiency-mc} will accurately represent our likelihood and detection efficiency.
In Appendix~\ref{appendix:Neff} we will further assess the accuracy of these Monte Carlo averages, and in Appendix~\ref{appendix:kde} describe how to bypass them completely.

For this study, we consider the subset of black holes among GWTC-3 with false alarm rates below $1\,\mathrm{yr}^{-1}$~\citep{LIGOScientific:2021djp}.
We do not include GW190814~\citep{GW190814} or GW190917~\citep{the_ligo_scientific_collaboration_gwtc-21_2021}, both of which have secondary masses $m_2 < 3\,M_\odot$ and are outliers with respect to the binary black hole population~\citep{O3a-pop,O3b-pop}.
This leaves us with a total of 69 binary black holes in our sample.
We use parameter estimation samples made publicly available through the Gravitational-Wave Open Science Center\footnote{https://www.gw-openscience.org/}~\citep{LIGOScientific:2019lzm,losc}.
For events first published in GWTC-1\footnote{GWTC-1 samples available at https://dcc.ligo.org/LIGO-P1800370/public}~\citep{the_ligo_scientific_collaboration_gwtc-1:_2018} we use the ``\texttt{Overall\_posterior}'' parameter estimation samples.
We adopt the ``\texttt{PrecessingSpinIMRHM}'' samples  for events first published in GWTC-2\footnote{GWTC-2 samples available at https://dcc.ligo.org/LIGO-P2000223/public}~\citep{LIGOScientific:2020ibl} and GWTC-2.1\footnote{GWTC-2.1 samples available at https://zenodo.org/record/5117703}\citep{the_ligo_scientific_collaboration_gwtc-21_2021}, and for new events in GWTC-3~\citep{LIGOScientific:2021djp} use the ``\texttt{C01:Mixed}'' samples\footnote{GWTC-3 samples available at https://zenodo.org/record/5546663}.
These choices correspond to a union of samples obtained with different waveform families.
All samples include spin precession effects, while the \texttt{PrecessingSpinIMRHM} and \texttt{C01:Mixed} samples from GWTC-2, GWTC-2.1, and GWTC-3 additionally include the effects of higher order modes (parameter estimation accounting for higher order modes was not available in GWTC-1).
We evaluate the detection efficiency using the set of successfully recovered binary black hole injections, provided by the LIGO-Virgo-KAGRA collaborations, spanning their first three observing runs\footnote{https://zenodo.org/record/5636816}.

\section{Model Checking}
\label{appendix:ppc}

\new{
In order to trust physical conclusions drawn from phenomenological models, we must be sure that the models themselves provide a reasonable fit to observation.
A particularly powerful means of model-checking is to perform predictive tests: comparing the \textit{predictions} made by a fitted model to our actual observed data.
In this section, we subject each of the effective and component spin models used in this paper to such predictive tests.
}

\begin{figure*}
    \centering
    \includegraphics[width=0.82\textwidth]{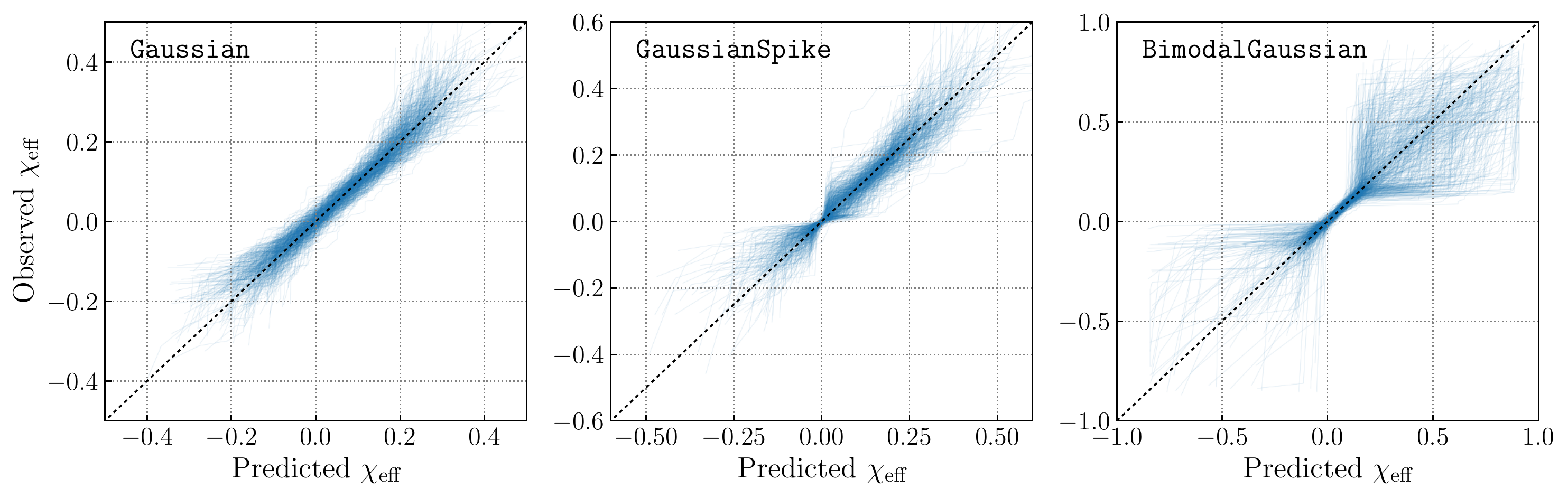}
    \caption{
    \new{
    Predictive checks of the three effective spin models explored in Sect.~\ref{sec:spike}.
    Under each model, we repeatedly generate sets of ``Observed'' $\chi_\mathrm{eff}$ values drawn from our binary black hole observations, and sets of ``Predicted'' values according to the fitted models; see Appendix~\ref{appendix:ppc} for further details.
    Each trace among the three subplots represents one such realization, with the spread among traces reflecting both our uncertainty in the properties of each observed binary as well as uncertainty in model hyperparameters.
    A model that provides a good fit to observation will yield an ensemble of traces centered symmetrically around the diagonal, whereas any systematic departure from the diagonal would indicate model failure.
    No such departures are seen, and so all three effective spin models are able to successfully reproduce observation.
    }
    }
    \label{fig:effective_ppd}
\end{figure*}

\new{
Figure~\ref{fig:effective_ppd} compares predicted and observed $\chi_\mathrm{eff}$ values under the \texttt{Gaussian}, \texttt{GaussianSpike}, and \texttt{BimodalGaussian} models explored in Sect.~\ref{sec:spike}.
The ensemble of traces in each panel reflects the uncertainties in the hyperparameters of each model, as well as our uncertainties in the observed properties of each event.
Specifically, these figures are generated via the following algorithm:
    \begin{enumerate}
        \setcounter{enumi}{-1}
        \item Perform hierarchical inference using the model in question, as described in Appendix~\ref{appendix:inference}, to obtain posteriors on the hyperparameters $\Lambda$ of the model.
        \item From the posterior on $\Lambda$, draw a single hyperparameter sample $\Lambda_i$.
        \item Use this $\Lambda_i$ to define a new prior $p(\lambda|\Lambda_i)$ on the properties of individual events.
        Reweight the posterior of each observed binary to this new prior, as in e.g.~\citet{Miller2020}, and randomly select a single sample $\lambda_j$ from each event.
        The resulting set $\{\lambda\}_\mathrm{obs}$ constitutes one realization of ``observed'' values.\footnote{Note that reweighting single-event posteriors following Steps 1 and 2 avoids any ``double-counting'' of information~\citep{Callister-reweighting}.}
        \item Similarly, reweight the set of successfully found injections described above to the proposed prior $p(\lambda|\Lambda_i)$.
        Randomly select $N_\mathrm{obs}$ values from these injections; the resulting set $\{\lambda\}_\mathrm{inj}$ is one realization of ``predicted'' values.
        Drawing predicted values from found injections, rather than directly from the proposed population $\Lambda_i$, serves to accurately capture relevant selection effects.
        \item Independently sort $\{\lambda\}_\mathrm{obs}$ and $\{\lambda\}_\mathrm{inj}$ and plot them against one another, yielding a single ``Observed vs. Predicted'' trace as in Fig.~\ref{fig:effective_ppd}.
        \item Repeat Steps 1-4.
    \end{enumerate}
    }

\new{
A model that accurately captures features in our observed data will yield an ensemble of traces centered on the $\mathrm{Predicted}\,\chi_\mathrm{eff}=\mathrm{Observed}\,\chi_\mathrm{eff}$ diagonal, shown as a dashed black line.
Systematic departures away from the diagonal, on the other hand, would indicate a failure of the fitted model to reflect true features in the data.
All three effective spin models show good predictive power in Fig.~\ref{fig:effective_ppd}, yielding traces distributed symmetrically about the diagonal.
Note that the large variance in the \texttt{BimodalGaussian} results reflects our correspondingly large uncertainty about the $\chi_\mathrm{eff}$ distribution at very negative and very positive values.
The elevated variance is symmetric about the diagonal, though, and so is not a sign of model failure.
}

\begin{figure*}[h!]
    \centering
    \includegraphics[height=4.55in]{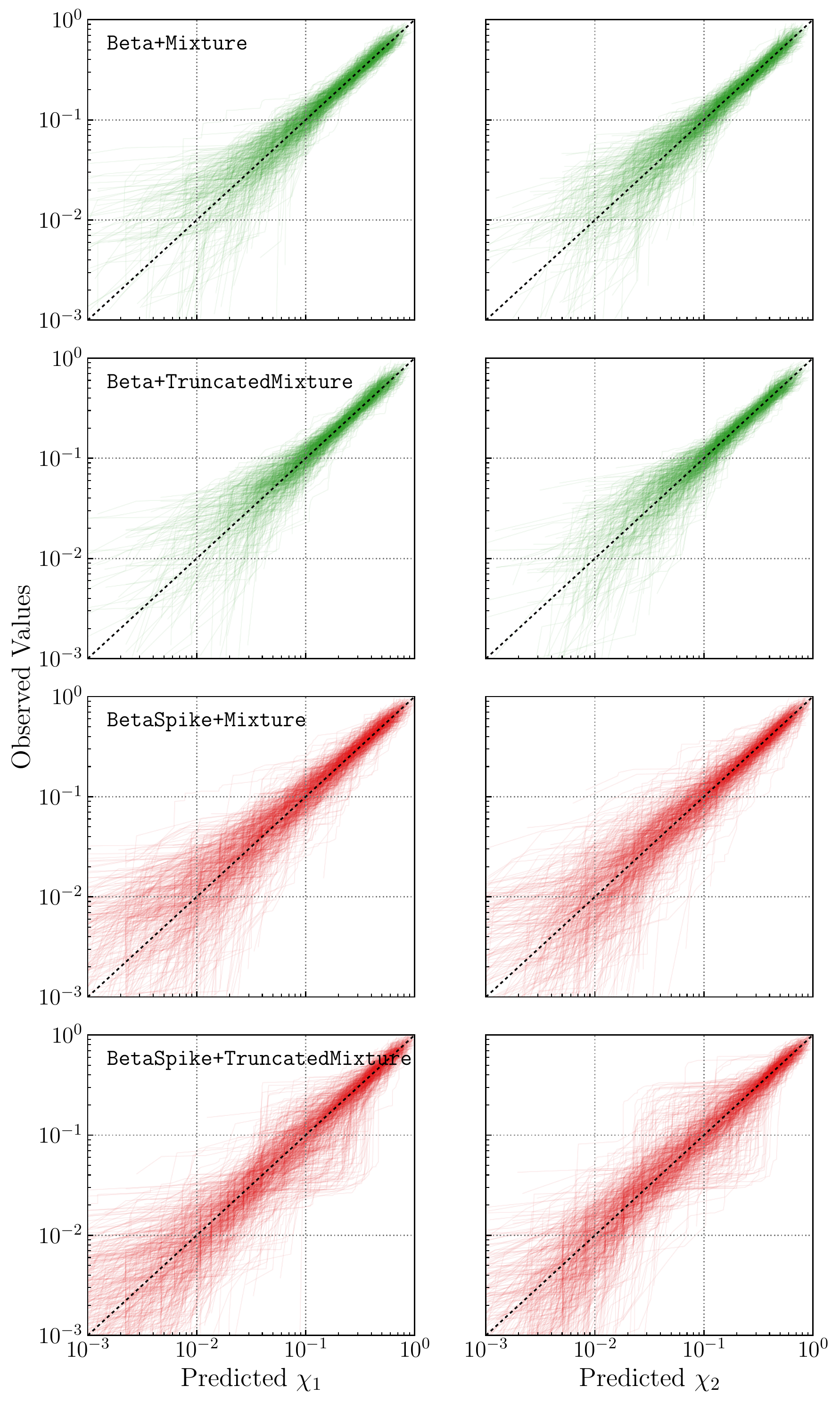}
    \includegraphics[height=4.55in]{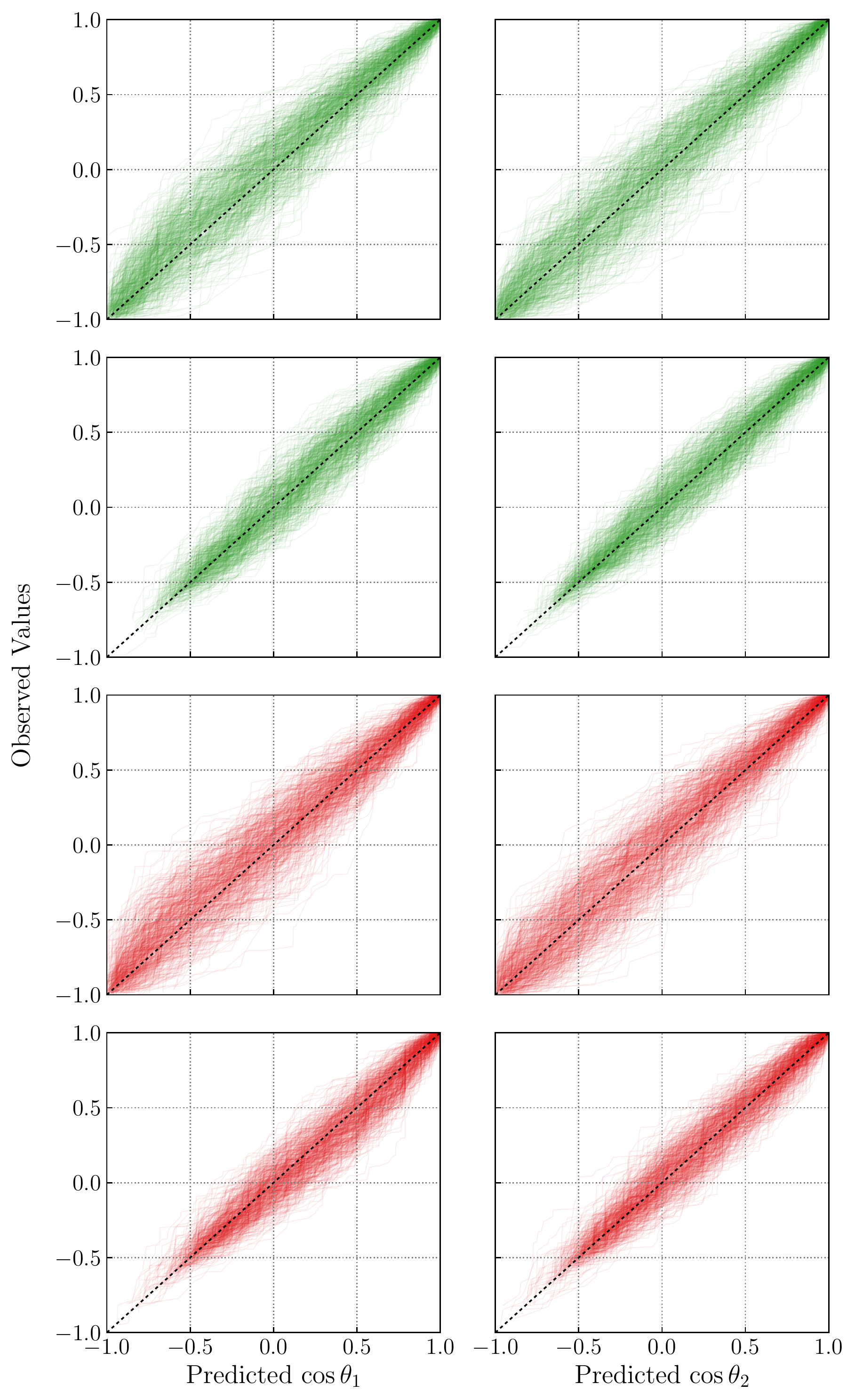}
    \includegraphics[height=4.55in]{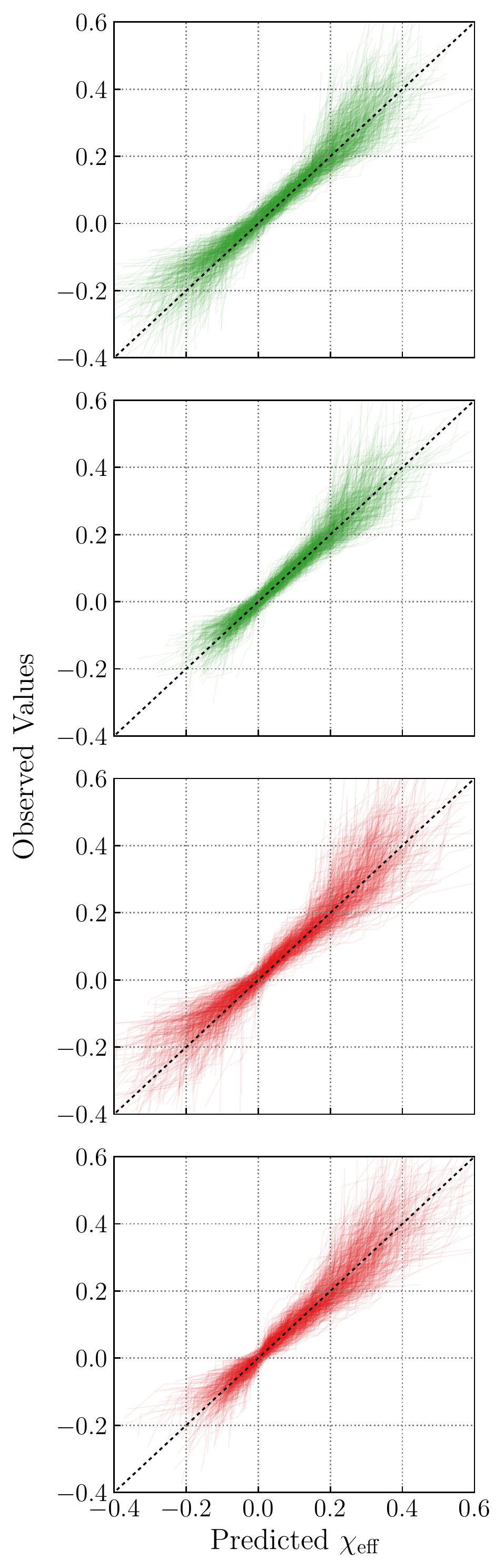}
    \caption{
    \new{
    As in Fig.~\ref{fig:effective_ppd} above, but for the component spin models studied in Sect.~\ref{sec:results}.
    We explore each model's ability to predict the observed spin magnitudes (first and second columns), spin tilts (third and fourth columns), and effective spins (fifth column).
    Again, we see no clear departures from the diagonal, such that none of the four models can be rejected as a poor descriptor of our data.
    The \texttt{Beta+Mixture} and \texttt{BetaSpike+Mixture} models may exhibit slight tension with observation, possibly over-predicting the occurrence of very negative $\cos\theta_1$ and $\chi_\mathrm{eff}$, but additional data are required to confirm the significance of this tendency.
    }
    }
    \label{fig:component_ppd}
\end{figure*}

\new{
Figure~\ref{fig:component_ppd} shows analogous predictive checks using the four component spin models explored in Sect.~\ref{sec:results}.
We investigate each model's ability to predict observed spin magnitudes (first and second columns), spin tilts (third and fourth columns), and effective spins (final column).
Each model shows good predictive power, with no significant departures from the diagonal that would indicate model failure.
The \texttt{Beta+Mixture} and \texttt{BetaSpike+Mixture} models, though, may show slight signs of tension in their ability to predict $\cos\theta_1$ and $\chi_\mathrm{eff}$.
For each model, 70\% of traces lie above the diagonal at $\mathrm{Predicted}\,\cos\theta_1 = -0.5$, possibly indicating a tendency to over-predict the occurrence of large and negative $\cos\theta$ values.
Accordingly, both models slightly over-predict the prevalence of large negative $\chi_\mathrm{eff}$, with $\sim75\%$ of traces rising above the diagonal at $\mathrm{Predicted}\,\chi_\mathrm{eff} = -0.2$.
This behaviour is consistent with the preference for a truncation at $z_\mathrm{min} \approx -0.5$ found using the \texttt{Beta+TruncatedMixture} and \texttt{BetaSpike+TruncatedMixture} models.
These tendencies are slight, however, and so cannot yet be taken as an indicator of model failure; more data will be required to better resolve the underlying spin distributions and determine which (if any) models can no longer provide a good descriptor of observation.
}

\section{Effective samples}
\label{appendix:Neff}

In order to identify and mitigate possible issues due to such finite sampling effects when estimating Eq.~\eqref{eq:likelihood-mc}, we track the number of ``effective samples'' informing the Monte Carlo estimates of the likelihood for every event.
Given a set of $N_i$ posterior samples $\{\lambda_{i,j}\}_{j=1}^{N_i}$ for each event $i$, the number of ``effective samples'' under a proposed population $\Lambda$ is
    \begin{equation}
        N_{\mathrm{eff},i}(\Lambda) \equiv \frac{\left[ \sum_{j=1}^{N_i} w_{i,j}(\Lambda)\right]^2}{\sum_{j=1}^{N_i} \left[ w_{i,j}(\Lambda)\right]^2}\,,
        \label{eq:Neff}
    \end{equation}
where $w_{i,j}(\Lambda) = p(\lambda_{i,j}|\Lambda)/p_\mathrm{pe}(\lambda_{i,j})$.
Small $N_{\mathrm{eff},i}(\Lambda)$ indicates that the given event is sparsely sampled in the region where the population $p(\lambda|\Lambda)$ is concentrated, and hence the likelihood may be dominated by sampling variance.
In these regions, we should not necessarily trust the results of Monte-Carlo-based hierarchical inference.
To avoid such regions, \citet{O3b-pop} impose a data-dependent \textit{prior cut}, rejecting those $\Lambda$ for which $\mathrm{min}\left[N_{\mathrm{eff},i}(\Lambda)\right]$ falls below some threshold value.
We avoid imposing such a cut, which amounts to an implicit (and often stringent) prior on $\Lambda$.
Instead, we compute and track $\mathrm{min}\left[N_{\mathrm{eff},i}(\Lambda)\right]$ for each component spin population model we consider.
Additionally, for models with distinct ``spike'' and ``bulk'' spin sub-populations (i.e. the half-Gaussian at zero and the Beta distribution, respectively), we track the minimum effective sample count within each of these sub-populations.

\begin{figure*}
    \centering
    \includegraphics[width=\textwidth]{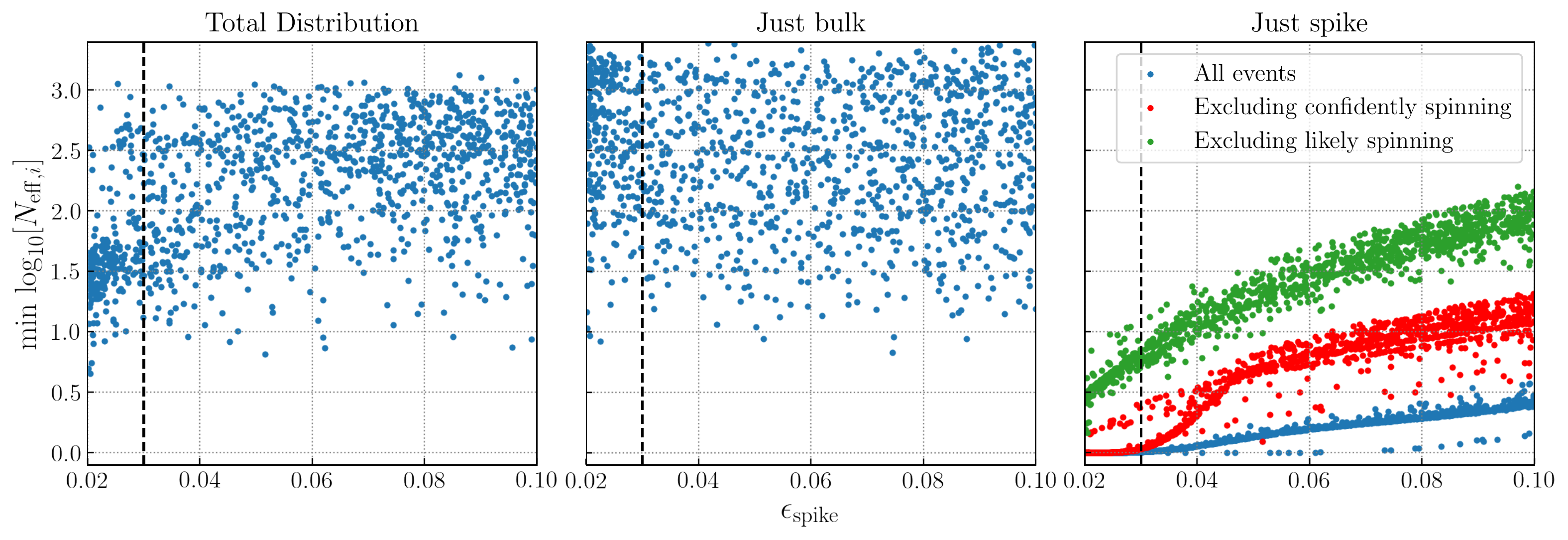}
    \caption{Scatter plot of the minimum (across events) number of effective samples $\mathrm{min}\left[N_{\mathrm{eff},i}\right]$ and the width of the zero-spin spike $\epsilon_{\mathrm{spike}}$ for the \texttt{BetaSpike+TruncatedMixture} model.
    Each point represents a single draw from the posterior shown in Fig.~\ref{fig:component_cornerplot}.
    Left panel: The minimum number of effective samples as computed across the entire population model.
    Middle panel: The minimum number of effective samples in the ``bulk" (Beta distribution) spin magnitude sub-population.
    Right panel: The minimum number of effective samples in the ``spike" (zero-spin half-Gaussian) spin magnitude sub-population.
    Dashed vertical lines denote $\epsilon_{\mathrm{spike}}=0.03$, the lower limit we impose on this parameter in our prior.
    Although the minimum sample count in the spike appears unacceptably low, the minimum is driven by events that are confidently spinning and which therefore do not impact our inference regarding the zero-spin spike (see the left panel of Fig.~\ref{fig:good_versus_bad_Neff_events}).
    Red points illustrate the minimum effective samples when excluding events these events that are confidently spinning.
    Green points show the minimum effective sample count if we further exclude events that are ``likely'' spinning (e.g. middle panel of Fig.~\ref{fig:good_versus_bad_Neff_events}), focusing only on those events that are well-described by zero-spin.
    To ensure a reasonable number of effective samples among these remaining events, we bound $\epsilon_\mathrm{spike}>0.03$, and note that the region $\epsilon_\mathrm{spike}\lesssim0.04$ may be subject to elevated Monte Carlo variance.
    }
    \label{fig:Neff_total_bulk_spike}
\end{figure*}

Our primary concern in tracking $N_\mathrm{eff}$ is to calibrate the allowed width $\epsilon_\mathrm{spike}$ of a possible zero-spin spike in the component spin distribution.
Due to finite sampling effects, we cannot let this spike width be arbitrarily small, but must bound $\epsilon_\mathrm{spike}$ to sufficiently large values to ensure reasonable $N_\mathrm{eff}$.
\citet{Essick2022} argue that $N_\mathrm{eff}\sim 10$ per event is sufficiently high to ensure accurate marginalization over each event's parameters.
Figure~\ref{fig:Neff_total_bulk_spike} illustrates how $\mathrm{min}\left[N_{\mathrm{eff},i}\right]$ varies as a function of $\epsilon_\mathrm{spike}$; each point represents a posterior sample drawn from the \texttt{BetaSpike+TruncatedMixture} results in Fig.~\ref{fig:component_cornerplot}.
The left panel shows $N_\mathrm{eff}$ as computed across the full spin distribution, while the center panel shows the effective sample count taken just over the ``bulk'' Beta distribution (via fixing $f_\mathrm{spike}=0$ when computing Eq.~\eqref{eq:Neff}).
The effective sample counts across the total population and in the bulk are largely
uncorrelated with $\epsilon_{\mathrm{spike}}>0.03$ and are everywhere large, with $\mathrm{min}\,N_\mathrm{eff} \gtrsim 10$.

\begin{figure*}
    \centering
    \includegraphics[width=\textwidth]{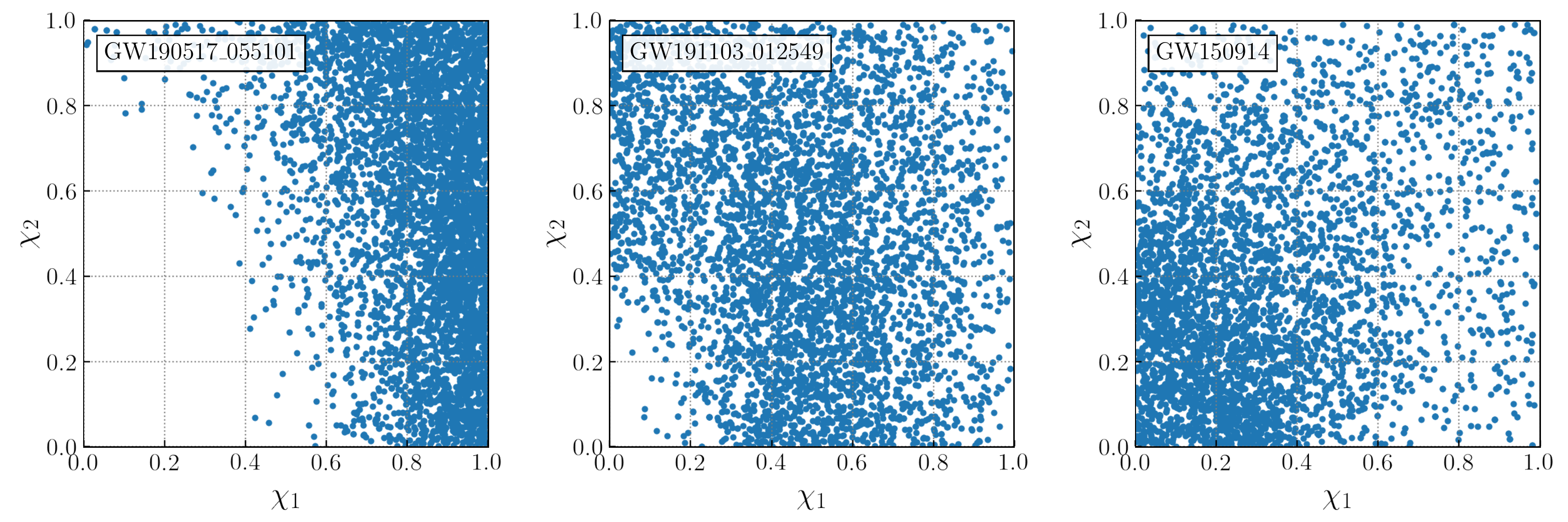}
    \caption{Spin magnitude posterior samples demonstrating the three classes of event discussed in Fig.~\ref{fig:Neff_total_bulk_spike} above.
    The left-hand panel illustrates a confidently spinning binary which is clearly identified as a member of the spinning ``bulk'' population, and whose low effective sample count in the spike is therefore unimportant.
    The middle panel illustrates a ``likely'' spinning event, defined by having less than a fraction $1/200$ of its posterior samples in the region $\chi_1,\chi_2<0.1$.
    The right panel, finally, shows an event belonging to neither category; these remaining events are used to generate the green points in Fig.~\ref{fig:Neff_total_bulk_spike} above.}
    \label{fig:good_versus_bad_Neff_events}
\end{figure*}

The right panel of Fig.~\ref{fig:Neff_total_bulk_spike}, meanwhile, shows the number of effective samples contained in the zero-spin spike (obtained via fixing $f_\mathrm{spike}=1$ in Eq.~\eqref{eq:Neff}).
The number of effective samples in the spike is extremely correlated with $\epsilon_{\mathrm{spike}}$.
This is expected, as a wider spike will encompass more posterior samples and thus have a larger $\mathrm{min}\left[N_{\mathrm{eff},i}\right]$.
The blue points show $\mathrm{min}\left[N_{\mathrm{eff},i}\right]$ taken across all events.
This minimum sample count is unacceptably low~\citep{Essick2022}, with every population sample yielding at least one binary with only $\sim1$ effective sample.
However, we argue that this is not in fact concerning.\footnote{If one wishes to avoid this argument altogether, the alternative method discussed in Appendix \ref{appendix:kde} and used to infer the binary $\chiEff$ distribution is accurate in the presence of narrow population features, and produces results that agree with those discussed in this section.}
The events with a very low number of effective samples in the spike are the same events that are unambiguously spinning, and hence confidently belong in the bulk and not the spike.
Their extremely low number of effective samples in the spike, therefore, does not affect inference.
The left-hand panel in Fig.~\ref{fig:good_versus_bad_Neff_events} illustrates the component spin magnitude posterior for one such ``confidently spinning'' event that unambiguously rules out $\chi_1 = \chi_2 = 0$.
The red dots in Fig.~\ref{fig:Neff_total_bulk_spike} show the minimum effective sample count if we now exclude these confidently-spinning events (GW190517, GW190412, GW151226, and GW191204, specifically).
This minimum count still contains events that are likely (but not necessarily unambiguously) non-spinning, such as the event shown in the center panel of Fig.~\ref{fig:good_versus_bad_Neff_events}.
The most critical measure of stability is whether $N_\mathrm{eff}$ is large for events such as GW150914 (right panel of Fig.~\ref{fig:good_versus_bad_Neff_events}) whose posteriors are finite up to and including $\chi_1 = \chi_2 = 0$.
The green points in Fig.~\ref{fig:Neff_total_bulk_spike}, therefore, show the minimum sample count when we further exclude ``likely nonspinning" events, defined as any event with less than $1/200$ of its samples in the region $\chi_1, \chi_2 < 0.1$.
The minimum number of effective samples across remaining events now rises to $\sim5$ at $\epsilon_{\mathrm{spike}}=0.03$ and $\gtrsim 10$ at $\epsilon_{\mathrm{spike}}=0.04$.
Given this, we choose to limit $\epsilon_\mathrm{spike}\geq 0.03$, and caution that the region $\epsilon_{\mathrm{spike}} < 0.04$ may be subject to increased Monte Carlo averaging error.

\section{Avoiding samples with likelihood KDEs}
\label{appendix:kde}

Our ordinary population likelihood, Eq.~\eqref{eq:likelihood-mc}, relies on Monte Carlo averages over posterior samples, an approach that can behave poorly when evaluating narrow population features as described above.
Within the \texttt{GaussianSpike} effective spin model, for example, in the limit that $\epsilon\to 0$ there will be precisely \textit{no} posterior samples falling in the zero-spin spike, and hence we will be unable to precisely evaluate the model's likelihood.
This limitation is not physical, but purely due to our representation of each event's posterior as a set of discrete samples.
As an alternative approach, we can abandon samples altogether and instead represent the likelihoods of individual events as Gaussian kernel density estimates (KDEs).
This alternative representation remains well-behaved even when the population features of interest are narrow, provided that there are enough samples in the neighborhood of the narrow feature to estimate the KDE.

For convenience, let $\tilde \lambda$ represent all individual-event parameters \textit{except} $\chiEff$.
Under the discrete sample representation, the ordinary Monte Carlo average in Eq.~\eqref{eq:likelihood-mc} is obtained by identifying individual events' likelihoods $p(d_i|\tilde \lambda_i,\chi_{\mathrm{eff},i})$ as a sum of delta-functions located at each posterior sample $j$:
    \begin{equation}
    p(d_i|\tilde\lambda_i,\chi_{\mathrm{eff},i}) 
        \propto \frac{1}{N_i}\sum_{j=1}^{N_i}
        \frac{\delta(\tilde\lambda_i-\tilde\lambda_{i,j}) \delta(\chi_{\mathrm{eff},i}-\chi_{\mathrm{eff},i,j})}
        {p_\mathrm{pe}(\tilde\lambda_{i,j},\chi_{\mathrm{eff},i,j})}\,.
    \label{eq:deltas}
    \end{equation}
In the KDE approach, we impart a finite ``resolution'' to each posterior sample, replacing the delta-functions with Gaussians of width $\sigma_\mathrm{kde}$:
\begin{equation}
\begin{aligned}
p(d_i|\tilde\lambda_i,\chi_{\mathrm{eff},i}) 
    \propto  \frac{1}{N_i}\sum_{j=1}^{N_i}
    \frac{\delta(\tilde\lambda_i-\tilde\lambda_{i,j})
    N(\chi_{\mathrm{eff},i}|\chi_{\mathrm{eff},i,j},\sigma_\mathrm{kde})}
    {p_\mathrm{pe}(\tilde\lambda_{i,j},\chi_{\mathrm{eff},i,j})}\,.
\end{aligned}
\label{eq:kde}
\end{equation}
Equation~\eqref{eq:kde} is now a continuous function of $\chi_{\mathrm{eff},i}$, and hence allows us to evaluate the likelihood of a population with arbitrarily narrow features.
This comes at a cost.
As shown in Eq.~\eqref{eq:likelihood}, evaluation of a population likelihood requires evaluation of the integral $\int d\tilde\lambda\, d\chiEff\, p(d_i|\tilde \lambda, \chiEff) p(\tilde\lambda,\chiEff|\Lambda)$.
This integration is trivial when the likelihoods are composed purely of delta functions, as in Eq.~\eqref{eq:deltas}, but not necessarily straightforward using Eq.~\eqref{eq:kde}.

Fortunately, both the individual-event likelihoods $p(d_i|\tilde \lambda_i, \chi_{\mathrm{eff},i})$ and our population models $p(\tilde\lambda,\chiEff|\Lambda)$ are mixtures of Gaussians in $\chiEff$, and so we can analytically carry out the required integration over $\chiEff$.
For the {\tt GaussianSpike} model, for example, the result is
    \begin{equation}
    \begin{aligned}
    p(\{d\}|\Lambda)
    &\propto \xi(\Lambda)^{-N_\mathrm{obs}} \prod_i \frac{1}{N_i} \sum_{j=1}^{N_i}         
        \frac{p(\tilde\lambda_{i,j}|\Lambda)}{p_\mathrm{pe}(\tilde\lambda_{i,j},\chi_{\mathrm{eff},i,j})} \\
            &\hspace{2cm}\times\int d\chi_{\mathrm{eff},i}\,
                N(\chi_{\mathrm{eff},i}|\chi_{\mathrm{eff},i,j},\sigma_\mathrm{kde})
                \bigg[
                    \zeta_\mathrm{spike} N_{[-1,1]}(\chi_{\mathrm{eff},i}|0,\epsilon)
                    + (1-\zeta_\mathrm{spike})N_{[-1,1]}(\chi_{\mathrm{eff},i}|\mu_\mathrm{eff},\sigma_\mathrm{eff})
                \bigg] \\
    &\propto \xi(\Lambda)^{-N_\mathrm{obs}} \prod_i \frac{1}{N_i} \sum_{j=1}^{N_i}         
        \frac{p(\tilde\lambda_{i,j}|\Lambda)}{p_\mathrm{pe}(\tilde\lambda_{i,j},\chi_{\mathrm{eff},i,j})} \\
        &\hspace{2cm}\times \bigg[
            \mathcal{A}_{i,j} \,\zeta_\mathrm{spike}\,
                \mathrm{exp}\left(-\frac{(\chi_{\mathrm{eff},i,j})^2}{2(\epsilon^2+\sigma_\mathrm{kde}^2)}\right)
            + \mathcal{B}_{i,j} \,(1-\zeta_\mathrm{spike})\, \mathrm{exp}\left(-\frac{(\chi_{\mathrm{eff},i,j}-\mu_\mathrm{eff})^2}{2(\sigma_\mathrm{eff}^2+\sigma_\mathrm{kde}^2)}\right)
            \bigg]\,,
    \end{aligned}
    \label{eq:likelihood-kde}
    \end{equation}
with
    \begin{equation}
    \mathcal{A}_{i,j} =
    \frac{
        \mathrm{Erf}\left(
            \frac{(1-\chi_{\mathrm{eff},i,j})\epsilon^2
                    + \sigma_\mathrm{kde}^2}
                {\sqrt{2\epsilon^2\sigma_\mathrm{kde}^2
                    (\epsilon^2+\sigma_\mathrm{kde}^2)}}
            \right)
        + \mathrm{Erf}\left(
            \frac{(1+\chi_{\mathrm{eff},i,j})\epsilon^2
                    + \sigma_\mathrm{kde}^2}
                {\sqrt{2\epsilon^2\sigma_\mathrm{kde}^2
                    (\epsilon^2+\sigma_\mathrm{kde}^2)}}
            \right)
    }{
    2\sqrt{2\pi(\epsilon^2+\sigma_\mathrm{kde}^2)}
        \mathrm{Erf}\left(
                \frac{1}{\sqrt{2\epsilon^2}}
                \right)
    }\,,
    \end{equation}
and
    \begin{equation}
    \mathcal{B}_{i,j} =
    \frac{
        \mathrm{Erf}\left(
            \frac{(1-\chi_{\mathrm{eff},i,j})\sigma_\mathrm{eff}^2
                    + (1-\mu_\mathrm{eff})\sigma_\mathrm{kde}^2}
                {\sqrt{2\sigma_\mathrm{eff}^2\sigma_\mathrm{kde}^2
                    (\sigma_\mathrm{eff}^2+\sigma_\mathrm{kde}^2)}}
            \right)
        + \mathrm{Erf}\left(
            \frac{(1+\chi_{\mathrm{eff},i,j})\sigma_\mathrm{eff}^2
                    + (1+\mu_\mathrm{eff})\sigma_\mathrm{kde}^2}
                {\sqrt{2\sigma_\mathrm{eff}^2\sigma_\mathrm{kde}^2
                    (\sigma_\mathrm{eff}^2+\sigma_\mathrm{kde}^2)}}
            \right)
    }{
    \sqrt{2\pi(\sigma_\mathrm{eff}^2+\sigma_\mathrm{kde}^2)}
        \left[
            \mathrm{Erf}\left(
                \frac{1-\mu_\mathrm{eff}}{\sqrt{2\sigma_\mathrm{eff}^2}}
                \right)
            +\mathrm{Erf}\left(
                \frac{1+\mu_\mathrm{eff}}{\sqrt{2\sigma_\mathrm{eff}^2}}
                \right)
            \right]
    }\,.
    \end{equation}
In an exactly analogous fashion, when computing the detection efficiency $\xi(\Lambda)$ we can replace a Monte Carlo average over successfully found injections [as in Eq.~\eqref{eq:detection-efficiency-mc}] with a KDE over injections:
    \begin{equation}
    \xi(\Lambda) \propto \sum_{i=1}^{N_\mathrm{found}}  
    \frac{p(\tilde\lambda_i|\Lambda)}{p_\mathrm{inj}(\tilde\lambda_i,\chi_{\mathrm{eff},i})}
    \int d\chiEff\,N(\chiEff|\chi_{\mathrm{eff},i},\sigma_\mathrm{kde})\, p(\chiEff|\Lambda)\,.
    \label{eq:detection-efficiency-kde}
    \end{equation}
As above, the integration can be carried out analytically when (as in our case) the population model $p(\chiEff|\Lambda)$ is a mixture of Gaussians.

\begin{figure}
    \centering
    \includegraphics[width=0.5\textwidth]{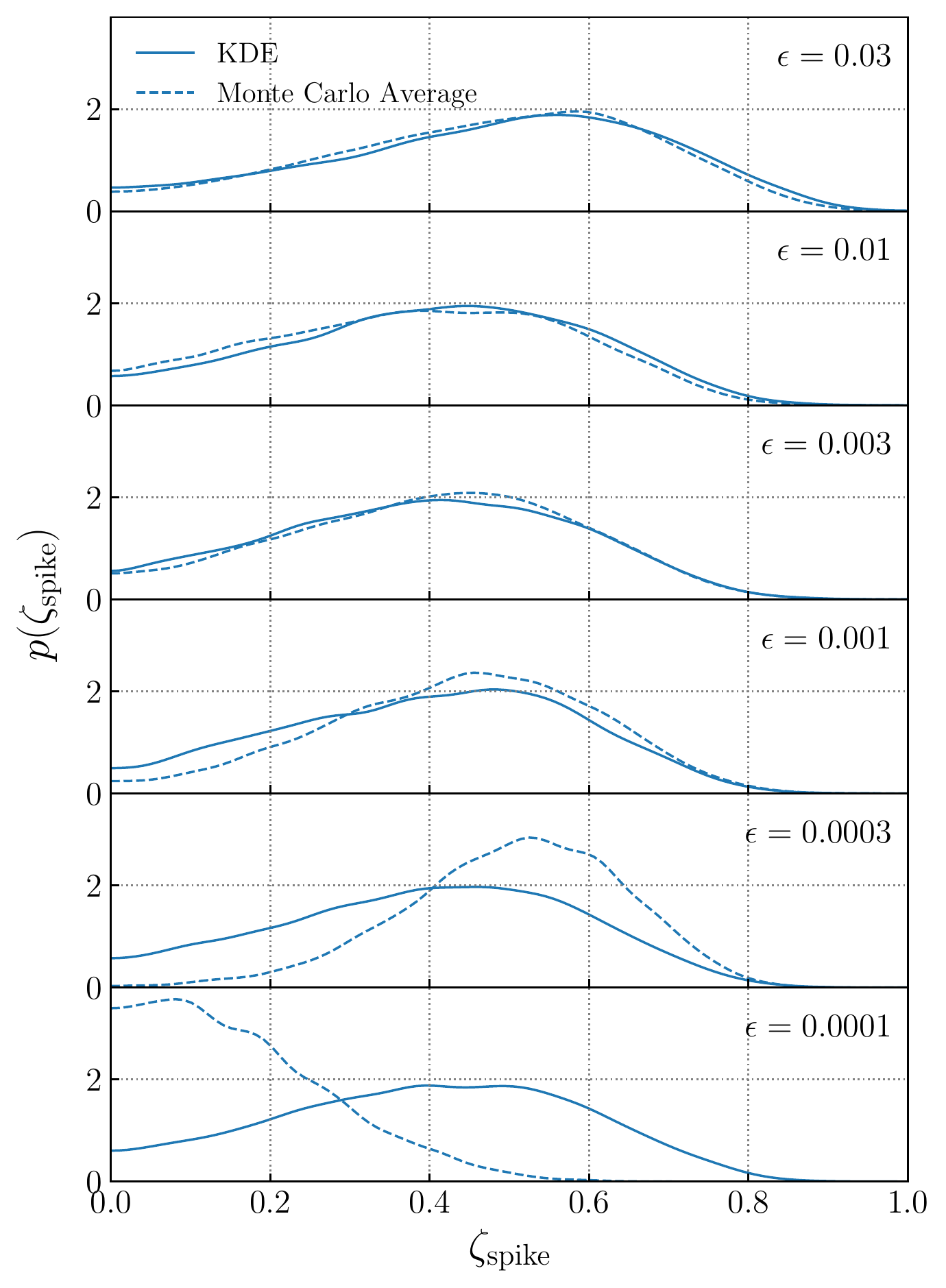}
    \caption{
    Validation of the KDE likelihood approach discussed in Appendix~\ref{appendix:kde}.
    Each subplot shows the marginalized posterior for the fraction of events contained within a narrow spike at $\chiEff=0$.
    Solid lines show posteriors computed using the KDE method, while dashed lines show posteriors obtained using ordinary Monte Carlo averaging (Appendix~\ref{appendix:inference}).
    Results are shown for a variety of spike widths, from broad spikes with standard deviation $\epsilon=0.03$ to narrow spikes with $\epsilon=10^{-4}$.
    We see that the KDE method yields consistent and convergent results as $\epsilon$ approaches zero, while the Monte Carlo averaging gives increasingly divergent results for small spike widths due to growing Monte Carlo errors.}
    \label{fig:kde-comparison}
\end{figure}

As a demonstration and validation of this approach, we revisit inference of the GWTC-3 effective spin distribution via the {\tt GaussianSpike} model, but now fix the ``spike" at $\chiEff=0$ to have a finite width $\epsilon >0 $.
We repeat the analysis in two ways: (i) using the ordinary Monte Carlo representation of Eqs.~\eqref{eq:likelihood-mc} and \eqref{eq:detection-efficiency-mc}, and (ii) using the KDE representation of Eqs.~\eqref{eq:likelihood-kde} and \eqref{eq:detection-efficiency-kde}.
Figure~\ref{fig:kde-comparison} shows the marginal posterior for $\zetaSpike$ for different choices of $\epsilon$.
When $\epsilon$ is large (i.e., the ``spike'' is broad compared to the typical inter-sample spacing) the two methods give identical results.
As we let $\epsilon$ approach zero, however, the Monte Carlo average produces divergent results, while the KDE likelihood representation yields stable and converging posteriors.

\section{Mock injection campaign}
\label{appendix:mock-injections}

\begin{figure}
    \centering
    \includegraphics[width=0.32\textwidth]{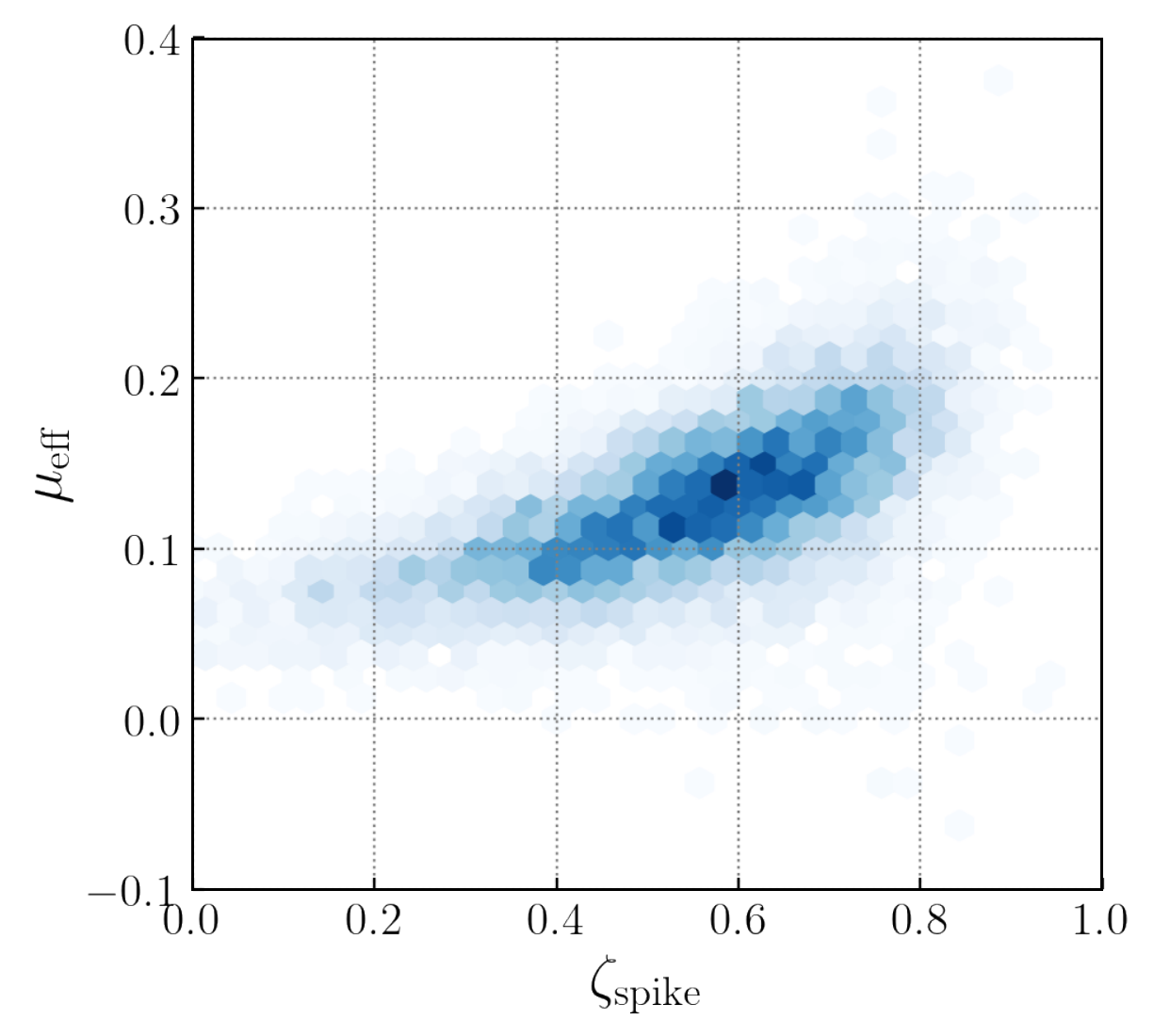}
    \includegraphics[width=0.32\textwidth]{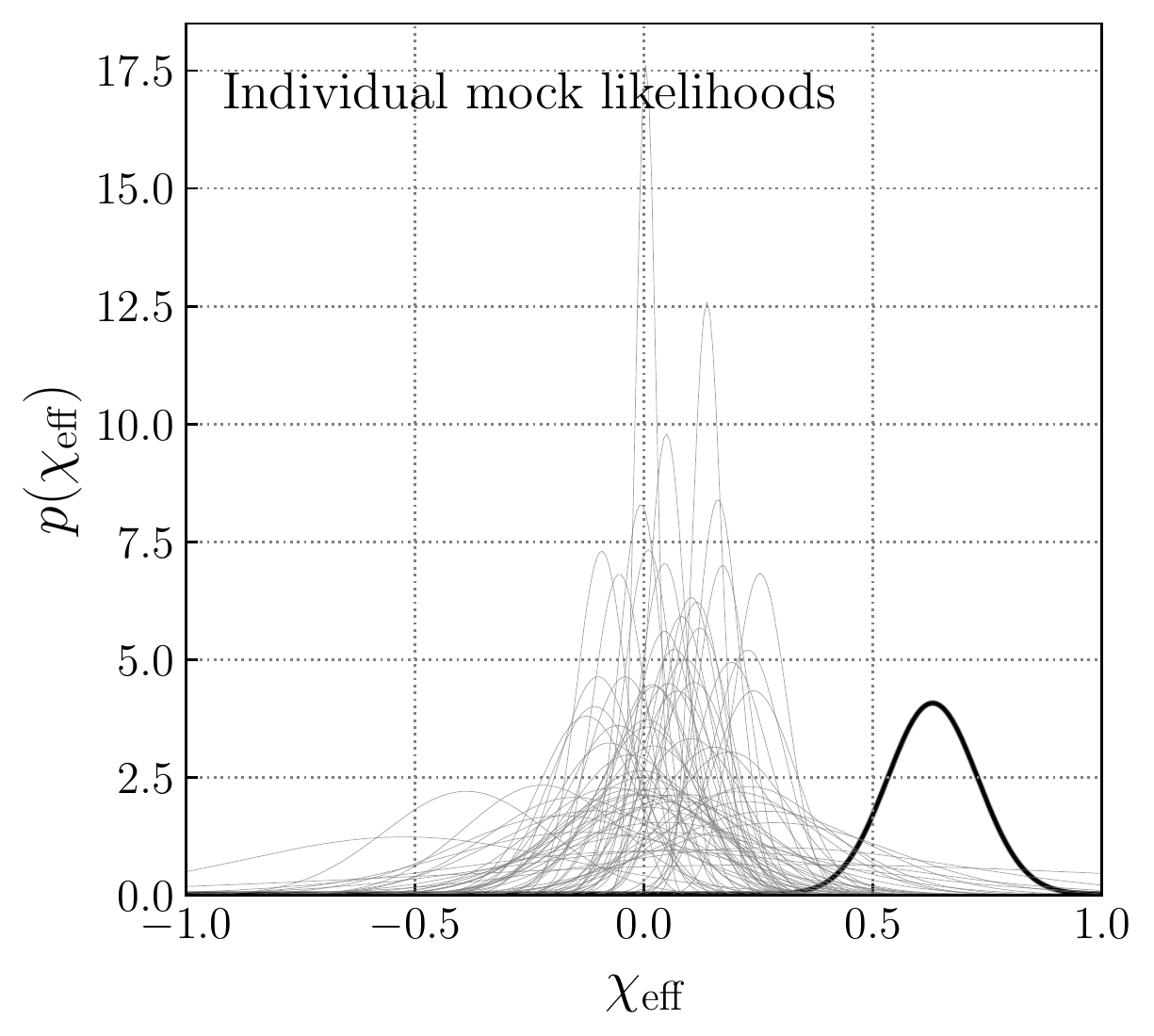}
    \includegraphics[width=0.32\textwidth]{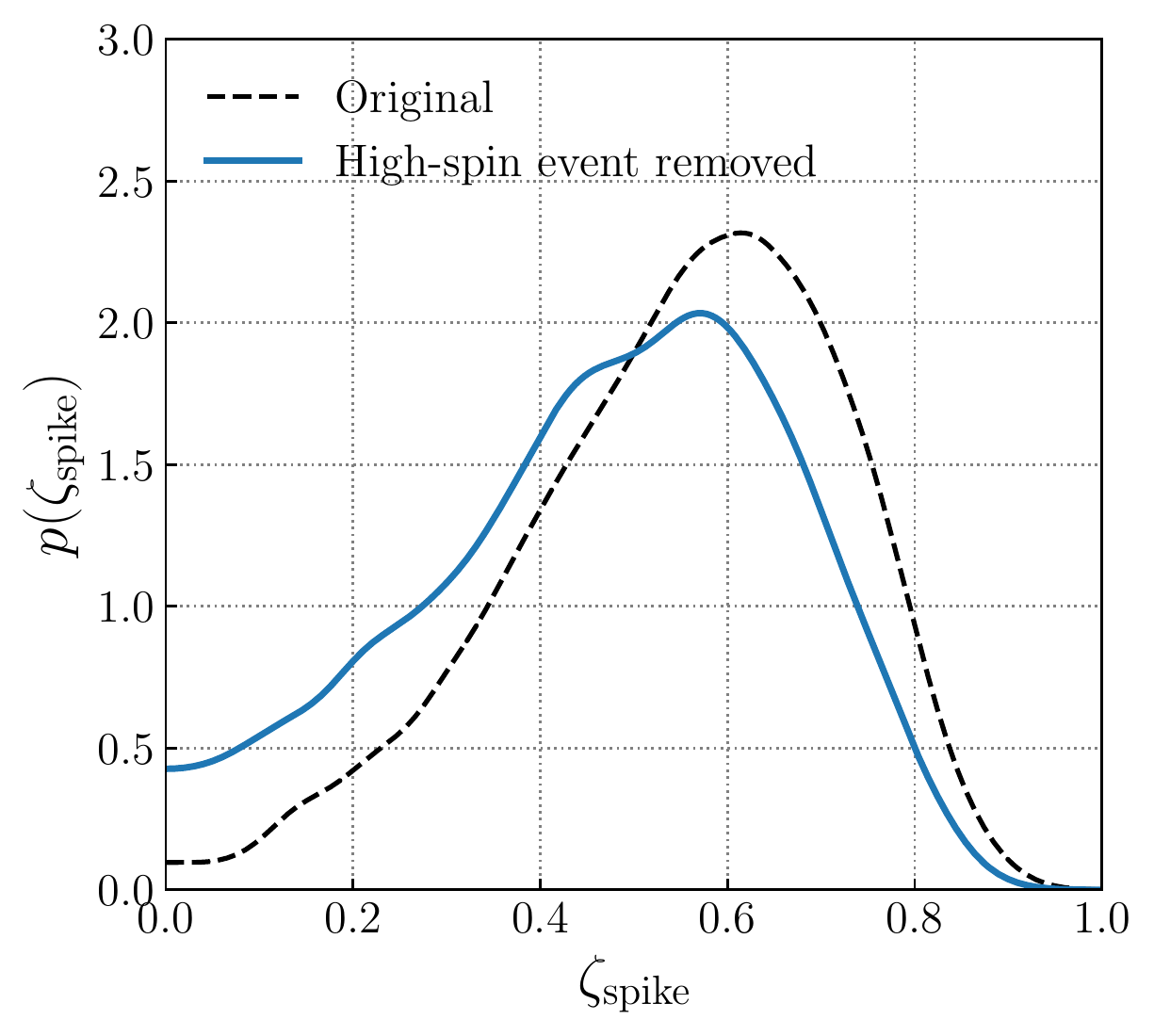}
    \caption{
    Follow-up exploration of the injected mock catalog that most strongly (and incorrectly) excludes $\zetaSpike=0$ in Fig.~\ref{fig:spike-posterior}.
    Left: The joint posterior on $\zetaSpike$ and $\mu_\mathrm{eff}$ from the selected catalog.
    In general, we find significant correlations between $\zetaSpike$ and the mean $\mu_\mathrm{eff}$ of the spinning ``bulk'' population.
    Mock catalogs that most strongly disfavor $\zetaSpike=0$ are generally those that have the largest inferred $\mu_\mathrm{eff}$.
    Middle: The individual events comprising the selected catalog.
    The event with the largest measured $\chiEff$ (and hence the event that most strongly influences $\mu_\mathrm{eff}$) is highlighted.
    Right: The solid blue curve shows the updated $\zetaSpike$ posterior after removing the highlighted high-spin event.
    Compared to the original $\zetaSpike$ measurement (dashed black curve), the updated posterior is now more (correctly) consistent with $\zetaSpike=0$.
    }
    \label{fig:inj-investigation}
\end{figure}

In Fig.~\ref{fig:spike-posterior} in the main text, we showed posteriors on $\zetaSpike$ obtained by analyzing mock catalogs of binary black holes drawn from an intrinsically spike-less population.
In this appendix we describe our procedure for generating and analyzing these mock catalogs.

To generate mock $\chiEff$ measurements, we assume an underlying astrophysical distribution following the \texttt{Gaussian} effective spin model, with mean $\mu_\mathrm{eff} = 0.06$ and standard deviation $\sigma_\mathrm{eff} = 0.09$.
These values are chosen to match the median measurements of $\mu_\mathrm{eff}$ and $\sigma_\mathrm{eff}$ obtained when hierarchically analyzing GWTC-3 with the \texttt{Gaussian} effective spin model.
For each mock catalog, we draw 69 ``true'' spin values from this Gaussian population:
    \begin{equation}
    \chi_\mathrm{eff,true} \sim N(0.06,0.09).
    \end{equation}
We then add a random measurement uncertainty to each mock event.
Assuming Gaussian likelihoods for each $\chiEff$ measurement, ``observed'' maximum-likelihood values are generated via
    \begin{equation}
    \chi_\mathrm{eff,obs} \sim N(\chi_\mathrm{eff,true},\sigma_\mathrm{obs}),
    \label{eq:gen-obs}
    \end{equation}
where $\sigma_\mathrm{obs}$ is the standard deviation of each event's likelihood.
These uncertainties are themselves randomly distributed to match the range of uncertainties exhibited by real measurements.
The $\chiEff$ likelihoods among binary black holes in GWTC-3 have log standard deviations that are approximately normally distributed, with a mean $\log_{10}\sigma_\mathrm{obs}$ of $-0.9$ and a standard deviation of 0.3.
When drawing mock observed values in Eq.~\eqref{eq:gen-obs}, we therefore assign each event a random measurement uncertainty according to
    \begin{equation}
    \log_{10}\sigma_\mathrm{obs} \sim N(-0.9,0.3).
    \end{equation}
Since both our population model and our mock likelihoods are Gaussian, the hierarchical likelihood for $\mu_\mathrm{eff}$ and $\sigma_\mathrm{eff}$ can be calculated analytically, exactly as was done in Appendix~\ref{appendix:kde} above.
Specifically, the likelihood $p({\{d}\}|\mu_\mathrm{eff},\sigma_\mathrm{eff})$ of our mock catalog is of the same form as Eq.~\eqref{eq:likelihood-kde}, with the observed $\chi_\mathrm{eff,obs}$ values in place of $\chi_{\mathrm{eff},i,j}$, the measurement uncertainties $\sigma_\mathrm{obs}$ in place of $\epsilon$, and neglecting the summation over $j$.

As highlighted in Sec.~\ref{sec:spike}, many of our mock catalogs exhibit $\zeta_\mathrm{spike}$ posteriors qualitatively similar to the posterior obtained using real data, encompassing zero but peaking at $\zeta_\mathrm{spike}\approx 0.5$.
Some mock catalogs even more confidently (and incorrectly) appear to rule out the injected value of $\zeta_\mathrm{spike}=0$.
We find that this elevated ``false-alarm probability'' can be explained by a degeneracy between $\zeta_\mathrm{spike}$ and $\mu_\mathrm{eff}$ that occurs when analyzing relatively small number of individually uncertain measurements.
To illustrate this behavior, we choose the mock catalog whose $\zetaSpike$ posterior \textit{most strongly} rules out zero in Fig.~\ref{fig:spike-posterior}.
The left-hand panel in Fig.~\ref{fig:inj-investigation} shows the joint posterior on $\zetaSpike$ and $\mu_\mathrm{eff}$ given by this catalog.
These parameters are correlated: if the Gaussian ``bulk'' is shifted to a larger mean, events located at negative or near-zero $\chiEff$ values are left behind and must be absorbed into the zero-spin spike.
To test this intuition, we show in the middle panel of Fig.~\ref{fig:inj-investigation} the individual likelihoods that comprise this mock catalog.
The event with the largest $\chiEff$ (and hence the event pulling $\mu_\mathrm{eff}$ to large values) is highlighted.
If we remove this high-spin event and redo the hierarchical inference on this catalog, we obtain the blue $\zetaSpike$ posterior shown in the right-hand panel of Fig.~\ref{fig:inj-investigation}.
After removing the high spin event, we now find $\zetaSpike$ to be considerably more consistent with zero.

\section{Bayes factors through the Savage Dickey density ratio}
\label{appendix:BFs}

%
\begin{figure}
    \centering
     \includegraphics[width=0.32\textwidth]{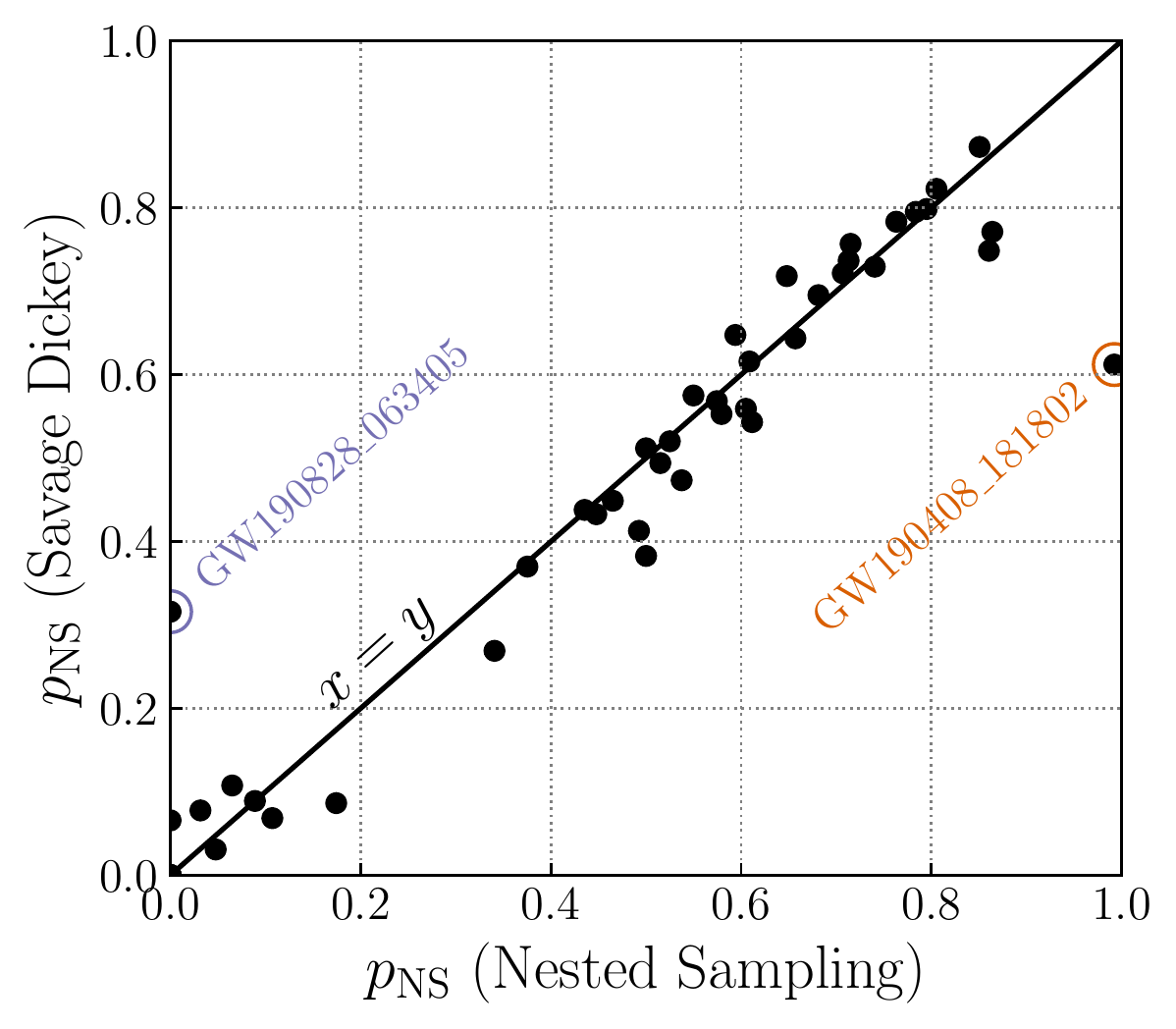}
    \includegraphics[width=0.32\textwidth]{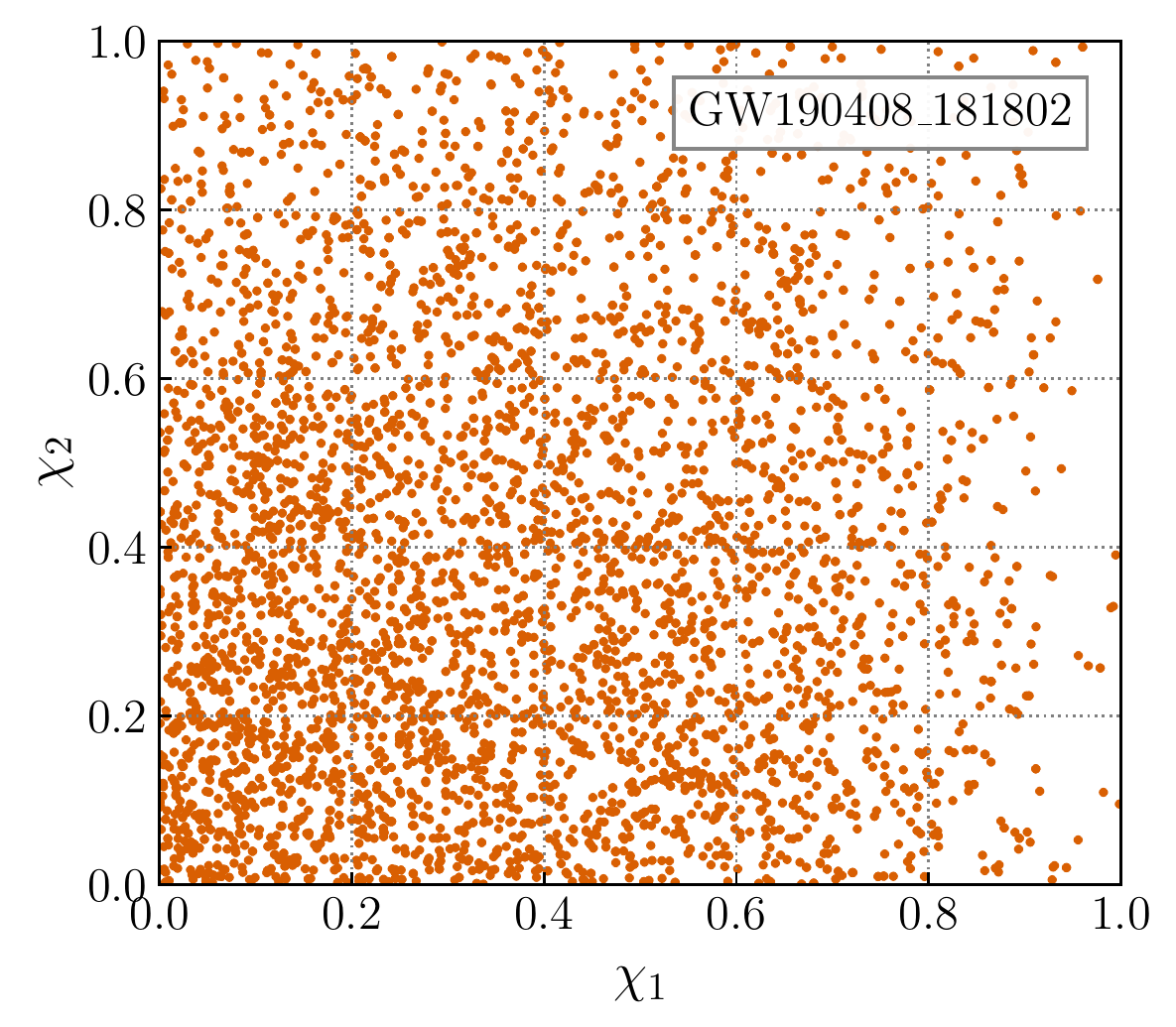}
    \includegraphics[width=0.32\textwidth]{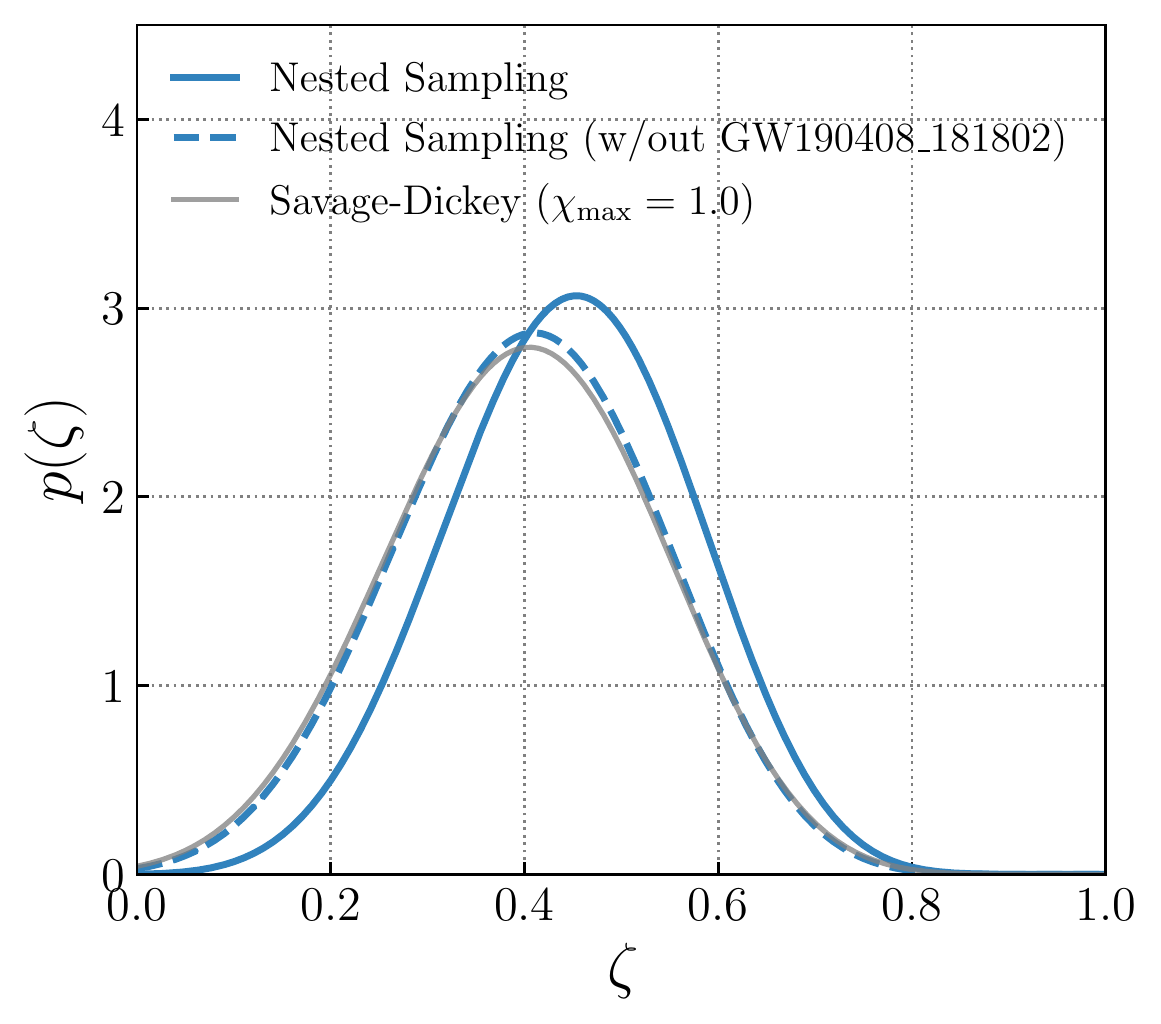}
    \caption{
    Left panel: The probabilities $p_{\mathrm{NS}}$ for each binary black hole in GWTC-2 to be non-spinning, as computed in two different ways: through a Savage Dickey density ratio and using fully-marginalized likelihoods computed via nested sampling and reported in~\citet{LIGOScientific:2020ibl}.
    The circles denote GW190408\_181802 and GW190828\_063405, whose nested sampling results appear to be outliers.
    Center panel: The spin magnitude posterior for GW190408\_181802.
    Nested sampling yields a Bayes factor of $\mathcal{B}^\mathrm{NS}_\mathrm{S}\sim 130$ in favor of the hypothesis that this event is non-spinning.
    This would require the spin magnitude posterior to be $\sim 130$ times larger than the (flat) prior at $\chi_1=\chi_2=0$.
    This is inconsistent, though, with the true posterior on the spin magnitudes of GW190408\_181802, suggesting that the nested sampling Bayes factor in favor of the non-spinning hypothesis is significantly overestimated.
    Right panel: As a consistency check, we repeat the counting experiment from Sec.~\ref{sec:counting} using the nested sampling Bayes factors serving as input to \citet{Galaudage2021}, rather than our Savage-Dickey estimates.
    We show posteriors obtained using the entire GWTC-2 catalog (solid blue curve) and results after excluding GW190408\_181802 from our sample (dashed blue).
    As mentioned in Sec.~\ref{sec:counting}, we obtain much closer agreement with the Savage-Dickey result (grey curve) after excluding GW190408\_181802.
    }
    \label{fig:chiposterior}
\end{figure}

Our initial counting experiment in Sec.~\ref{sec:counting} uses only the Bayes factors for each event in GWTC-2 between non-spinning ($\chi_1=\chi_2=0$) and spinning ($\chi_1,\chi_2\geq 0$) hypotheses.
Such Bayes factors also act as inputs in the analysis of \citet{Galaudage2021}.
\citet{Galaudage2021} form Bayes factors via the ratios of fully-marginalized likelihoods computed for each event via nested sampling under non-spinning and spinning priors~\citep{LIGOScientific:2020ibl,2021ApJ...915L..35K}.
Instead, we compute Bayes factors directly from the posterior samples for each event.
For nested models where the simpler model (e.g. the non-spinning model) is contained within a more complex model (the spinning model with $\chi_1=\chi_2=0$), the Bayes factor between models is given through the Savage-Dickey density ratio~\citep{SDDR}
\begin{equation}
    \mathcal{B}^\mathrm{NS}_\mathrm{S} = \frac{p(\chi_1=0,\chi_2=0|d)}{p(\chi_1=0,\chi_2=0)}\,,
\end{equation}
where $p(\chi_1=0,\chi_2=0|d)$ and $p(\chi_1=0,\chi_2=0)$ are the posterior and prior densities at $\chi_1=\chi_2=0$, respectively.
See Appendix B of~\citet{Chatziioannou:2014bma} for a proof of this equation when the more complex model has additional parameters (here, the spin angles) that are absent from the simpler model.
For every binary black hole in GWTC-2, the left panel of Fig.~\ref{fig:chiposterior} shows the probability that the event is non-spinning,
    \begin{equation}
    \begin{aligned}
    p_{\mathrm{NS}} &= \frac{p_\mathrm{NS}}{p_\mathrm{S} + p_\mathrm{NS}} \\
        &=\frac{\mathcal{B}^\mathrm{NS}_\mathrm{S}}{1+\mathcal{B}^\mathrm{NS}_\mathrm{S}},
    \end{aligned}
    \end{equation}
as implied by both sets of Bayes factors.
We find general agreement, within numerical uncertainties, between our Bayes factors computed via Savage Dickey density ratios and those computed via nested sampling~\citep{LIGOScientific:2020ibl}. We note that Savage Dickey density ratios, which rely on KDEs over posterior samples, may be less reliable for events with very little support at $(\chi_1=0,\chi_2=0)$. In these cases, however, a Savage Dickey density ratio would still conclude that $p_{\mathrm{NS}} \rightarrow 0$ consistently.

There are two significant outliers, however.
Nested sampling returns $\mathcal{B}^\mathrm{NS}_\mathrm{S}\sim 130$ in favor of the non-spinning hypothesis for the event GW190408\_181802.
If this Bayes factor were correct, then GW190408\_181802 is almost certainly non-spinning, guaranteeing that a non-zero number of events will be placed in a non-spinning sub-population.
This is indeed the conclusion drawn by~\citet{Galaudage2021} with the nested sampling Bayes factors.
Such a large preference for vanishing spins is not supported by this event's spin magnitude posteriors (right panel); it would require that the posterior is $\sim 130$ times larger than the prior at $(\chi_1=\chi_2=0)$.
A Savage Dickey density ratio instead gives an agnostic $\mathcal{B}^\mathrm{NS}_\mathrm{S} \approx 1.6$ for GW190408\_181802.
The opposite situation is encountered for GW190828\_063405, which is reported to have $\mathcal{B}^\mathrm{NS}_\mathrm{S}\sim 1.7\times 10^{-5}$ from nested sampling.
Inspection of the posterior again shows that the posterior and the Bayes factor are inconsistent, as the former remains finite at $(\chi_1=\chi_2=0)$.
Although these two outliers were identified by comparison of nested sampling and Savage Dickey density ratio results, the conclusion that the nested sampling Bayes factors are misestimated can be drawn through visual inspection of the posterior samples alone.

\section{Different Catalogs and Model Variations}
\label{appendix:gwtc2}

\begin{figure}
    \centering
    \includegraphics[width=0.48\textwidth]{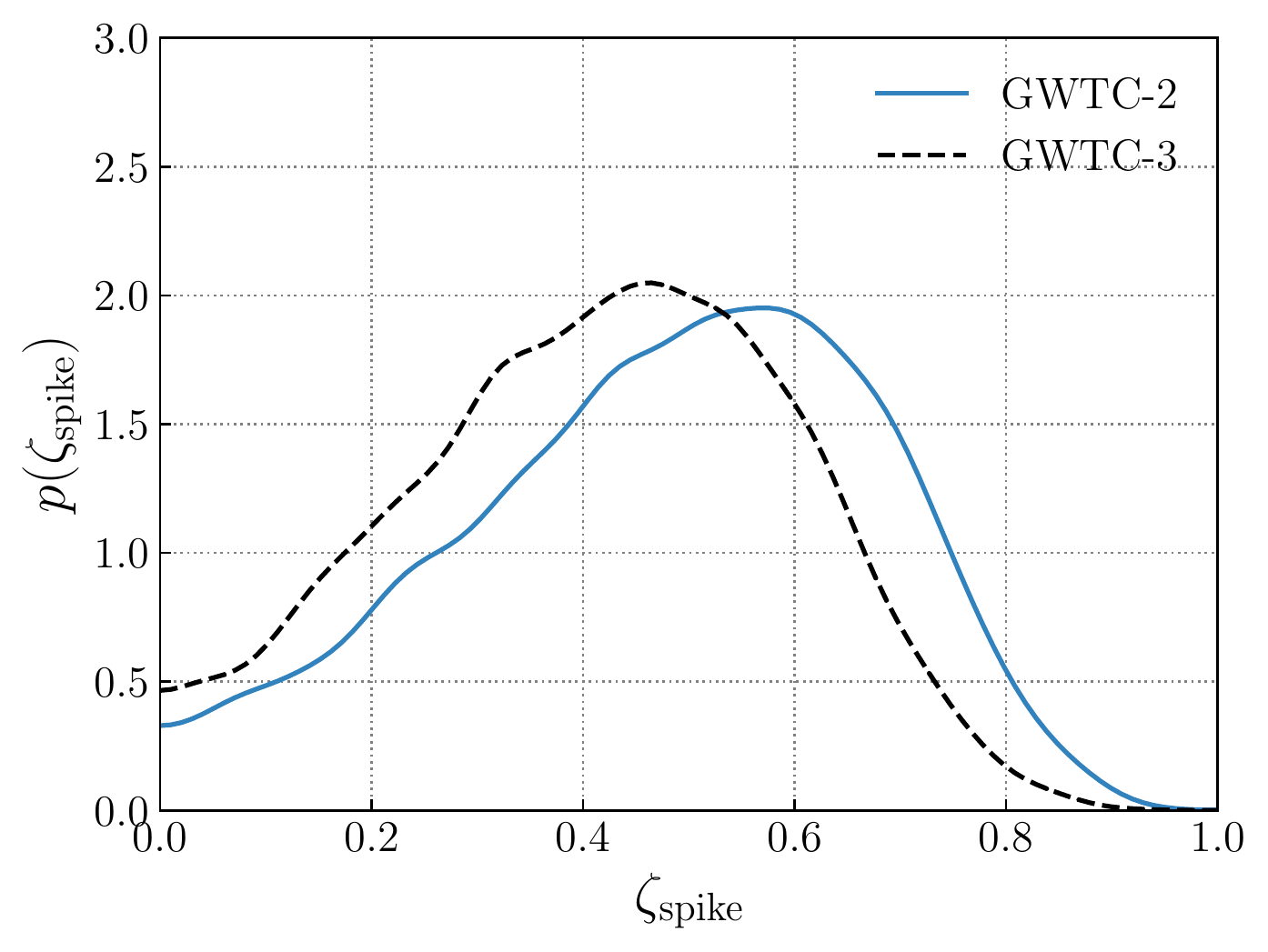}
    \caption{
    The fraction of binary black holes among GWTC-2~\citep[solid blue line;][]{LIGOScientific:2020ibl} with $\chiEff=0$, inferred using the \texttt{GaussianSpike} effective spin model.
    For comparison, the dashed black curve shows the result obtained using GWTC-3~\citep{LIGOScientific:2021djp}; this is the same result shown in Fig.~\ref{fig:spike-posterior} in the main text.
    Both results are consistent with one another, showing no evidence for a distinct sub-population of events with $\chiEff=0$, although the GWTC-3 result more strongly favors $\zetaSpike=0$.
    }
    \label{fig:gwtc2vs3_chieff}
\end{figure}
\begin{figure}[h!]
    \centering
    \includegraphics[width=0.9\textwidth]{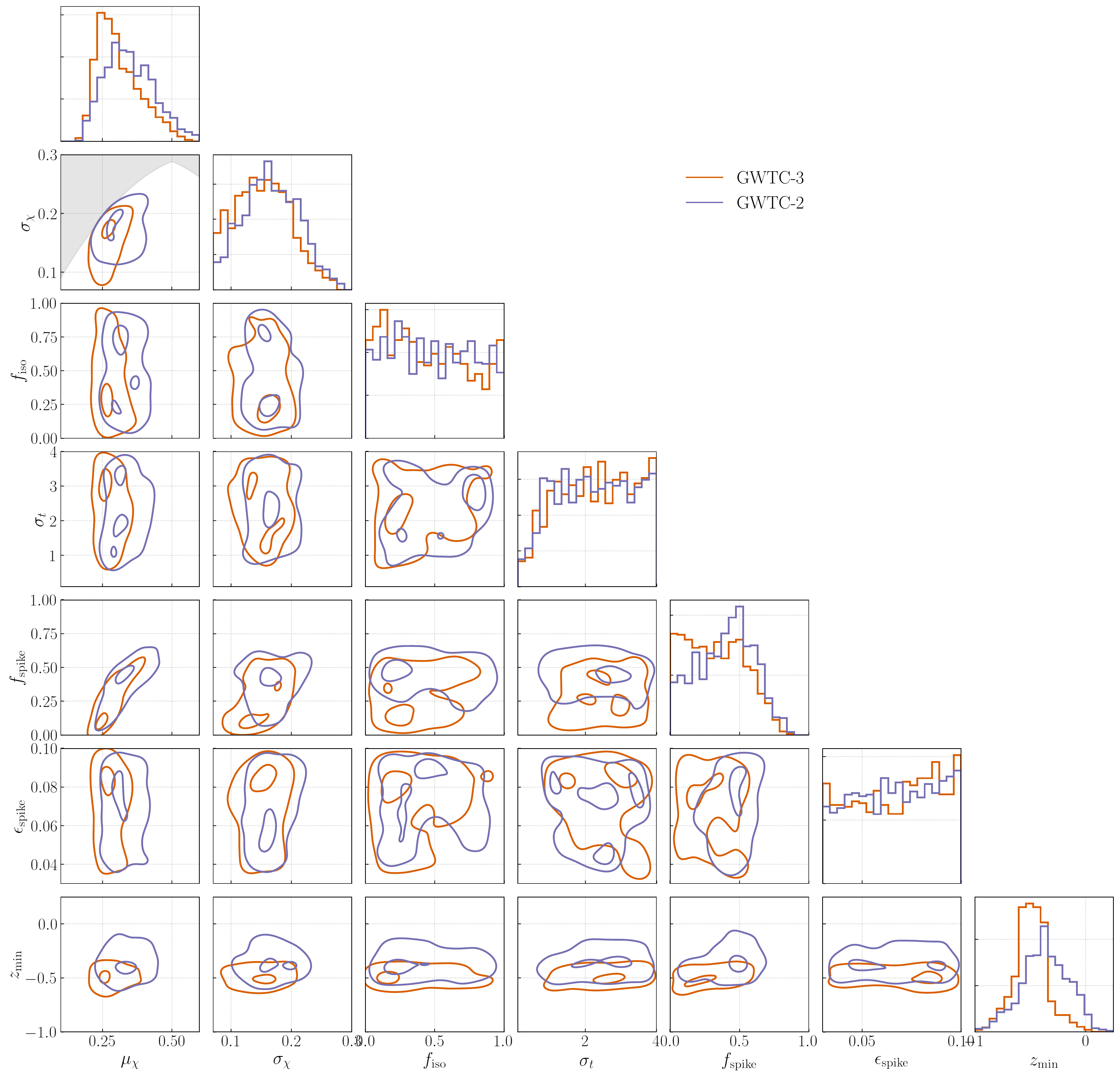}
    \caption{
    Posterior on the parameters of the {\tt BetaSpike+TruncatedMicture} component spin model, now using only GWTC-2 events (purple).
    The shaded region in the $\mu_\chi$--$\sigma_\chi$ two-dimensional posterior is the region excluded by the prior cut on the shape parameters of the beta distribution $\alpha, \beta > 1$.
    For reference, results using the full GWTC-3 event list are shown in orange; the GWTC-3 result is identical to the one presented in Fig.~\ref{fig:component_cornerplot}.
    We find broadly consistent results between the two catalogs, including no requirement for a zero-spin sub-population as well as a preference for spins misaligned by more than $90^\circ$ relative to binaries' Newtonian orbital angular momentum.}
    \label{fig:gwtc2vs3}
\end{figure}
\begin{figure}[h!]
    \centering
    \includegraphics[width=0.9\textwidth]{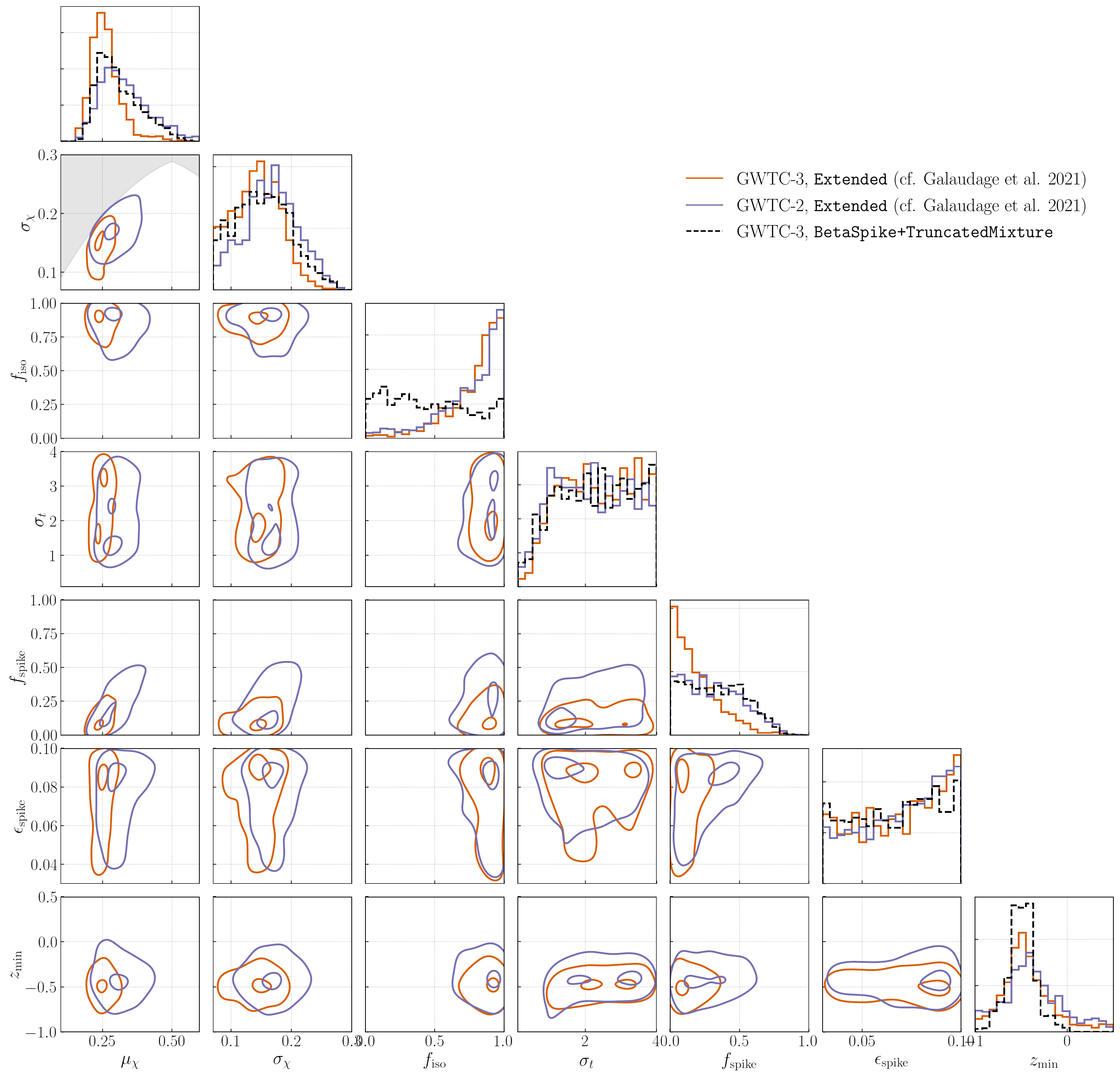}
    \caption{Posterior on the parameters of the {\tt Extended} component spin model [Eqs.~\eqref{eq:alternate_pairing_chi} and~\eqref{eq:alternate_pairing_cost}] using GWTC-3 data (orange) and just GWTC-2 data (purple).
    The one dimensional marginalized posteriors for the {\tt BetaSpike+TruncatedMicture} model for GWTC-3 are shown in black dashed lines for comparison; this result is identical to that shown in Figs.~\ref{fig:component_cornerplot} and~\ref{fig:gwtc2vs3}. 
    The {\tt Extended} model infers a population with smaller $f_\mathrm{spike}$ and larger $f_\mathrm{iso}$ than {\tt BetaSpike+TruncatedMicture}; it also has more support for $z_\mathrm{min} > 0$ than {\tt BetaSpike+TruncatedMicture}, although the $z_\mathrm{min}$ posterior peaks at a negative value for both cases.
    The posteriors for the remaining population parameters are qualitatively similar between the two models.}
    \label{fig:iid_vs_noniid}
\end{figure}

All results in Secs.~\ref{sec:spike} and \ref{sec:results} of the main text rely on the latest GWTC-3 catalog of gravitational-wave detections~\citep{LIGOScientific:2021djp}, as detailed in Appendix~\ref{appendix:inference}. 
However, the results of~\citet{Galaudage2021} (to which we frequently compare) instead made use of GWTC-2~\citep{LIGOScientific:2020ibl}, since GWTC-3 results were published only after their study.
Moreover, our  model differs slightly  from ours in how component spin magnitudes and tilt angle are paired. 
We assume that spin magnitudes $\chi_1$ and $\chi_2$ are independently and identically distributed (IID), as are tilt angles $\cos\theta_1$ and $\cos\theta_2$.
In our \texttt{BetaSpike+TruncatedMixture} model, for example, one component spin could lie in the zero-spin spike while its companion spin could lie in the Beta distribution ``bulk.''
\citet{Galaudage2021}, on the other hand, adopt preferential pairing and require that both component spins in a given binary occupy the same mixture component (bulk or spike; aligned or isotropic).
In order to enable a more direct comparison between our results, in this appendix we show results obtained using only binary black holes among GWTC-2, as well as results generated with an alternative pairing model that directly matches that of~\citet{Galaudage2021}.

Figure~\ref{fig:gwtc2vs3_chieff} illustrates the fraction of binary black holes inferred to be non-spinning, using the \texttt{GaussianSpike} effective spin model (see Secs.~\ref{sec:spike} and \ref{app:effectivemodels}) and analyzing data from GWTC-2.
Like our results with GWTC-3, results using only GWTC-2 are consistent with $\zetaSpike=0$, indicating no evidence for a distinct sub-population of non-spinning events.
Similarly, Fig.~\ref{fig:gwtc2vs3} shows results when analyzing GWTC-2 binary black holes with the {\tt BetaSpike+TruncatedMixture} component spin model.
We again find consistent results between GWTC-2 and GWTC-3 events.
When analyzing only events in GWTC-2, we infer the fraction $f_\mathrm{spike}$ of approximately non-spinning systems to be consistent with zero, and find no requirement for the width of this possible sub-population to be narrow (small $\epsilon_\mathrm{spike}$).
Using GWTC-2, we also conclude that the distribution of spin-orbit misalignment angles extends beyond $90^\circ$, with the truncation $z_\mathrm{min}$ on the $\cos\theta$ distribution inferred to be negative  (\gwtctwozminpercentile credibility), although with slightly decreased certainty compared to GWTC-3  (\gwtcthreezminpercentile credibility).

Finally, Fig.~\ref{fig:iid_vs_noniid} shows results obtained under an alternative version of our \texttt{BetaSpike+TruncatedMixture} model that directly matches the pairing function used by~\citet{Galaudage2021} in their \texttt{Extended} model.
Explicitly, spin magnitudes are jointly distributed as
    \begin{equation}
    \begin{aligned}
    p(\chi_1,\chi_2 |\alpha, \beta, f_\mathrm{spike}, \epsilon_\mathrm{spike}) &= 
    f_\mathrm{spike} \,\mathcal{N}_{[0,1]}(\chi_1 | 0,\epsilon_\mathrm{spike})\,\mathcal{N}_{[0,1]}(\chi_2 | 0,\epsilon_\mathrm{spike}) \\
     &\hspace{1cm}+ (1-f_\mathrm{spike})\frac{\chi_1^{1-\alpha} \, (1-\chi_1)^{1-\beta}}{c(\alpha, \beta)}\frac{\chi_2^{1-\alpha} \, (1-\chi_2)^{1-\beta}}{c(\alpha, \beta)},
    \end{aligned}
    \label{eq:alternate_pairing_chi}
    \end{equation}
and the joint tilt angle distribution is
\begin{equation}
    \label{eq:alternate_pairing_cost}
     p(\cos\theta_1, \cos\theta_2 |f_\mathrm{iso},\sigma_t) = \frac{f_\mathrm{iso}}{(1-z_\mathrm{min})^2} + (1-f_\mathrm{iso})
     \mathcal{N}_{[z_\mathrm{min},1]}(\cos\theta_1|1,\sigma_t) \mathcal{N}_{[z_\mathrm{min},1]}(\cos\theta_2|1,\sigma_t) \, .
    \end{equation} 
Figure~\ref{fig:iid_vs_noniid} shows results from this alternative model using both GWTC-2 and GWTC-3.
For reference, the dashed black distributions show results from our usual \texttt{BetaSpike+TruncatedMixture} model.
Compared to \texttt{BetaSpike+TruncatedMixture}, the \texttt{Extended} model has \textit{more} support at $f_\mathrm{spike} = 0$, strengthening the conclusion that the fraction of black holes with negligible spin is consistent with zero.
For the tilt angle distribution, we find that data still has a strong preference for negative $z_\mathrm{min}$ under the \texttt{Extended} model, but with decreased confidence (\gwtctwozminpercentilenoniid quantile for GWTC-2 and \gwtcthreezminpercentilenoniid quantile for GWTC-3) than with \texttt{BetaSpike+TruncatedMixture}.
Perhaps the starkest difference between results generated with the \texttt{Extended} and \texttt{BetaSpike+TruncatedMixture} models is that under the \texttt{Extended}, the data even more strongly prefer isotropically distributed tilt angles over aligned, as seen by the $f_\mathrm{iso}$ peaking strongly at unity.
This again indicates no strong preference, given current data, that black hole spins are collectively nearly aligned with their orbital angular momenta.

{\fontsize{10}{10}\selectfont
\bibliography{References}
}

\end{document}

%% file: macros.tex
\newcommand{\betaPlusMixturezOne}{-0.96^{+0.07}_{-0.02}}
\newcommand{\betaPlusMixturechiOne}{0.02^{+0.05}_{-0.02}}
\newcommand{\betaPlusMixturechiNinetyNine}{0.58^{+0.18}_{-0.16}}
\newcommand{\betaPlusMixturechieffOne}{-0.20^{+0.10}_{-0.13}}
\newcommand{\betaPlusMixturechieffNinetyNine}{0.27^{+0.12}_{-0.09}}
\newcommand{\betaPlusTruncatedMixturezmindata}{-0.55^{+0.19}_{-0.22}}
\newcommand{\betaPlusTruncatedMixturezOne}{-0.53^{+0.18}_{-0.21}}
\newcommand{\betaPlusTruncatedMixturechiOne}{0.03^{+0.05}_{-0.03}}
\newcommand{\betaPlusTruncatedMixturechiNinetyNine}{0.57^{+0.19}_{-0.14}}
\newcommand{\betaPlusTruncatedMixturechieffOne}{-0.10^{+0.06}_{-0.10}}
\newcommand{\betaPlusTruncatedMixturechieffNinetyNine}{0.27^{+0.12}_{-0.09}}
\newcommand{\betaSpikePlusMixturefspike}{0.22^{+0.31}_{-0.22}}
\newcommand{\betaSpikePlusMixturezOne}{-0.96^{+0.08}_{-0.02}}
\newcommand{\betaSpikePlusMixturechiOne}{0.004^{+0.010}_{-0.003}}
\newcommand{\betaSpikePlusMixturechiNinetyNine}{0.64^{+0.20}_{-0.19}}
\newcommand{\betaSpikePlusMixturechieffOne}{-0.21^{+0.10}_{-0.13}}
\newcommand{\betaSpikePlusMixturechieffNinetyNine}{0.30^{+0.13}_{-0.10}}
\newcommand{\betaSpikePlusTruncatedMixturefspike}{0.33^{+0.27}_{-0.33}}
\newcommand{\betaSpikePlusTruncatedMixturezmindata}{-0.49^{+0.26}_{-0.26}}
\newcommand{\betaSpikePlusTruncatedMixturezOne}{-0.47^{+0.27}_{-0.25}}
\newcommand{\betaSpikePlusTruncatedMixturechiOne}{0.003^{+0.008}_{-0.002}}
\newcommand{\betaSpikePlusTruncatedMixturechiNinetyNine}{0.67^{+0.20}_{-0.21}}
\newcommand{\betaSpikePlusTruncatedMixturechieffOne}{-0.10^{+0.08}_{-0.11}}
\newcommand{\betaSpikePlusTruncatedMixturechieffNinetyNine}{0.30^{+0.15}_{-0.10}}

\newcommand{\betaSpikePlusTruncatedMixturechieffOnePercentile}{98.8\%}

\newcommand{\gwtcthreezminpercentile}{99.7\%\,}
\newcommand{\gwtctwozminpercentile}{97.1\%\,}
\newcommand{\gwtcthreezminpercentilenoniid}{90.4\%\,}
\newcommand{\gwtctwozminpercentilenoniid}{84.6\%\,}

%% file: spin_summary_macros.txt
\newcommand{\gaussianChiEffMean}{ 0.05^{+ 0.03}_{- 0.03} }
\newcommand{\gaussianChiEffStd}{ 0.09^{+ 0.03}_{- 0.04} }
\newcommand{\doubleGaussianChiEffMean}{ 0.06^{+ 0.04}_{- 0.03} }
\newcommand{\doubleGaussianChiEffStd}{ 0.12^{+ 0.04}_{- 0.08} }